\documentclass[useAMS,usenatbib,onecolumn]{mnras}
\usepackage{graphicx}
\usepackage{amsmath}
\usepackage{amssymb}
\usepackage{xspace}
\usepackage{commath}
\usepackage{times}
\usepackage{bm} 
\usepackage{hyperref}
\hypersetup{dvips, colorlinks=true, linkcolor=black, citecolor=black, filecolor=blue, urlcolor=black}

\newcommand{\bx}{\mathbf{x}}
\newcommand{\bvel}{\mathbf{v}}
\newcommand{\bT}{\bm{\theta}}
\newcommand{\bO}{\bm{\Omega}}
\newcommand{\bJ}{\mathbf{J}}
\newcommand{\bk}{\mathbf{k}}
\newcommand{\bp}{\mathbf{p}}

\newcommand{\bF}{\mathbf{F}}
\newcommand{\bD}{\mathbf{D}}
\newcommand{\bC}{\mathbf{C}}
\newcommand{\bE}{\mathbf{E}}
\newcommand{\bP}{\mathbf{P}}

\newcommand{\tF}{\widetilde{F}}
\newcommand{\teta}{\widetilde{\eta}}
\newcommand{\tphi}{\widetilde{\phi}}
\newcommand{\tbP}{\widetilde{\mathbf{P}}}
\newcommand{\tbT}{\widetilde{\mathbf{T}}}
\newcommand{\tbE}{\widetilde{\mathbf{E}}}
\newcommand{\tbS}{\widetilde{\mathbf{S}}}
\newcommand{\bI}{\mathbf{I}}

\newcommand{\wbM}{\widehat{\mathbf{M}}}
\newcommand{\wbC}{\widehat{\mathbf{C}}}
\newcommand{\wbP}{\widehat{\mathbf{P}}}
\newcommand{\wbE}{\widehat{\mathbf{E}}}
\newcommand{\wC}{\widehat{C}}

\newcommand{\pprime}{{\prime\prime}}

\newcommand{\bxp}{{\bx^\prime}}
\newcommand{\bvelp}{{\bvel^\prime}}
\newcommand{\bTp}{{\bT^\prime}}
\newcommand{\bJp}{{\bJ^\prime}}
\newcommand{\bJpp}{{\bJ^\pprime}}

\newcommand{\bkp}{{\bk^\prime}}
\newcommand{\bkpp}{{\bk^\pprime}}

\newcommand{\tp}{{t^{\prime}}}
\newcommand{\tpp}{{t^{\pprime}}}
\newcommand{\kp}{{k^{\prime}}}

\newcommand{\phip}{\phi^{\prime}}
\newcommand{\vp}{v^{\prime}}
\newcommand{\mBp}{\mathcal{B}^{\prime}}
\newcommand{\kappap}{\kappa^{\prime}}

\newcommand{\rb}{\mathrm{b}}
\newcommand{\rc}{\mathrm{c}}
\newcommand{\rd}{\mathrm{d}}
\newcommand{\rD}{\mathrm{D}}
\newcommand{\re}{\mathrm{e}}
\newcommand{\ri}{\mathrm{i}}
\newcommand{\rp}{\mathrm{p}}
\newcommand{\rP}{\mathrm{P}}
\newcommand{\rt}{\mathrm{t}}
\newcommand{\mt}{m_\rt}
\newcommand{\mb}{m_\rb}
\newcommand{\ext}{{\mathrm{ext}}}

\newcommand{\ophi}{\overline{\phi}}
\newcommand{\oM}{\overline{M}}
\newcommand{\oH}{\overline{H}}
\newcommand{\oJ}{\overline{J}}

\newcommand{\Nbody}{$N$-body\xspace}

\newcommand{\ccdot}{\! \cdot \!}
\newcommand{\veps}{\varepsilon}

\usepackage{xargs}
\newcommandx\dint[2][usedefault, addprefix=\global, 1=, 2=]{\!\!\int_{#1}^{#2}\!\! \rd}

\newcommand\deltaD{\delta_{\rD}}

\providecommand{\dfd}[3][]{\ensuremath{\mathinner{
\dfrac{\delta{^{#1}}#2}{\delta{#3^{#1}}}
}}}

\usepackage{interval}
\intervalconfig{soft open fences}

\newcommand{\eq}{equation}
\newcommand{\eqs}{equations}
\newcommand{\Eq}{Equation}
\newcommand{\Eqs}{Equations}

\usepackage{acronym}
\newacro{sma}{semi-major axis}
\newacroplural{sma}[sma's]{semi-major axis}

\newacro{MBH}{super massive black hole}
\newacroplural{MBH}[MBHs]{super massive black holes}

\newacro{BH}{black hole}
\newacroplural{BH}[BHs]{black holes}

\newacro{HMF}{Hamiltonian mean field}
\newcommand{\HMF}{\ac{HMF}}
\newacro{DC}{diffusion coefficient}
\newacroplural{DCs}{diffusion coefficients}
\newcommand{\DC}{diffusion coefficient}
\newcommand{\DCs}{diffusion coefficients}
\newacro{DF}{distribution function}
\newacroplural{DFs}{distribution functions}
\newcommand{\DF}{\ac{DF}}

\newacro{PDF}{probability distribution function}
\newacroplural{PDFs}{probability distribution functions}
\newcommand{\PDF}{\ac{PDF}}

\newacro{FP}{Fokker-Planck}
\newcommand{\FP}{\ac{FP}}
\newacro{BL}{Balescu-Lenard}
\newcommand{\BL}{\ac{BL}}

\begin{document}

\title[Relaxation in self-gravitating systems]{Relaxation in self-gravitating systems}

\author[J.-B. Fouvry \& B. Bar-Or]{Jean-Baptiste Fouvry$^{1,\star}$ and Ben Bar-Or$^{1}$
\vspace*{6pt}\\
\noindent
$^{1}$ Institute for Advanced Study, Princeton, NJ, USA\\
$^{\star}$ Hubble Fellow
}

\maketitle

\begin{abstract}
 The long timescale evolution of a self-gravitating system is generically
 driven by two-body encounters. In many cases, the motion of the particles is
 primarily governed by the mean field potential. When this potential is
 integrable, particles move on nearly fixed orbits, which can be described in
 terms of angle-action variables. The mean field potential drives fast orbital
 motions (angles) whose associated orbits (actions) are adiabatically conserved
 on short dynamical timescales. The long-term stochastic evolution of the
 actions is driven by the potential fluctuations around the mean field and in
 particular by ``resonant two-body encounters'', for which the angular
 frequencies of two particles are in resonance. We show that the stochastic
 gravitational fluctuations acting on the particles can generically be
 described by a correlated Gaussian noise. Using this approach, the so-called
 $\eta$-formalism, we derive a diffusion equation for the actions in the test
 particle limit. We show that in the appropriate limits, this diffusion
 equation is equivalent to the inhomogeneous Balescu-Lenard and Landau
 equations. This approach provides a new view of the resonant diffusion
 processes associated with long-term orbital distortions. Finally, by
 investigating the example of the Hamiltonian Mean Field Model, we show how the
 present method generically allows for alternative calculations of the
 long-term diffusion coefficients in inhomogeneous systems.
\end{abstract}


\section{Introduction}
\label{sec:intro}

The long-term evolution of self-gravitating systems has been
a long-standing subject of interest in many astrophysical contexts. Long-range
interacting systems such as stellar disks, globular clusters, and nuclear star
clusters, share indeed some fundamental
similarities. First, they are inhomogeneous systems: the dynamics of their
individual components is intricate. Second, owing to their relatively short dynamical
timescales compared to their age, these systems are also generically
dynamically relaxed so that their mean field distribution may be assumed to be
quasi-stationary. Third, these systems are also perturbed, either via external sources
(e.g., satellite infall in stellar disks or tidal forcing via an external
potential) or via self-induced fluctuations (e.g., finite${-N}$
effects). Finally, these systems are self-gravitating so that any
perturbation, either internal or external, can be amplified. These various
effects contribute to the long-term dynamics of self-gravitating systems.

A first source of diffusion is external potential
fluctuations. Neglecting collective effects (i.e.\ the ability of the system to
respond to perturbations),~\cite{BinneyLacey1988} computed the first- and
second-order \DCs\ in orbital
space.~\cite{Weinberg1993,Weinberg2001a,Weinberg2001b} emphasized the
importance of collective effects and studied the impact of the properties of
the noise processes in shaping the \DCs\@.~\cite{PichonAubert2006} presented
a time-decoupling approach to
solve the collisionless Boltzmann equation in the presence of external
perturbations, and studied the long-term evolution dark matter halos. A similar
method was presented in~\cite{FouvryPichonPrunet2015} in the context of
stellar disks. The effects of stochastic forces on long-range interacting
systems were also investigated in~\cite{Chavanis2012EPJP,Nardini2012}.

Even without external perturbations, isolated self-gravitating systems can
undergo a long-term diffusion caused by potential fluctuations arising from
their finite number of particles. Early studies of the long-term effects of
self-induced internal potential fluctuations~\citep[e.g.,][]{Chandrasekhar1942,
 Spitzer1987} relied essentially on various simplifying assumptions~\citep[see
e.g.,][for a detailed historical
account]{NelsonTremaine1999,Bar-Or+2013,Chavanis2013,Heyvaerts2017}. First, the
system was assumed to be spatially infinite and homogeneous so that individual
trajectories become simple straight lines. Second, stellar encounters are
assumed to be instantaneous and uncorrelated so that the stochastic evolution of the
stellar velocities can be assumed to be a Markov process. Third, changes in
velocity are assumed to result from a sequence of many small changes so that, by the
central limit theorem, the distribution of velocity changes can be assumed to be
Gaussian. Finally, collective effects (i.e.\ the ability of the system to
amplify perturbations via its self-gravity) were neglected. Recent progresses
in the kinetic theory of inhomogeneous self-gravitating systems have been able
to relax some of these assumptions. These developments rely on the use of
angle-action coordinates~\citep[e.g.,][]{BinneyTremaine2008} to account for the
intricate trajectories, and the use of linear response theory and the associated
matrix method~\citep{Kalnajs1976II} to account for self-gravity~\citep[see
Section~5.3.\ in][]{BinneyTremaine2008}. This led to the so-called
inhomogeneous Landau equation~\citep{PolyachenkoShukman1982,Chavanis2013} when
collective effects are neglected, and the inhomogeneous \BL\
equation~\citep{LucianiPellat1987,Mynick1988,Heyvaerts2010,Chavanis2012} when
collective effects are accounted for. Various derivations of these kinetic
equations were proposed in the
literature.~\cite{LucianiPellat1987,Chavanis2012} obtained the inhomogeneous
\BL\ equation starting from the Klimontovich equation~\citep{Klimontovich1967}.
The same kinetic equations were derived from the direct solution of the two
first equations of the BBGKY equations truncated at the order ${1/N}$
by~\cite{Heyvaerts2010} and by direct computation of the first- and
second-order \DCs\ following the Fokker-Planck
approach~\citep{Mynick1988,Chavanis2012,Heyvaerts2017}.

Subsequent studies illustrated the relevance of these kinetic theories to
various astrophysical systems.
~\cite{FouvryPichonChavanis2015,FouvryPichonMagorrianChavanis2015,FouvryPichonChavanisMonk2017},
applied the \BL\ equation to razor-thin and thickened stellar disks. These
studies emphasized in particular the importance of collective effects to hasten
the (resonant) relaxation of dynamically cold self-gravitating systems. The
same kinetic equation was also specialized to quasi-Keplerian (degenerate)
systems, such as galactic centers,
in~\cite{SridharTouma2016I,SridharTouma2016II,SridharTouma2017III,FouvryPichonMagorrian2017,FouvryPichonChavanis2017}.
Finally,~\cite{BenettiMarcos2017} used the same framework to compute the \DCs\
of the one-dimensional inhomogeneous \HMF\ model~\citep{AntoniRuffo1995}, both
with and without collective effects. In Section~\ref{sec:HMF}, we will consider
the same \HMF\ model to illustrate the new paradigm introduced in the present work. In
all these contexts, the inhomogeneous \BL\ equation was put forward as a
powerful new kinetic equation allowing for detailed and quantitative
descriptions of the long-term self-induced evolution of self-gravitating
systems. This framework offers in particular alternative probes of complex
long-term regimes in complement to traditional \Nbody\ methods. Such kinetic
theories may indeed be applied to a wide range of astrophysical scales from the
cusp-core transformation of dark halos, through the processes of radial
migration or thickening in stellar disks, all the way down to the resonant
relaxation of galactic nuclei~\citep{RauchTremaine1996, Hopman+2006,
 Bar-Or+2016}.

The inhomogeneous Landau equation and the more general inhomogeneous \BL\
equation are, to date, the most general diffusion equations to describe the
long-term evolution of inhomogeneous self-gravitating multi-component
systems. Nevertheless, in practice, calculating the associated \DCs\ is not an
easy task. The \DCs\ are expressed as an infinite sum over resonances and
involve an integral over action space. As a result, in many systems these \DCs\
have to be evaluated numerically to some finite order. An alternative approach
was put forward in~\cite{BarOr2014} to study the resonant relaxation of stars
around a massive black hole, in a regime where general relativistic effects
play an important role.~\cite{BarOr2014} introduced the
$\eta$-formalism~\citep[see the review by][]{Alexander2015}, in which the
intricate orbital motion of a test star is perturbed by an external stochastic
noise acting on it. The orbital evolution of the test star is then described by
a \FP\ equation where the \DCs\ depend on the power density of the noise. In
fact, as we will show here, these \DCs\ are equivalent to the Landau or \BL\
ones in the appropriate regime. Even within this approach, the associated \DCs\
remain hard to evaluate, and the difficulty lies here in the evaluation of the
noise term. However, because this noise term has a physical meaning, it can be
approximated via considerations on the typical timescales of the system. This
approach was used in~\cite{BarOr2014} to explain the so-called ``Schwarzschild
Barrier'', which was observed in \Nbody\ simulations~\citep{Merritt+2011}. The
same method was also used in~\cite{Bar-Or+2016} to estimate the rate with which
compact objects, like stellar black holes, are driven to strongly interact with
a central massive black hole to produce gravitational waves.

In the present paper, we revisit these different kinetic equations (for both
external and self-induced evolution), and emphasize their strong connections.
To do so, we follow and generalize the $\eta$-formalism. Our method underlines
especially the importance of the stochasticity of potential fluctuations in
sourcing the long-term diffusion of self-gravitating systems. We show in
particular how both external and self-induced potential fluctuations can be
reconciled within the same framework. The paper is organized as follows. In
Section~\ref{sec:stochastic_h}, we present the key ingredients of the
$\eta$-formalism. In Section~\ref{sec:diffusion}, we derive the inhomogeneous
\DCs\ describing the orbital diffusion of a test particle
induced by an external stochastic bath. In Section~\ref{sec:Recovery}, we show
how these approaches allow for the explicit and exact recovery of the
inhomogeneous Landau and \BL\ \DCs\@. In Section~\ref{sec:DynFric}, we briefly recover the so-called
friction force by polarization by accounting for the back-reaction of the test
particle on the bath particles, an essential component of any self-consistent
diffusion equation. In Section~\ref{sec:dFPeq}, we detail how the external
dressed diffusion equation from~\cite{BinneyLacey1988,Weinberg2001a} may also
be recovered.In Section~\ref{sec:HMF}, we demonstrate and emphasize the
practical relevance of this method by recovering the inhomogeneous \DCs\
of the \HMF\ model. Finally, we summarize our results in
Section~\ref{sec:conclusion}.

\section{Stochastic Hamiltonian}
\label{sec:stochastic_h}

In this first section, we consider the long-term evolution of a test particle
embedded in an external ``bath'' of ${ N \gg 1 }$ particles. As discussed
later, here, external means that there are no back-reactions of the test particle
onto the evolution of the bath particles. It is a completely deterministic
system, that is the evolution of the test particle depends only on the
positions ${\bx_i}$ and velocities ${\bvel_i}$ of all the ${N}$ bath particles
at some time $t_0$. Nevertheless, the complex motion of the bath particles
exerts a force on the test particle that in the large $N$ limit can be regarded
as a random process. This is one of the key considerations of the upcoming
calculations. As a result, the potential induced by the bath, which depends on
the exact motion of the bath particles, can be replaced by a stochastic
potential characterized by its statistical correlations. The motion of the test
particle may then be described by a stochastic Hamiltonian.

First, let us specify the properties of the considered bath. We assume that it
takes the form of a generic \Nbody\ system governed by a long-range pairwise
interaction potential ${ \psi(\bx, \bxp)\propto 1/{|\bx - \bxp|}^\alpha
}$. This interaction is taken to be long-range, so that ${\alpha \leq d}$, with
$d$ the dimensionality of the system~\cite[e.g,][]{Campa+2009}. In the
gravitational context, one has ${ \psi(\bx, \bxp)= -G/|\bx - \bxp| }$, and
therefore ${\alpha=1}$. The Hamiltonian driving the evolution of the bath
particles is then given by
\begin{equation}
  \label{eq:hamiltonian_rs}
  H_{\rb} (\{\bx_i\},\{\bp_i\}) = 
  \sum_{i=1}^{N} \frac{{|\bp_i|}^2}{2 m_i}\,\,
  + \!\!\sum_{1\le i < j \le N}\!\!\!\!\!\! m_i \, m_j \, \psi(\bx_i,\bx_j) 
  + \sum_{i=1}^{N} m_i \, U_\ext(\bx_i),
\end{equation}
where $m_{i}$ is the mass of the $i$-th bath particle. The location in phase
space of this particle at time $t$ is given by the canonical coordinates
${ (\bx_{i}, \bp_{i} = m_{i} \bvel_i) }$. In addition to pairwise
interactions, the Hamiltonian in \eq~\eqref{eq:hamiltonian_rs} can also
involve an external stationary time-independent potential, $U_{\rm ext}$. For
simplicity, we will assume that the bath is of total mass $M$, and that all the
bath particles have the same individual mass ${ \mb = M/N }$.

In the present work, we are interested in the long-term behavior of one given
test particle of mass $\mt$, orbiting in the potential induced by the
bath. The evolution of this test particle is governed by the time-dependent
specific Hamiltonian
\begin{equation}
  \label{eq:hamiltonian_1rs}
  H_{\rt} (\bvel,\bx,t) = \frac{{|\bvel|}^2}{2} 
  + \sum_{n = 1}^N \mb \, \psi(\bx,\bx_n(t)) 
  + U_\ext(\bx),
\end{equation}
where the location of the test particle in phase space is given by
${ (\bx,\bvel) }$, and the sum over $n$ runs over all the bath particles. By
averaging the Hamiltonian over all possible realizations of the bath, we obtain
the mean Hamiltonian of the test particle
\begin{equation}
  \label{eq:H_0}
  H_0(\bx,\bvel,t) =
  \frac{{|\bvel|}^2}{2}
  + \!\!\int\!\!  \rd \bxp \rho_{\rb} (\bxp,t) \, \psi(\bx,\bxp)  
  + U_\ext(\bx),
\end{equation}
where $\rho_{\rb}$ is the mean mass density of the bath, which satisfies
${ \int\!\rd \bxp \rho_{\rb} (\bxp,t) = M }$.

Here, we assume that mean Hamiltonian $H_{0}$ is integrable and therefore there
exist angle-action coordinates
${ (\bT, \bJ) }$~\citep{BinneyTremaine2008}, such that ${ H_0 = H_0 (\bJ) }$
depends only on the action $\bJ$. Let us recall that the actions $\bJ$ are
integrals of motion for $H_0$, while the angles $\bT$ are ${2\pi}$-periodic and
increase linearly in time with the frequency
${ \bO (\bJ) = \partial H_{0} / \partial \bJ }$. In situations where $U_\ext$
is the dominant component of the mean Hamiltonian $H_{0}$, the angle-action
coordinates ${ (\bT, \bJ) }$ are essentially imposed by $U_\ext$. This occurs
for example in quasi-Keplerian systems such as galactic nuclei, where most of
the potential is imposed by the central super massive black hole. In such a
case, it might be that $H_0$ has a small $\bT$ dependence, that we ignore at
this point. Finally, following Jeans theorem, we assume that the mean
distribution of the bath particles can be characterized by the \DF\
${ F_{\rb} (\bJ) }$ which depends only on actions. In all the subsequent
calculations, we follow the normalization convention
${ \int \! \rd \bx \rd \bvel \, F_{\rb} \!=\! M}$.

Following these assumptions, the test particle's specific Hamiltonian from
\eq~\eqref{eq:hamiltonian_1rs} can be rewritten as
\begin{equation}
  \label{eq:hamiltonian}
  H_{\rt} (\bT, \bJ, t) =
  H_0(\bJ)
  + \eta (\bT, \bJ, t),
\end{equation}
where ${ \eta (\bT, \bJ, t) }$ accounts for the potential fluctuations around the mean
Hamiltonian, which depend on time through the complicated motion of the bath particles. In the
statistical limit, where ${ N \gg 1 }$, these can be considered as stochastic potential
fluctuations, which satisfy ${ \langle \eta \rangle \!=\! 0 }$, where
${ \langle \; \cdot \; \rangle }$ stands for the ensemble average over
all possible realizations of the bath. Following the normalization convention of $F_{\rb}$,
the ensemble average amounts here to
\begin{equation}
  \label{eq:definition_ensemble_average}
  \langle X \rangle 
  \equiv 
  \! {(N\mb)}^{-N} \dint \bT_1 \rd \bJ_1 \, F_{\rb} (\bJ_1) \, 
  \ldots \, \rd \bT_{N} \rd \bJ_N \, F_{\rb} (\bJ_{N}) \, X.
\end{equation}
If $X$ depends only on one single particle, then the ensemble average reduces
to
${ \langle X \rangle = \int \! \rd \bJ \rd \bT F_{\rb} (\bJ) X /(m_\rb N) }$.

Relying on the ${2\pi}$-periodicity of the angles $\bT$, the pairwise
interaction potential $\psi$ can be decomposed into Fourier elements so that
\begin{equation}
  \label{eq:psi_exp}
  \psi(\bT, \bJ, \bTp, \bJp) =
  \sum_{\bk,\bkp} \psi_{\bk\bkp}(\bJ,\bJp)
  \, \re^{ \ri (\bk\cdot\bT-\bkp\cdot\bTp)}.
\end{equation}
The Fourier coefficients ${ \psi_{\bk\bkp} (\bJ, \bJp) }$ are called the bare
susceptibility coefficients~\citep{LyndenBell1994,Pichon1994,Chavanis2012}, and
are given by
\begin{equation}
  \label{eq:psi_basis}
  \psi_{\bk\bkp}(\bJ,\bJp) =
  \!\! \int \!\! \frac{\rd \bT}{{(2 \pi)}^{d}} \frac{\rd \bTp}{{(2 \pi)}^{d}} \,
  \psi(\bT,\bJ,\bTp,\bJp) \, \re^{- \ri (\bk\cdot\bT-\bkp\cdot\bTp)}
 ,
\end{equation}
with ${ \bk, \bkp \in \mathbb{Z}^{d} }$. Since $\psi$ is real, the bare
susceptibility coefficients satisfy
${ \psi_{\bk\bkp}^*(\bJ,\bJp) \!=\! \psi_{-\bk-\bkp}(\bJ,\bJp) \!=\!
  \psi_{\bkp\bk}(\bJp,\bJ) }$. Following this decomposition, the test
particle's Hamiltonian in \eq~\eqref{eq:hamiltonian} may finally be written
as
\begin{equation}
  \label{eq:H_expn}
  H_{\rt} (\bT , \bJ , t)  = H_0 (\bJ) + \sum_\bk \re^{ \ri \bk\cdot\bT} \, \eta_\bk (\bJ, t),
\end{equation}
where we introduced the stochastic potential fluctuations
${ \eta_{\bk} (\bJ, t) }$ as
\begin{align}
  \label{eq:eta_def}
  \eta_\bk(\bJ, t) \equiv {} 
  &  
    \!\! \int \!\! \frac{\rd \bT}{{(2 \pi)}^d} \,
    \eta (\bT, \bJ, t) \, \re^{- \ri \bk \cdot \bT} 
    \nonumber  \\
  = {} 
  &
    \sum_{n=1}^{N}  \mb \sum_{\bkp}
    \psi_{\bk\bkp}(\bJ,\bJ_n(t)) \, \re^{ - \ri \bkp\cdot\bT_n(t)} - \dint
    \bJp F_\rb(\bJp) \psi_{\bk0}(\bJ,\bJp),
\end{align}
which all satisfy ${ \langle \eta_\bk(\bJ, t) \rangle = 0 }$. Since
\eq~\eqref{eq:eta_def} involves the angles ${ \bT_{n} (t) }$ of the bath
particles, these perturbations fluctuate on the dynamical timescale associated
with the mean field potential. The dynamics of the test particle is governed by
the Hamiltonian in \eq~\eqref{eq:H_expn} and the associated evolution equations
are given by Hamilton's equations, which take the simple form
\begin{equation}
  \label{eq:J_dot}
  \frac{\rd \bJ}{\rd t} = - \dpd{H_{\rt}}{\bT} = 
  - \ri\sum_\bk \bk \, \re^{ \ri \bk\cdot\bT} \eta_\bk(\bJ, t),
\end{equation}
and
\begin{equation}
  \label{eq:theta_dot}
  \frac{\rd \bT}{\rd t} = \dpd{H_{\rt}}{\bJ} = \dpd{}{\bJ}\!
  H_0(\bJ) + \sum_\bk \re^{ \ri \bk\cdot\bT} \dpd{}{\bJ} \eta_\bk(\bJ, t).
\end{equation}
\Eq~\eqref{eq:J_dot} describes the long-term evolution of the test particle's
action $\bJ$. As the mean field Hamiltonian $H_0$ is integrable, the evolution
in $\bJ$ is only sourced by the fluctuations in the potential and will be the
starting point of the derivation of the associated diffusion equation. This is
considered in the next section.

\section{Diffusion coefficients}
\label{sec:diffusion}

In this section, we investigate how the action, $\bJ$, of a zero mass test
particle, diffuses under the effect of the potential fluctuations induced by the
bath particles. For now, we consider an external bath. The test particle
is of zero mass, so that the motion of the bath particles is independent of the test
particle. In Section~\ref{sec:DynFric}, we will briefly relax this assumption,
and recover how a test particle with a finite mass perturbs the orbits of the
bath particles, which in turn back-reacts on the test particle itself giving
rise to the friction force by polarization.

For a given realization of bath particles, i.e.\ for a given set of
trajectories ${ \{ \bT_n (t), \bJ_n (t) \} }$, the motion of the test particle
is uniquely determined by the equations of motion (\eqs~\eqref{eq:J_dot}
and~\eqref{eq:theta_dot}), and the test particle's initial conditions
${ (\bT (0), \bJ (0)) }$. Here however, we will not consider the motion of the
test particle for one specific bath realization, but will rather try to
describe the averaged evolution of the test particle over many bath
realizations. As a consequence, the bath is not described by a set of exact
trajectories but by their associated statistical properties. To do so, we
assume that the smooth mean distribution of the bath particles
${ F_{\rb} (\bJ) }$ and the statistics of the potential fluctuations are time
independent. In such a limit, \eqs~\eqref{eq:J_dot} and~\eqref{eq:theta_dot}
can be treated as a set of Langevin-type stochastic equations, where the stochastic
potential fluctuations, ${ \eta_{\bk} (\bJ, t) }$, act as a noise
terms. Following \eq~\eqref{eq:eta_def}, this noise is of zero mean and
corresponds to the joint contribution from the $N$ bath particles. We assume that
${ \eta_{\bk} (\bJ, t) }$ are random stationary Gaussian noise terms, which can
be uniquely characterized by their correlation functions
\begin{equation}
  \label{eq:C_def}
  C_{\bk\bkp}(\bJ, \bJp, t-\tp)  \equiv  \left\langle \eta_\bk(\bJ,t) \,
  \eta_\bkp^{*} (\bJp,\tp)
  \right\rangle .
\end{equation}
We already note that
${ C_{\bk\bkp}^{*} (\bJ,\bJp,t - \tp) = C_{\bkp\bk}(\bJp,\bJ,\tp - t) }$. As
will be emphasized in the upcoming calculations, the correlation functions of
the stochastic potential fluctuations determine the long-term diffusion of the
test particle's action $\bJ$. One key assumption in the derivation
of a Fokker-Planck type diffusion equation is that ${ \eta_{\bk} (\bJ, t) }$
follows a Gaussian distribution. The associated diffusion equation then only
involves the first two moments of ${ \eta_{\bk} (\bJ, t)}$.  We note that the expression
of ${ \eta_{\bk} (\bJ, t)}$ (see \eq~\eqref{eq:eta_def}) requires to sum over
all the bath particles. When collective effects are ignored (i.e.\ bath particles
only interact through the mean field), the
trajectories of the different bath particles are independent, as illustrated in \eq~\eqref{eq:deltapsi_Landau}.
Yet, when collective
effects are accounted for (i.e.\ bath particles interact with one another),
their individual trajectories are correlated. Even in that dressed regime,
as shown in \eq~\eqref{eq:deltapsi_BL}
(to compare with \eq~\eqref{eq:deltapsi_Landau}),
using the dressed pairwise interaction potential, ${ \eta_{\bk} (\bJ, t) }$ can be
rewritten as a sole function of the initial phase-space coordinates of the bath
particles, which are statistically independent. As a result, in both cases, provided that
${ \langle \eta_{\bk} (\bJ, t) \, \eta_{\bk^\prime}^* (\bJ^\prime, t) \rangle }$ is
finite, the noise ${\eta_{\bk} (\bJ, t)}$ can be seen as the sum of many independent random
variables of zero mean and finite variance. Owing to the central limit theorem,
this noise can then be assumed to be Gaussian, justifying the derivation of a Fokker-Planck
type diffusion equation.
Let us however point that, for the standard two-body
relaxation of stellar systems,
${ \langle \eta_{\bk} (\bJ, t) \, \eta_{\bk^\prime}^* (\bJ^\prime, t) \rangle}$ is
diverges logarithmically, as manifested by the Coulomb logarithm.
This is a sign that higher order moments of the noise can be important,
and that the Fokker-Plank approximation may break down on timescales
shorter than the relaxation time~\citep{Bar-Or+2013}. These effects will not be considered
in our present inhomogeneous approach.

We now follow and generalize the derivation of~\cite{BarOr2014} to obtain the
diffusion equation associated with the Langevin \eqs~\eqref{eq:J_dot}
and~\eqref{eq:theta_dot}.  For a given realization of the bath, the trajectory
of the test particle in action space can be formally described by
\begin{equation}
  \label{eq:varphi_def}
  \varphi (\bJ, t) \equiv \deltaD(\bJ \!-\! \bJ (t)),
\end{equation}
where ${ \bJ (t) }$ is the action of the test particle at time $t$,
and $\deltaD$ stands for a Dirac delta.
The function ${ \varphi (\bJ, t) }$ satisfies the continuity (Liouville) equation
\begin{align}
  \label{eq:phi_dot}
  \dpd{}{t} \varphi (\bJ,t) = {}
  & 
    -\dpd{}{\bJ} \ccdot \bigg[ \dot{\bJ} (\bJ,\bT(t), t) \, \varphi(\bJ,t) \bigg]
    \nonumber \\
  = {}
  & 
    \ri \dpd{}{\bJ} \ccdot \bigg[ \sum_\bk \bk \,
    \re^{ \ri \bk\cdot\bT(t)} \, \eta_\bk(\bJ, t) \, \varphi(\bJ,t) \bigg] ,
\end{align}
where we used \eq~\eqref{eq:J_dot} to obtain the value of $\dot{\bJ}$.

Rather than investigating the particular trajectory of a test particle in
action space, one can consider the statistical evolution of a collection of
test particles with different initial $\bT$ and different bath
realizations. Let us therefore describe the statistical distribution of the
actions $\bJ$ as a function of time by the \PDF\
\begin{equation}
  \label{eq:definition_PDF}
  P (\bJ,t) \equiv \langle \varphi(\bJ,t) \rangle ,
\end{equation}
where ${ \langle \; \cdot \; \rangle }$ is the average over the initial
conditions of the bath particles (ensemble average) and over the initial
conditions $\bT_{0}$ of the test particle.
As given by \eq~\eqref{eq:phi_dot}, the
evolution of ${ P (\bJ,t) }$ is governed by
\begin{equation}
  \label{eq:F_dot}
  \dpd{}{t} P (\bJ,t) = \ri \dpd{}{\bJ} \ccdot \bigg[ \sum_\bk \bk
  \left\langle \eta_\bk(\bJ, t) \, \re^{ \ri \bk\cdot\bT(t)}  \, \varphi(\bJ,t)
  \right\rangle \bigg] .
\end{equation}
Evaluating the ensemble averaged term appearing in the r.h.s.\ of
\eq~\eqref{eq:F_dot} is the purpose of the next sections.

\subsection{Novikov's theorem}
\label{sec:Novikov}

The r.h.s.\ of \eq~\eqref{eq:F_dot} involves not only the stochastic potential
fluctuations ${ \eta_{\bk} (\bJ, t) }$, but also the detailed trajectory
${ (\bT(t), \bJ (t)) }$ of the test particle in angle-action space. As shown by
\eqs~\eqref{eq:J_dot} and~\eqref{eq:theta_dot}, the exact trajectory of the
test particle in phase space is itself a function of the stochastic noise
${\eta_{\bk} (\bJ, t)}$. One should therefore interpret the r.h.s.\ of
\eq~\eqref{eq:F_dot} as being sourced by the correlation of the noise
${ \eta_{\bk} (\bJ, t) }$ with ${ \re^{\ri\bk\cdot\bT (t)}\varphi(\bJ,t) }$,
which is a functional of the noise of the generic form
${ R [ \eta](\bJ, t) }$. The difficulty here amounts to evaluating the
correlation of a noise with a functional of itself. This
calculation is made all the more intricate because the noise
${ \eta_{\bk} (\bJ, t) }$ is spatially extended~\citep{Garcia1999}, i.e.\ it
depends on both time and location in action space.

Fortunately, relying on the assumption that the ${ \eta_{\bk} (\bJ, t) }$ are random
stationary Gaussian processes of zero mean, correlations of the form
${ \left\langle \eta_{\bk} (\bJ, t) \, R [\eta](\bJp, \tp) \right\rangle }$
can be computed by Novikov's theorem~\citep{Novikov1965} generalized for
spatially extended noises~\citep{Garcia1999}. Novikov's theorem generically
allows us to write
\begin{equation}
  \label{eq:novikov}
  \left\langle \eta_\bk(\bJ, t) \, R[\eta](\bJp, \tp) \right\rangle =
  \sum_{\bkpp}
  \int_{0}^{t} \!\! \rd \tpp \dint \bJpp \, \langle \eta_\bk(\bJ,
  t) \, \eta_{\bkpp}^*(\bJpp, \tpp)  \rangle
  \left \langle
    \dfd{ R[\eta](\bJp, \tp)}{\eta_{\bkpp}^*(\bJpp, \tpp)}
  \right \rangle ,
\end{equation}
where the conjugate was introduced for later convenience and ${ \delta R
  [\eta](\bJp, \tp) / \delta \eta_{\bkpp}^{*} (\bJpp, \tpp) }$ stands for the
functional derivative of ${ R[\eta](\bJp,\tp) }$ w.r.t.\ the noise ${
  \eta_{\bkpp}^{*} (\bJpp, \tpp) }$. \Eq~\eqref{eq:novikov} should be
understood as follows. The l.h.s.\ of \eq~\eqref{eq:novikov} aims at computing
the correlation between the noise ${ \eta_{\bk} (\bJ, t) }$, evaluated at the
location $\bJ$ and time $t$, with a functional of the noise ${ R [\eta](\bJp,
  \tp) }$ evaluated at the location $\bJp$ and time $\tp$, which can depend on
the noise at any past time ${ \tpp < \tp }$ and any location $\bJpp$. Novikov's
theorem states then that this correlation is given by the joint contributions
from all the different noise terms ${ \eta_{\bkpp}^{*} (\bJpp, \tpp) }$ (via
the sum $\sum_{\bkpp}$) for all past values (via the integration ${ \int \rd
  \tpp }$) and for all locations (via the integration ${ \int \rd \bJpp }$) of
the correlation between ${ \eta_{\bk} (\bJ, t) }$ and ${ \eta_{\bkpp}^{*}
  (\bJpp, \tpp) }$ (via the correlation ${ \left\langle \eta_{\bk} (\bJ, t)
  \, \eta_{\bkpp}^{*} (\bJpp, \tpp) \right\rangle }$) multiplied by the
functional gradient ${ \delta R [\eta] (\bJp, \tp) / \delta \eta_{\bkpp}^{*}
  (\bJpp, \tpp) }$. This functional gradient describes how much the value of
${ R [\eta] (\bJp, \tp) }$ varies as a result of a modification of the noise
${ \eta_{\bkpp}^{*} (\bJpp, \tpp) }$ at the time $\tpp$ and location $\bJpp$.
As a summary, Novikov's theorem states that the correlation between the noise
and a functional of itself, scales qualitatively like the product of the noise
correlation function and the response of the functional ${ R [ \eta](\bJ, t) }$
to changes in the noise.

When applied to
the r.h.s.\ of \eq~\eqref{eq:F_dot}, Novikov's theorem yields terms of the form
\begin{equation}
  \label{eq:eta_phi}
  \left\langle
    \eta_\bk(\bJ, t) \, \re^{ \ri \bk\cdot\bT(t)} \, \varphi(\bJ,t)
  \right\rangle =
  \sum_\bkp
  \int_0^t \!\! \rd \tp \dint \bJp \,
  C_{\bk\bkp}(\bJ,\bJp,t-\tp)
  \left\langle \re^{ \ri \bk\cdot\bT(t)} \left[
      \ri \bk\ccdot \dfd{\bT(t)}{\eta_\bkp^*(\bJp, \tp)} -
      \dfd{\bJ(t)}{\eta_\bkp^*(\bJp, \tp)}  \ccdot
      \dpd{}{\bJ}
    \right ]
    \varphi(\bJ,t)
  \right\rangle ,
\end{equation}
where the correlation function ${ C_{\bk\bkp} (\bJ,\bJp, t - \tp) }$ has been
introduced in \eq~\eqref{eq:C_def}. To obtain \eq~\eqref{eq:eta_phi}, we relied
on the relation ${ \partial \varphi (\bJ, t) / \partial \bJ (t) \!=\! -
  \partial \varphi (\bJ, t) / \partial \bJ }$, for ${ \varphi (\bJ, t) }$
given by \eq~\eqref{eq:varphi_def}. We also used the chain rule for
functional derivatives to obtain
\begin{equation}
  \label{eq:d_phi}
  \dfd{\varphi(\bJ,t)}{\eta_{\bkp}^{*} (\bJp, \tp)} =
  - \dfd{\bJ(t)}{\eta_{\bkp}^{*} (\bJp, \tp)} 
  \ccdot  \dpd{\varphi(\bJ,t)}{\bJ} ,
\end{equation}
and
\begin{equation}
  \label{eq:d_etheta}
  \dfd{\big[ \re^{\ri \bk \cdot \bT (t)} \big]}{\eta_{\bkp}^{*} (\bJp, \tp)} = 
  \ri \re^{\ri \bk \cdot \bT (t)} \, \bk 
  \ccdot \dfd{\bT(t)}{\eta_{\bkp}^{*}(\bJp, \tp)} .
\end{equation}
We note that \eq~\eqref{eq:eta_phi} involves the so-called response
functions ${ \delta \bJ(t) / \delta \eta_\bkp^*(\bJp, \tp) }$ and
${ \delta \bT(t) / \delta \eta_\bkp^*(\bJp, \tp) }$, which describe how the
position ${ (\bT (t), \bJ (t)) }$ of the test particle at time $t$ changes as
one varies the noise term ${ \eta_{\bkp}^{*} (\bJp, \tp) }$ felt by the test
particle as it arrived at $\bJp$ at time $\tp$. In the next section we
proceed to compute these response functions.

\subsection{Response functions}
\label{sec:responsefunctions}

The equations of motion (\eqs~\eqref{eq:J_dot} and~\eqref{eq:theta_dot})
describe the evolution of the test particle in angle-action space. These
equations can be explicitly integrated in time to obtain
\begin{equation}
  \label{eq:J_t}
  \bJ(t) = \bJ_0  - \ri \sum_\bk \bk \dint[0][t]s \, 
  \re^{\ri \bk\cdot\bT(s)} \, \eta_\bk(\bJ(s), s) ,
\end{equation}
and
\begin{equation}
  \label{eq:theta_t}
  \bT(t) = 
  \bT_0 + \dint[0][t]s \dint \bJ\,\deltaD(\bJ-\bJ(s)) \dpd{H_0 (\bJ)}{\bJ}
  + \sum_\bk \dint[0][t]s \, \re^{ \ri \bk\cdot\bT(s)}
  \dint \bJ\, \deltaD(\bJ-\bJ(s)) \dpd{}{\bJ}
  \eta_\bk(\bJ, s) ,
\end{equation}
where ${ (\bT_{0}, \bJ_{0}) }$ is the initial position of the test
particle at time ${ t \!=\! 0 }$. In \eq~\eqref{eq:theta_t}, we introduced
terms of the form
\begin{equation}
\label{explicit_time_derivative}
\int \!\! \rd \bJ \, \deltaD (\bJ - \bJ(s)) \, \frac{\partial X (\bJ)}{\partial \bJ} = {\bigg[ \frac{\partial X (\bJ)}{\partial \bJ} \bigg]}_{\bJ = \bJ (s)} ,
\end{equation}
to express explicitly the derivatives w.r.t.\ a time-dependent variable.
The response function of the action ${ \bJ (t) }$ is then calculated by taking the
functional derivative of \eq~\eqref{eq:J_t}, and one gets
\begin{equation}
  \label{eq:res_J_temp}
  \dfd{\bJ (t)}{\eta_{\bkp}^{*} (\bJp, \tp)} =
  - \ri \sum_{\bk} \bk\dint[\tp][t] s \, 
  \dfd{\big[ \re^{\ri \bk \cdot \bT (t)} \big] }{\eta_{\bkp}^{*} (\bJp, \tp)} \eta_{\bk} (\bJ (s), s)
  - \ri \sum_{\bk} \bk\dint[\tp][t] s \, \re^{\ri \bk \cdot \bT (s)}
  \dfd{\eta_{\bk} (\bJ (s), s)}{\eta_{\bkp}^{*} (\bJp, \tp)}
 ,
\end{equation}
where the limits of the time integration illustrate that the noise at time $\tp$
can only affect the system at later times ${ s \!\geq\! \tp }$.

Applying the chain rule to the last term in \eq~\eqref{eq:res_J_temp}, we obtain
\begin{equation}
  \label{eq:calc_grad_2}
  \dfd{\eta_{\bk} (\bJ (s), s)}{\eta_{\bkp}^{*} (\bJp, \tp)} =
  \dfd{\bJ (s)}{\eta_{\bkp}^{*} (\bJp, \tp)} \ccdot
  {\bigg[\dpd{\eta_{\bk} (\bJ, s)}{\bJ}\bigg]}_{\bJ = \bJ (s)} \!\! +
  {\bigg[ \dfd{\eta_{\bk} (\bJ, s)}{\eta_{\bkp}^{*} (\bJp, \tp)} \bigg]}_{\bJ=\bJ (s)} ,
\end{equation}
where in \eq~\eqref{eq:calc_grad_2}, the first term comes from the variation of
the location ${ \bJ (s) }$ of the test particle at time $s$ as one varies the
noise ${ \eta_{\bkp}^{*} (\bJp, \tp) }$. The second term comes from the
variations of the stochastic noise term ${\eta_{\bk} (\bJ, t) }$ itself, as
one varies the noise ${ \eta_{\bkp}^{*} (\bJp, \tp) }$. This second term may
be explicitly computed using the fundamental relation
\begin{equation}
  \label{eq:delta_eta}
  \dfd{\eta_{\bk}(\bJ, t)}{\eta_\bkp^*(\bJp,
    \tp)} = \deltaD(t-\tp) \, \deltaD(\bJ-\bJp) \, \delta_{-\bk \bkp} ,
\end{equation}
which is a direct consequence of Novikov's theorem, when applied to
${ R[\eta] (\bJp,\tp) = \eta_\bkp^*(\bJp,\tp) }$. Combining these results,
one can finally rewrite \eq~\eqref{eq:res_J_temp} as
\begin{align}
  \label{eq:res_J}
  \dfd{\bJ (t)}{\eta_{\bkp}^{*}(\bJp, \tp)} = {} 
  & 
    \ri \bkp \, \re^{- \ri \bkp \cdot \bT (\tp)} \, \deltaD(\bJp \!-\! \bJ (\tp))
    \nonumber  \\
  & 
    - \ri \sum_{\bk} \bk\dint[\tp][t] s \dint \bJ \, \deltaD(\bJ-\bJ(s)) \, 
    \re^{\ri \bk \cdot \bT (s)}
    \bigg[  
    \, \ri \bk \ccdot \dfd{\bT(s)}{\eta_{\bkp}^{*} (\bJp, \tp)} \, 
    + \dfd{\bJ(s)}{\eta_{\bkp}^{*} (\bJp, \tp)} \ccdot
    \dpd{}{\bJ}
    \bigg]\eta_{\bk} (\bJ, s) .
\end{align}

Starting from \eq~\eqref{eq:theta_t}, one may follow a similar procedure to
compute the response function
${ \delta \bT (t) / \delta \eta_{\bkp}^{*} (\bJp, \tp) }$, where in addition
to \eqs~\eqref{eq:d_etheta} and~\eqref{eq:delta_eta}, we also use the fact that
the functional derivative of ${ \!\int\! \rd \bJ \, \deltaD(\bJ-\bJ(t)) X(\bJ)}$
w.r.t.\ the noise is obtained by the chain rule and an integration by parts,
so that
\begin{equation}
  \label{eq:deltaD_eta}
  \dint \bJ \dfd{\big[\deltaD(\bJ-\bJ(t))\big]}{\eta_\bkp^*(\bJp,\tp)} X(\bJ)
  =\dint \bJ \, \deltaD(\bJ-\bJ(t)) \dfd{\bJ(t)}{\eta_\bkp^*(\bJp,\tp)} \ccdot
  \dpd{X(\bJ)}{\bJ}
  .
\end{equation}
Thus, the functional derivative of \eq~\eqref{eq:theta_t} w.r.t.\ the noise
is
\begin{align}
  \label{eq:res_theta}
  \dfd{\bT (t)}{\eta_{\bkp}^{*} (\bJp, \tp)} = {} 
  & -  \re^{-\ri \bkp \cdot \bT (\tp)} \dpd{}{ \bJp} \deltaD  (\bJp - \bJ (\tp))
    \nonumber \\
  & +\dint[\tp][t] s \dint\bJ\,\deltaD(\bJ-\bJ(s)) \,
    \dfd{\bJ (s)}{\eta_{\bkp}^{*} (\bJp, \tp)}\ccdot
    \dpd{}{\bJ}\!
    \dpd{H_0 (\bJ)}{\bJ}
    \nonumber
  \\
  & + \sum_{\bk}\dint[\tp][t] s \dint\bJ\,\deltaD(\bJ-\bJ(s)) \,
    \re^{\ri \bk \cdot \bT (s)} \,
    \bigg[
    \ri \bk \ccdot\dfd{\bT (s)}{\eta_{\bkp}^{*} (\bJp, \tp)}   +
    \dfd{\bJ (s)}{\eta_{\bkp}^{*} (\bJp, \tp)}
    \ccdot\dpd{}{\bJ}
    \bigg]\dpd{\eta_{\bk} (\bJ, s)}{\bJ} .
\end{align}
\Eqs~\eqref{eq:res_J} and~\eqref{eq:res_theta} are the important results
of this section. These response functions express how the location
${ (\bT (t), \bJ (t)) }$ of the test particle at time $t$ in angle-action space
is affected by changes in the stochastic perturbation
${ \eta_{\bkp}^{*} (\bJp, \tp) }$. The first term in each of these equations is
the variation of the trajectory due to variation of the noise itself along the
trajectory. The other integral terms describe the variations of the trajectory due
to the fact that the test particle sees a different noise along the modified
trajectory. One should note that these equations are not closed, as they
depend on the noise, both directly and indirectly through the response
functions and the trajectory, ${ (\bT (t), \bJ (t)) }$, of the test
particle. In the upcoming section, we will show how one can truncate these
expressions in some specific regimes. First, in Section~\ref{sec:markov}, we will
consider the Markovian limit, for which the noise is assumed to be uncorrelated
in time. Then, in Section~\ref{sec:correlated}, we will consider the regime where the
bath particles' evolution is dominated by a fast evolution of the angles,
defining therefore the dynamical timescale of the system.

\subsection{Markovian limit}
\label{sec:markov}

One standard way to deal with the closure problem of the response functions
in \eqs~\eqref{eq:res_J} and~\eqref{eq:res_theta} is to assume that the noise
terms can be approximated as Markovian, that is to assume that they are
uncorrelated on any relevant timescale. As we emphasize in the next sections,
this Markovian limit is, generally, inconsistent with our assumption that the
angles of the test and bath particles are driven by a mean field
potential, for which changes in ${ \eta_{\bk} (\bJ, t) }$ are on the same
timescale than changes in $\bT$. Let us nevertheless pursue here the
calculation of the diffusion equation in the Markovian limit, as it provides a
simple illustration of how one can derive a diffusion equation from Novikov's
theorem.

On long timescales (${ t \to \infty }$), the correlation
${ C_{\bk\bkp}(\bJ,\bJp,t) }$ generically decays\footnote{When ${ C_{\bk\bkp}(\bJ,\bJp,t) }$ is
oscillating, one considers the decay of the envelope. See for example the upper panels in
Figures~\ref{fig:hmf_acf}.} to some constant value
${C_{\bk\bkp}^\infty(\bJ,\bJp)}$.
The Markovian limit then amounts to the assumption
that the timescale for this decay is shorter than the timescales for
significant changes in $\bT$ and $\bJ$.
In that limit, one can then treat the $\langle \; \cdot \; \rangle$ term in \eq~\eqref{eq:eta_phi}
as constant over the decay timescale of ${ C_{\bk\bkp}(\bJ,\bJp,t) }$.
If ${ C_{\bk\bkp}^\infty(\bJ,\bJp) = 0 }$, the integrand in
\eq~\eqref{eq:eta_phi} is then dominated by the correlation function and the term
inside the angle brackets can be evaluated for ${ \tp \to t }$. Such a regime is
equivalent to the assumption that the noise terms are Markovian, i.e.\ that
they are uncorrelated in time. In that limit, one can write
\begin{equation}
  \label{eq:C_markov}
  C_{\bk\bkp}(\bJ,\bJp, t - \tp) =
  \deltaD (t - \tp) \int_{-\infty}^\infty \!\!\!\! \rd s \,
  C_{\bk\bkp}(\bJ,\bJp,s) .
\end{equation}
Because of the Dirac delta ${ \deltaD (t - \tp) }$, the response functions
appearing in \eq~\eqref{eq:eta_phi} may be evaluated for ${t = \tp}$.
\Eqs~\eqref{eq:res_J} and~\eqref{eq:res_theta} become
\begin{equation}
  \label{eq:res_J_short}
  {\bigg[\dfd{\bJ(t)}{\eta_{\bkp}^{*} (\bJp, \tp)}\bigg]}_{\tp=t} = 
  \ri \bkp \, \re^{ - \ri \bkp\cdot\bT(t)} \, \deltaD  (\bJp - \bJ(t)) ,
\end{equation}
and
\begin{equation}
  \label{eq:res_theta_short}
  {\bigg[\dfd{\bT(t)}{\eta_{\bkp}^{*} (\bJp,
      \tp)}\bigg]}_{\tp=t} = - \re^{ - \ri \bkp\cdot\bT(t)} \dpd{}{\bJp} \deltaD
  (\bJp - \bJ(t)) .
\end{equation}
These expressions may then be used in \eq~\eqref{eq:eta_phi}, so that the
diffusion \eq~\eqref{eq:F_dot} becomes
\begin{equation}
  \label{eq:F_dot_markov_tmp}
  \dpd{P(\bJ,t)}{t} = 
  \frac{1}{2}\dpd{}{\bJ} \ccdot
  \sum_{\bk,\bkp} \bk
  \int \!\! \rd \bJp
  \!\! \int_{-\infty}^\infty \!\!\!\! \rd s \,
  C_{\bk\bkp}(\bJ,\bJp,s)
  \left[
    \bk\ccdot\dpd{}{\bJp}
    +
    \bkp\ccdot\dpd{}{\bJ}
  \right]
  \left[
    \deltaD  (\bJp - \bJ)
    \left\langle
      \re^{ - \ri (\bkp-\bk)\cdot\bT(t)}
      \, \varphi(\bJ,t)
    \right\rangle
  \right] ,
\end{equation}
where the $1/2$ factor results from the integration
${\int_0^t \rd \tp \, \deltaD(t-\tp)=1/2}$.

The ensemble average ${ \langle \; \cdot \; \rangle }$ implies averaging over
${ \bT(t) }$, which is assumed to be distributed uniformly in its definition
domain. As a consequence, one has
${ \langle \re^{ - \ri (\bkp-\bk)\cdot\bT(t)} \varphi(\bJ,t) \rangle =
  \delta_{\bk\bkp}\langle\varphi(\bJ,t)\rangle }$. The differential operation
in the square brackets applied to ${ \deltaD(\bJp-\bJ) }$ gives a term of the
form
\begin{equation}
  \label{eq:F_dot_intermidet}
  \left[
    \bk\ccdot\dpd{}{\bJp}
    +
    \bkp\ccdot\dpd{}{\bJ}
  \right]\deltaD  (\bJp - \bJ) = (\bk - \bkp) \ccdot \dpd{}{\bJp}
  \deltaD(\bJp-\bJ) ,
\end{equation}
which vanishes when contracted with
${ \langle \re^{ - \ri(\bkp-\bk)\cdot\bT(t)} \varphi(\bJ,t) \rangle}$. In
\eq~\eqref{eq:F_dot_markov_tmp}, the only remaining term is of the form
\begin{equation}
\label{eq:F_dot_interm_calc}
\deltaD(\bJp-\bJ) \, \bkp \ccdot \dpd{}{\bJ} \left\langle \re^{ - \ri
    (\bkp-\bk)\cdot\bT(t)} \,\varphi(\bJ,t) \right\rangle = \deltaD(\bJp-\bJ) \, 
\bk \ccdot \dpd{}{\bJ} \, \delta_{\bk\bkp}\langle\varphi(\bJ,t)\rangle  ,
\end{equation}
so that \eq~\eqref{eq:F_dot_markov_tmp} becomes
\begin{equation}
  \label{eq:F_dot_markov}
  \dpd{P(\bJ,t)}{t}
  =
  \frac{1}{2}\dpd{}{\bJ} \ccdot
  \sum_\bk \bk
  \int_{-\infty}^\infty \!\!\!\! \rd s \,
  C_{\bk\bk}(\bJ,\bJ,s)
  \, \bk \ccdot
  \dpd{}{\bJ}
  \left\langle
    \varphi(\bJ,t)
  \right\rangle .
\end{equation}
Here, it is important to note that following the ensemble average, the test
particle's \DC\ depends only on the value of
${ C_{\bk\bkp} (\bJ, \bJp, t - \tp) }$ for identical resonance vectors
(${ \bk = \bkp }$) and identical actions (${ \bJ = \bJp }$). Recalling that the
\PDF\ of the test particle is defined as
${ P (\bJ, t) = \langle \varphi (\bJ, t) \rangle }$, the associated diffusion
equation can finally be written as
\begin{equation}
  \label{eq:fp_markov}
  \dpd{P(\bJ,t)}{t} =
  \frac{1}{2}
  \dpd{}{J_i}
  D_{ij}(\bJ)\dpd{}{J_j}
  P(\bJ,t),
\end{equation}
with a \DC\
\begin{equation}
  \label{eq:D_markov}
  D_{ij}(\bJ) =
  \sum_\bk k_i k_j
  \!\! \int_{-\infty}^\infty \!\!\!\! \rd s \,
    C_{\bk\bk}(\bJ,\bJ,s) .
\end{equation}

\Eqs~\eqref{eq:fp_markov} and~\eqref{eq:D_markov} provide a closed
diffusion equation, where the action-space \DC\ depends only on the total power
of the noise at a given action $\bJ$ (and resonance vector $\bk$). These
equations were, however, obtained in the Markovian limit, i.e.\ assuming that
the noise is uncorrelated in time (see \eq~\eqref{eq:C_markov}). Such a limit
is valid if the motion of the test particle (which may be driven either by the
mean field or by the noise) is slower than the correlation time of the
noise.
This Markovian limit could also prove useful in cases where one or
more of the angles is degenerate, i.e.\ has zero frequency. This is for example
the case in isotropic galactic nuclei, during the process of vector resonant
relaxation~\citep{KocsisTremaine2011,KocsisTremaine2015}.
However, as discussed in the next section, in other generic regimes
such assumptions cannot be applied, as the angles of the test particle
typically evolve on the same timescale
than the correlation of the noise. This situation requires more precise considerations in
the derivation of the diffusion equation in order to account for the temporal
correlations of the noise, as we show in the next section.

\subsection{Correlated noise}
\label{sec:correlated}

Throughout the previous sections, we assumed that the test particle's mean
Hamiltonian, $H_{0}$ (see \eq~\eqref{eq:H_0}), is integrable. As a consequence,
the actions $\bJ$ are constants of motion of $H_0$, and the potential
fluctuations around it (characterized by the noise terms
$\eta_\bk(\bJ,t)$) are comparably small, i.e.\ ${ \eta \ll H_{0} }$. Between
the times $t$ and $\tp$, the motion of the test particle can then be written as
$ {\bT(\tp)} = {\bT(t)} + {\bO(\bJ(t))(\tp-t)} + {\delta\bT(t,\tp)} $, and
$ {\bJ(\tp)} = {\bJ(t)} + {\delta \bJ(t,\tp)} $, where
${\bO (\bJ) = \partial H_0(\bJ) / \partial \bJ \sim H_0/J}$ are the (fast)
frequencies of the mean field motion of the angles, and $\delta \bT$,
$\delta \bJ$ are the deviations of the trajectory due to the potential
fluctuations. If we assume that the motion of the bath particles is also driven
by the same mean Hamiltonian, the test particle and the bath particles evolve
on similar timescales and therefore the previous Markovian limit cannot be
applied. To proceed forward, we will use the small noise approximation and
expand the response functions w.r.t.\ noise to overcome the closure problem. As
we will show, the associated \DCs\ will then depend on the
full temporal properties of the noise, while in the Markovian limit they only
depended on ${ \int_{- \infty}^{\infty} \! \rd t \, C (t) }$, i.e.\ the total power of the noise.

Following \eqs~\eqref{eq:J_t} and~\eqref{eq:theta_t}, the changes in the test particle's trajectory
due to the fluctuations of the
potential around the mean field are given by
\begin{equation}
  \label{eq:dJ}
  \delta \bJ(t,\tp) = - \ri \sum_\bk \bk \dint[t][\tp]s \, 
  \re^{\ri \bk\cdot\bT(s)} \, \eta_\bk(\bJ(s), s),
\end{equation}
and
\begin{equation}
  \label{eq:dT}
  \delta \bT(t,\tp) = \dint[t][\tp]s \big[\bO (\bJ(s))-\bO(\bJ(t))\big]
                           + \sum_\bk \dint[t][\tp]s \, \re^{ \ri \bk\cdot\bT(s)}
                           \dint \bJ\, \deltaD(\bJ-\bJ(s)) \dpd{}{\bJ}
                           \eta_\bk(\bJ, s).
\end{equation}
These stochastic perturbations depend, to lowest order, linearly on the noise,
so that they will vanish when the noise vanishes.

Therefore, to obtain the lowest order expression of the response functions, we
substitute the unperturbed mean field motion ${ \bJ(\tp) = \bJ(t) }$ and
${ \bT(\tp) = \bT(t) + \bO(\bJ(t))(\tp-t) }$ into \eqs~\eqref{eq:res_J}
and~\eqref{eq:res_theta}, and ignore all terms which depend explicitly on the
noise. The response functions are then
\begin{equation}
  \label{eq:res_J_s}
  \dfd{\bJ(t)}{\eta_\bkp^*(\bJp, \tp)}
   =
   \ri \bkp \, \re^{ - \ri \bkp\cdot\bT(t)}\re^{ - \ri \bkp\cdot\bO(\bJp)(\tp-t)}
   \, \deltaD  (\bJp - \bJ(t)) ,
\end{equation}
and
\begin{align}
  \label{eq:res_theta_s}
  \dfd{\bT (t)}{\eta_{\bkp}^{*} (\bJp, \tp)}  = {} 
  &
    -  \re^{ - \ri \bkp\cdot\bT(t)}\re^{ - \ri \bkp\cdot\bO(\bJ(t))(\tp-t)}
    \dpd{}{\bJp}\bigg[\deltaD  (\bJp - \bJ (t)) \bigg]
    \nonumber \\
  &
    + \ri \, \re^{-\ri\bkp\cdot\bT(t)} \re^{-\ri\bkp\cdot\bO(\bJp)(\tp-t)} \, 
    \deltaD  (\bJp - \bJ(t))
    \dint[\tp][t] s \dint\bJ\,\deltaD(\bJ-\bJ(s))
    \bkp\ccdot\dpd{}{\bJ}\bO(\bJ) .
    \nonumber \\
  = {}
  &
    - \re^{ - \ri \bkp\cdot\bT(t)}\re^{ - \ri \bkp\cdot\bO(\bJp)(\tp-t)}
    \dpd{}{\bJp}\bigg[\deltaD  (\bJp - \bJ (t)) \bigg],
\end{align}
where the last step is done by using the integration over
$\deltaD(\bJp-\bJ(t))$ to replace $\bO(\bJ(t))$ by $\bO(\bJp)$. Let us now
pursue the calculation of the diffusion equation at this order, and postpone
the discussion of higher order contributions to later.

As \eqs~\eqref{eq:res_J_s} and~\eqref{eq:res_theta_s} are independent of the
noise, they can be used in \eq~\eqref{eq:eta_phi} together with
\eq~\eqref{eq:F_dot} to obtain the diffusion equation
\begin{align}
  \label{eq:F_dot_fast}
  \dpd{}{t}P(\bJ,t)
  &=
    \dpd{}{\bJ} \ccdot
    \sum_{\bk,\bkp} \bk
    \int_0^t \!\! \rd \tp \dint \bJp \,
    C_{\bk\bkp}(\bJ,\bJp,t-\tp)
    \, \re^{  \ri \bkp\cdot\bO(\bJp)(t-\tp)}
    \,
    \left[
    \bk \ccdot \dpd{}{\bJp}
    +
    \bkp \ccdot \dpd{}{\bJ}
    \right]
    \left\langle
    \re^{ - \ri (\bk-\bkp)\cdot\bT(t)}
    \, \deltaD(\bJp-\bJ) \, \varphi(\bJ,t)
    \right\rangle
    \nonumber \\
  &=
    \dpd{}{\bJ} \ccdot
    \sum_\bk \bk
    \int_0^t \!\! \rd s \,
    C_{\bk\bk}(\bJ,\bJ,s) \,
    \, \re^{  \ri \bk\cdot\bO(\bJ)s}
    \, \bk \ccdot \dpd{}{\bJ}
    \left\langle
    \varphi(\bJ,t)
    \right\rangle ,
\end{align}
where to obtain the second line, we used the same manipulations as in
\eq~\eqref{eq:F_dot_markov}. \Eq~\eqref{eq:F_dot_fast} describes the slow
diffusion of the test particle's action $\bJ$ on long-term timescales
${ {(J/\eta)}^2 \Omega \ll J/\eta \ll 1/\Omega }$, under the effect of the
potential fluctuations. Since in \eq~\eqref{eq:F_dot_fast}, the time $t$ is
much larger than the dynamical timescale ${ 1/ \Omega }$ on which
${ C_{\bk\bk}(t) }$ decays, one may take the limits of the time integration in
\eq~\eqref{eq:F_dot_fast} to ${ + \infty }$. \Eq~\eqref{eq:F_dot_fast}
becomes
\begin{equation}
  \label{eq:fp_non_markov}
  \dpd{P(\bJ,t)}{t} =
  \frac{1}{2}
  \dpd{}{J_i}
  D_{ij}(\bJ)\dpd{}{J_j}
  P(\bJ,t) ,
\end{equation}
where the anisotropic \DCs\ ${ D_{ij} (\bJ) }$ are given by
\begin{equation}
  \label{eq:D_non_markov}
  D_{ij}(\bJ) =
  2 \sum_\bk k_i k_j
   \int_{0}^\infty \!\!\!\! \rd t \,
   C_{\bk\bk}(\bJ,\bJ,t)
   \, \re^{ \ri \bk\cdot\bO(\bJ)t} .
\end{equation}
Here, let us emphasize that \eq~\eqref{eq:fp_non_markov} essentially
takes the form of the orbit-averaged Fokker-Planck equation for a zero mass
particle~\citep[see \eq~{(7.80)} in][]{BinneyTremaine2008}.
Relying on the property that
${ C_{\bk\bk} (\bJ,\bJ,-t) = C_{\bk\bk}^* (\bJ,\bJ,t) }$ and that the diffusion
tensor $D_{ij}$ is real, one can rewrite the \DCs\ as
\begin{equation}
  \label{eq:D_ft}
  D_{ij}(\bJ) = \sum_\bk k_i k_j \int_{-\infty}^\infty 
  \!\!\!\! \rd t \, C_{\bk\bk}(\bJ,\bJ,t) \, \re^{\ri \bk\cdot\bO(\bJ)t} .
\end{equation}
Introducing the temporal Fourier transform with the convention
\begin{equation}
  \label{eq:convention_time_FT}
  \widehat{f} (\omega) = \!\! \int_{- \infty}^{+ \infty} \!\!\!\! \rd t \, f
  (t) \, \re^{\ri \omega t} \;\;\; ; 
  \;\;\; 
  f (t) = \frac{1}{2 \pi} \!\! \int_{- \infty}^{+ \infty} 
  \!\!\!\! \rd \omega \, \widehat{f} (\omega) \, \re^{- \ri \omega t} ,
\end{equation}
\eq~\eqref{eq:D_ft} finally becomes
\begin{equation}
  \label{eq:diff_coeff_final}
  D_{ij} (\bJ) = 
  \sum_\bk k_{i} k_{j} \, \wC_{\bk\bk} (\bJ,\bJ,\bk\ccdot\bO(\bJ)) ,
\end{equation}
where ${ \wC_{\bk\bk} (\bJ,\bJ,\omega) }$ is the Fourier transform of the
temporal correlation function ${ C_{\bk\bk} (\bJ,\bJ,t) }$.

\Eq~\eqref{eq:D_ft} is the main result of this section and we recover here
the equivalent result from \eq~{(3.9a)} of~\cite{BinneyLacey1988}.
It shows that the
\DCs\ in action space are sourced by the Fourier transform of the noise
correlation function ${ \wC_{\bk\bk} (\bJ,\bJ,\omega) }$, evaluated at the
location $\bJ$ and dynamical frequency ${ \omega \!=\! \bk \ccdot \bO (\bJ) }$
of the test particle. In Section~\ref{sec:Recovery}, we will show how
\eq~\eqref{eq:diff_coeff_final} can be used to recover the \DCs\
of the inhomogeneous \BL\ and Landau
equations~\citep{Heyvaerts2010,Chavanis2012}, as well as the
dressed \DCs\ \citep{BinneyLacey1988,Weinberg2001a} in Section~\ref{sec:dFPeq}.

Let us now discuss the contributions from higher order noise terms to the
diffusion equation. As already mentioned, \eqs~\eqref{eq:res_J_s}
and~\eqref{eq:res_theta_s} are lowest order in the noise. The next order will
involve terms which depend linearly on the noise. These terms, when plugged
into \eq~\eqref{eq:eta_phi}, will produce terms of the form
${\int_0^t \rd \tp C(t-\tp) \int_\tp^t \rd s \langle \eta (\bJ(s), s)
 \, \varphi(\bJ,t) \rangle}$. One may then apply Novikov's theorem again, which
yields terms of the form
${\int_0^t \rd \tp C(t-\tp) \int_\tp^t \rd s \int_0^t \rd s' C(s-s') \langle
  \varphi(\bJ,t) \rangle}$. The correlation function ${ C_{\bk\bk} (\bJ,\bJ,t) }$ is,
generally, a decaying function in time, with the initial amplitude
${C_{\bk\bk} (\bJ,\bJ,0)=\langle | \eta_{\bk} (\bJ , 0) |^{2} \rangle}$. Let us define the
correlation time $T_{\rc}$ as the timescale on which ${ C(t) }$ decays. When
${ C(t) }$ is an oscillating function (see Figure~\ref{fig:hmf_acf}), $T_{\rc}$
will be the decay time of the envelope. The contribution of the higher noise
terms to the \DCs\ are typically of the smaller order
${\int_\tp^t \rd s \int_0^t \rd s' C(s-s')/J^2 \sim \eta^2 T_{\rc}^2/J^2}$.
Intuitively, the integrations over the response functions are multiplied by the
correlation function ${ C_{\bk\bkp}(\bJ, \bJp, t-\tp) }$ which decays on the
timescale $T_{\rc}$, so that temporal integrals can be evaluated up to
$T_{\rc}$. On this timescale, the test particle's action will change by
${ (\Delta J)/J \sim \eta T_{\rc}/J }$ because of the noise. Now, since $\eta$
is stochastic, correction to the response functions of order
$\mathcal{O}(\eta T_{\rc}/J)$ will result in a correction of order
${ \mathcal{O}(\eta^2 T_{\rc}^2/J^2) }$ to the \DCs\@. Therefore, as long as
${ \eta T_{\rc} /J \ll 1 }$, one can safely neglect higher order terms in the
noise, as was assumed in \eqs~\eqref{eq:res_J_s} and~\eqref{eq:res_theta_s} for
the response functions.

Fortunately, in the case considered here where the motion of the bath particles
is also governed by the mean Hamiltonian $H_{0}$, the assumption
${ \eta T_{\rc} / J \ll 1 }$ holds. As a result, in such a regime, the
randomization of the potential is mainly due to the randomization of the phases
$\bT$ of the bath particles. The correlation timescale of the potential is of
order ${T_{\rc} \sim 1/\Omega \sim H_0/J}$, which guarantees that
${\eta T_{\rc} /J \sim \eta/H_0 \ll 1}$. 

There can be cases (or specific values of $\bJ$) where the frequencies
${ \bO (\bJ) }$ of the test particle are smaller than correlation time of the
noise. In this case, the motion of the test particle will be slower than the
time to randomize the potential, and the Markovian approximation becomes
valid. Indeed, taking ${ \bO (\bJ) \to 0 }$ in \eq~\eqref{eq:D_ft}, one
recovers the Markovian \DCs, \eq~\eqref{eq:D_markov}.

In all the calculations above, we assumed that the bath particles were
completely independent from the test particle. Because of the absence of a
back-reaction from the test particle on the bath particle trajectories, we
recovered in \eq~\eqref{eq:fp_non_markov} a diffusion equation that only
contains a \DC\@. In such a configuration, the noise sourcing
the stochastic diffusion of the test particle is completely external. We note
in particular that these \DCs\ are independent of the mass of
the test particle, so that such a diffusion cannot induce any mass
segregation. In Section~\ref{sec:DynFric}, we will briefly illustrate how one can
adapt the previous calculations to account for the back-reaction of the test
particle on the bath, via the so-called friction force by
polarization~\cite[see e.g.,][and references therein]{Heyvaerts2017}. In
particular, the amplitude of this friction force will be proportional to the
mass of the test particle, so that it can lead to mass segregation.

\section{Recovering the Landau and Balescu-Lenard diffusion coefficients}
\label{sec:Recovery}

In the previous section, we emphasized how the \DCs\
describing the long-term orbital diffusion of the test particle are sourced by
the correlation function
${ C_{\bk\bk} (\bJ, \bJ, t - \tp) = \left\langle \eta_{\bk} (\bJ, t) \,
    \eta_{\bk}^{*} (\bJ, \tp) \right\rangle }$ of the stochastic potential
fluctuations induced by the bath. Fortunately, there exist various regimes in
which one can write explicit expressions for this correlation. In
Section~\ref{sec:LandauEq}, we will consider the case where the bath particles
only interact through the mean field, i.e.\ the individual orbits of the bath
particles are only driven by a smooth mean field potential. This will allow us
to recover straightforwardly the \DCs\ of the inhomogeneous
Landau equation~\citep{Chavanis2013}. In Section~\ref{sec:BLEq}, we will relax
this hypothesis and assume that the bath particles are fully interacting
with one another via the pairwise potential $\psi$. This will allow us to recover
the \DCs\ of the inhomogeneous \BL\
equation~\citep{Heyvaerts2010,Chavanis2012}.

\subsection{The Landau diffusion coefficients}
\label{sec:LandauEq}

Let us first assume that the bath is external to the test particle (i.e.\ no
back-reaction of the test particle on the bath particles). In addition, let us
also assume that bath itself is non-interacting so that the dynamics of a given
bath particle is imposed the specific Hamiltonian
\begin{equation}
  \label{eq:H_Landau}
  H_{\rb} (\bT, \bJ, t) = H_{0} (\bJ) ,
\end{equation}
where the mean field integrable Hamiltonian $H_{0}$ was introduced in
\eq~\eqref{eq:H_expn}. In \eq~\eqref{eq:H_Landau}, one can note that the bath
particles evolve only under the effect of the mean field potential: they do not
interact per se with one another. Such a regime amounts to neglecting
collective effects in the bath, and thus, the ability to amplify its
perturbations. In order to study the statistics of the perturbations induced by
the bath, we will rely on the Klimontovich equation~\citep{Klimontovich1967}. At
any given time, the state of the bath is fully characterized by the discrete
\DF, ${ F_{\rd} (\bT, \bJ, t) }$ given by
\begin{equation}
  \label{eq:properties_Fd}
  F_{\rd} (\bT, \bJ) = 
  \sum_{i = 1}^{N} \mb \, \deltaD  (\bT - \bT_{i} (t)) \, 
  \deltaD  (\bJ - \bJ_{i} (t)) ,
\end{equation}
where the sum over $i$ runs over the $N$ particles in the bath,
${ (\bT_{i} (t), \bJ_{i} (t)) }$ is the position in action space of the $i$-th
particle at time $t$, and ${ \mb = M/N }$ is the individual mass of the bath
particles. The evolution of the \DF\ $F_{\rd}$ is governed by the Klimontovich
equation
\begin{equation}
  \label{eq:Klimontovich_Eq}
  \dpd{F_{\rd}}{t} + \big[ F_{\rd}, H_{\rb} \big] = 0 ,
\end{equation}
where $H_{\rb}$ is the one-particle Hamiltonian introduced in
\eq~\eqref{eq:H_Landau}. In \eq~\eqref{eq:Klimontovich_Eq}, we introduced the
Poisson bracket as
\begin{align}
\label{eq:def_Poisson}
  \big[ F, H \big]  = {} 
  &
    \dpd{F}{\bx} \ccdot \dpd{H}{\bvel} - \dpd{F}{\bvel} \ccdot \dpd{H}{\bx}
    \nonumber  \\
  = {}
  &
    \dpd{F}{\bT} \ccdot \dpd{H}{\bJ} - \dpd{F}{\bJ} \ccdot \dpd{H}{\bT} .
\end{align}
Let us emphasize that \eq~\eqref{eq:Klimontovich_Eq} is an exact evolution
equation in phase space and that no approximations have been made yet. Let us
now assume that the bath's \DF\ can be decomposed in two components, so that
\begin{equation}
  \label{eq:decomposition_DF}
  F_{\rd} = F_{\rb} + \delta F \;\;\; \text{with} \;\;\; F_{\rb} 
  = F_{\rb} (\bJ) \;\; ; \;\; \left\langle \delta F \right\rangle = 0 ,
\end{equation}
where ${ F_{\rb} \!=\! \left\langle F_{\rd} \right\rangle }$ is the underlying
mean field \DF\ of the bath particles, and ${ \delta F }$ are the fluctuations
around it. Here the averaging ${ \left\langle \; \cdot \; \right\rangle }$ is
over the angle $\bT$ and different initial conditions of the bath.

Injecting this decomposition into the evolution
\eq~\eqref{eq:Klimontovich_Eq}, one can immediately write
\begin{align}
  \label{eq:Klim_Ld}
  0  = {}
  &
    \dpd{\delta F}{t} + \big[ \delta F, H_{0} \big]
      \nonumber  \\
  = {}
  &
    \dpd{\delta F}{t} + \dpd{\delta F}{\bT} \ccdot \bO(\bJ) ,
\end{align}
where ${ \bO (\bJ) = \partial H_{0} (\bJ) / \partial \bJ }$ are the mean field
orbital frequencies and we relied on the fact that
${ \big[ F_{\rb} (\bJ), H_{0} (\bJ) \big] = 0 }$. As shown in the previous
sections, the long-term evolution of a test particle is sourced by the bath
potential fluctuations ${ \eta_{\bk} (\bJ, t) }$ defined in
\eq~\eqref{eq:eta_def}. We are therefore interested in characterizing the
statistical properties of these potential fluctuations, which result from the
perturbations ${ \delta F }$ of the \DF\@. Following \eq~\eqref{eq:eta_def},
these fluctuations are directly related by
\begin{equation}
  \label{eq:link_fluc}
  \eta (\bx, t) =
  \dint \bx' \rd \bvel' \, 
  \psi (\bx, \bxp) \, \delta F (\bxp, \bvelp, t) ,
\end{equation}
where we recall that ${ \psi (\bx, \bxp) }$ is the pairwise interaction
potential. As emphasized by the
Hamiltonian from Eq.~\eqref{eq:H_Landau}, let us recall that in the Landau limit,
the pairwise interactions are suppressed between the bath particles.
The angle-action coordinates being canonical, and performing a
Fourier transform w.r.t.\ the angles, \eq~\eqref{eq:link_fluc} can be rewritten
as
\begin{equation}
  \label{eq:link_fluc_ft}
  \eta_{\bk} (\bJ, t) = 
  {(2 \pi)}^{d} \sum_{\bkp} \dint \bJp \, 
  \psi_{\bk\bkp} (\bJ, \bJp) \, \delta F_{\bkp} (\bJp, t) ,
\end{equation}
where the bare susceptibility coefficients ${ \psi_{\bk\bkp} (\bJ,\bJp) }$ were
introduced in \eq~\eqref{eq:psi_basis} and $ \delta F_{\bkp} (\bJp, t)$ are
the Fourier components of $\delta F$.

When Fourier transformed w.r.t.\ the angles (following
\eq~\eqref{eq:convention_time_FT}), the evolution \eq~\eqref{eq:Klim_Ld}
becomes
\begin{equation}
  \label{eq:Klim_Ld_Fourier}
  \dpd{\delta F_{\bk}}{t} 
  + \ri \bk \ccdot \bO (\bJ) \, \delta F_{\bk} (\bJ, t) = 0 .
\end{equation}
This equation describes the evolution of the fluctuations in the bath, and we
recover that in such a non-interacting bath, the bath particles are independent
one from one another because they strictly follow the mean field
motion. Assuming that the mean field orbital frequencies ${ \bO (\bJ) }$ are
independent of time, \eq~\eqref{eq:Klim_Ld_Fourier} can be integrated in time
to obtain
\begin{equation}
  \label{eq:integrated_Klim_Ld_Fourier}
  \delta F_{\bk} (\bJ, t) = \delta F_{\bk} (\bJ, 0) \, 
  \re^{- \ri \bk \cdot \bO (\bJ) t} ,
\end{equation}
where ${ \delta F_{\bk} (\bJ, 0) }$ correspond to the initial fluctuations in
the bath's \DF\ at ${ t = 0 }$. The potential
fluctuations in \eq~\eqref{eq:link_fluc_ft} then become
\begin{equation}
  \label{eq:deltapsi_Landau}
  \eta_{\bk} (\bJ, t) = 
  {(2 \pi)}^{d} \sum_{\bkp} \dint \bJp \, 
  \psi_{\bk\bkp} (\bJ,\bJp) \, \re^{- \ri \bkp \cdot \bO(\bJp) t} \, 
  \delta F_{\bkp} (\bJp, 0) ,
\end{equation}
which means that the initial fluctuations ${ \delta F_{\bkp} (\bJp, 0) }$ in
the bath's \DF\ uniquely determine the potential fluctuations
${ \eta_{\bk} (\bJ, t) }$. In Appendix~\ref{sec:FlucDF}, we briefly
characterize the properties of these initial fluctuations. In particular, we
show in \eq~\eqref{eq:Fluc_DF_Fourier} that one has
\begin{equation}
  \label{eq:repetitions_Fluc_DF_Fourier}
  \left\langle \delta F_{\bk} (\bJ, 0) \, \delta F_{\bkp} (\bJp, 0)  \right\rangle = 
  \frac{m_{\rb}}{{(2 \pi)}^{d}} \, \delta_{- \bk \bkp} \, \deltaD (\bJ - \bJp) \, F_{\rb} (\bJ) .
\end{equation}
Following \eq~\eqref{eq:C_def}, the correlation
${ C_{\bk\bk} (\bJ, \bJp, t - \tp) }$ of the potential fluctuations
subsequently takes the form
\begin{align}
  \label{eq:Correl_psi_Landau}
  C_{\bk \bk} (\bJ, \bJp, t - \tp) = {} 
  &
    \left\langle \eta_{\bk} (\bJ, t) \, \eta_{\bk}^{*} (\bJ, \tp) \right\rangle
    \nonumber \\
  = {} 
  &
    \mb \, {(2 \pi)}^{d} \, \sum_{\bkp} \! \int \!\! \rd \bJp \, {|
    \psi_{\bk\bkp} (\bJ,\bJp) |}^{2} \, F_{\rb} (\bJp) \, \re^{- \ri \bkp \cdot
    \bO (\bJp) (t - \tp)} .
\end{align}
Using \eq~\eqref{eq:diff_coeff_final}, one can finally compute the \DCs\ of a
test particle that would undergo such stochastic fluctuations. One gets
\begin{equation}
  \label{eq:Final_Diff_Coeff_Landau}
  D_{ij} (\bJ) = 
  m_{\rb} \, {(2 \pi)}^{d+1} \sum_{\bk, \bkp} k_{i} k_{j} \dint \bJp \, 
  \deltaD (\bk \ccdot \bO (\bJ) - \bkp \ccdot \bO (\bJp)) \, 
  {| \psi_{\bk\bkp} (\bJ, \bJp) |}^{2} \, F_{\rb} (\bJp) .
\end{equation}
In \eq~\eqref{eq:Final_Diff_Coeff_Landau}, we recovered exactly the \DCs\ of
the inhomogeneous Landau equation~\citep[see \eq~{(113)} in][]{Chavanis2013}. They
describe the diffusion component of the long-term orbital distortion undergone by
a test particle embedded in a bath made of $N$ bath particles in the limit
where collective effects are not accounted for, i.e.\ in the limit where the
bath particles only see the mean field potential (see \eq~\eqref{eq:H_Landau}).

The \DCs\ in \eq~\eqref{eq:Final_Diff_Coeff_Landau} can
qualitatively be understood as follows. A given test particle located on an
orbit of action $\bJ$ may diffuse on long-term timescales under the effects of
the finite-$N$ fluctuations induced by the discrete bath. This explains the
overall prefactor in ${ m_{\rb} \!=\! M / N }$ in
\eq~\eqref{eq:Final_Diff_Coeff_Landau}: more bath particles will lead to a
smoother potential and to a slower evolution. To diffuse, this test particle
has to resonantly couple with bath particles. In
\eq~\eqref{eq:Final_Diff_Coeff_Landau}, the integration over the dummy variable
$\bJp$ should therefore be seen as a scan of action space, looking for orbits
of bath particles, such that the resonant condition
${ \bk \cdot \bO (\bJ) = \bkp \cdot \bO (\bJp) }$ is satisfied. This resonance
condition is a direct consequence of \eq~\eqref{eq:diff_coeff_final}, where we
showed that the \DCs\ require the evaluation
of the noise correlation function at the test particle's local orbital
frequency ${ \omega = \bk \cdot \bO (\bJ) }$, for a noise created by
bath particles evolving with the frequencies ${ \bkp \cdot \bO (\bJp) }$. Each
resonant coupling is parametrized by a different pair of resonance vectors
${ (\bk, \bkp) }$, which determines which linear combinations of orbital
frequencies are matched on resonance. As can be seen from the factor
${ k_{i} }$ in \eq~\eqref{eq:Final_Diff_Coeff_Landau}, the resonance vector
$\bk$ also controls the direction in which the diffusion occurs in action
space. The diffusion is anisotropic not only because ${ D_{ij} (\bJ) }$ depends
on the action $\bJ$ of the test particle, but also because each resonance
vector leads to a preferential diffusion in a different direction in action
space. Finally, in the absence of collective effects, the strength of these
resonant couplings is controlled by the square of the bare susceptibility coefficient
${ {| \psi_{\bk\bkp} (\bJ, \bJp) |}^{2} }$. In Section~\ref{sec:HMF},
we will illustrate in detail how the present $\eta$-formalism allows us to
evaluate in a simple manner the bare \DCs\ from
\eq~\eqref{eq:Final_Diff_Coeff_Landau} for the one-dimensional inhomogeneous
\HMF\ model.

\subsection{The Balescu-Lenard diffusion coefficients}
\label{sec:BLEq}

In the previous section, we neglected collective effects, i.e.\ in
\eq~\eqref{eq:H_Landau} we assumed that the dynamics of the bath particles was
only governed by the mean field. This allowed for the recovery of the
inhomogeneous Landau \DCs\ in
\eq~\eqref{eq:Final_Diff_Coeff_Landau}. Let us now investigate how these
calculations have to be modified when one accounts for collective effects and
the associated self-gravitating amplification. In this context, collective
effects correspond to the fact that bath particles are influenced by the
perturbations they self-consistently generate. In such a regime, the specific
Hamiltonian in \eq~\eqref{eq:H_Landau} becomes
\begin{equation}
  \label{eq:H_BL}
  H_{\rb} (\bT, \bJ, t) = H_{0} (\bJ) + \eta (\bT, \bJ, t) ,
\end{equation}
where ${ \eta \!=\! \eta [\delta F] }$ was introduced in \eq~\eqref{eq:link_fluc}
and stands for the potential perturbations in the bath associated with the
perturbations of the bath's \DF\@. Keeping only linear terms in the
perturbations, the Klimontovich \eq~\eqref{eq:Klimontovich_Eq} becomes
\begin{align}
  \label{eq:Klim_BL}
  0  = {}
  &
    \dpd{\delta F}{t} + 
    \big[ \delta F, H_{0} \big] + \big[ F_{\rb}, \eta \big]
      \nonumber  \\
  = {}
  &
    \dpd{\delta F}{t} + 
    \dpd{\delta F}{\bT} \ccdot \bO (\bJ) - 
    \dpd{F_{\rb}}{\bJ} \ccdot \dpd{\eta}{\bT} .
\end{align}

Let us emphasize here that the linear approximation of the Klimontovich
equation is a crucial step of this calculation. It will allow us to characterize in detail
the self-gravitating amplification of fluctuations occurring in the bath.
Compared to the bare evolution \eq~\eqref{eq:Klim_Ld}, one can note in \eq~\eqref{eq:Klim_BL} the
additional presence of the potential fluctuations $\eta$. When Fourier
transformed w.r.t.\ the angles, \eq~\eqref{eq:Klim_BL} becomes
\begin{equation}
  \label{eq:Klim_BL_Fourier}
  \dpd{F_{\bk}}{t} + \ri \bk \ccdot \bO (\bJ) \, 
  \delta F_{\bk} (\bJ, t) - \ri \bk \ccdot \dpd{F_{\rb}}{\bJ} \, 
  \eta_{\bk} (\bJ, t) = 0 .
\end{equation}

The main difficulty with the evolution \eq~\eqref{eq:Klim_BL_Fourier} is that
both ${ \delta F_{\bk} }$ and $\eta_{\bk}$ depend on time. In addition, these
perturbations also satisfy the self-consistency requirement
${ \eta \!=\! \eta [\delta F] }$, as imposed by \eq~\eqref{eq:link_fluc_ft}. In
order to solve \eq~\eqref{eq:Klim_BL_Fourier}, we rely on the assumption of
timescale separation. We note that both the \DF's fluctuations, ${ \delta F }$,
and the potential perturbations, $\eta$, fluctuate on the dynamical timescale
${\sim 1 / \Omega }$, while the bath's mean \DF, $F_{\rb}$, as well as the mean
orbital frequencies ${ \bO (\bJ) }$ evolve on the slower long-term timescales
${\sim N / \Omega }$. As a result, we will assume that $F_{\rb}$ and $\bO$ can
be taken to be independent of time on dynamical timescales when solving
\eq~\eqref{eq:Klim_BL_Fourier} for ${ \delta F_{\bk} }$. It is easier to solve
the Laplace transform of \eq~\eqref{eq:Klim_BL_Fourier},
\begin{equation}
  \label{eq:Klim_BL_Laplace}
  \delta \tF_{\bk} (\bJ, \omega) = 
  - \frac{\bk \ccdot \partial F_{\rb} / \partial \bJ}{\omega - \bk \ccdot \bO (\bJ)} \, 
  \teta_{\bk} (\bJ, \omega) 
  - \frac{\delta F_{\bk} (\bJ, 0) }{\ri (\omega - \bk \ccdot \bO (\bJ))} ,
\end{equation}
where we defined the Laplace transform with the convention
\begin{equation}
  \label{eq:definition_Laplace}
  \widetilde{f} (\omega) = 
  \!\! \int_{0}^{+ \infty} \!\!\!\! \rd t \, f (t) \, 
  \re^{\ri \omega t} \;\;\; ; \;\;\; f (t) = 
  \frac{1}{2 \pi} \dint[\mathcal{B}] \omega \, 
  \widetilde{f} (\omega) \, \re^{- \ri \omega t},
\end{equation}
where the Bromwich contour $\mathcal{B}$ in the complex $\omega$-plane has to
pass above all the poles of the integrand, i.e.\ ${ \text{Im} [\omega] }$ has
to be large enough.

Following \eq~\eqref{eq:link_fluc_ft}, \eq~\eqref{eq:Klim_BL_Laplace} can
immediately be rewritten as a self-consistency relation involving only
${ \teta_{\bk} (\bJ, \omega) }$. To do so, one acts on both sides of
\eq~\eqref{eq:Klim_BL_Laplace} with the same operator as in the r.h.s.\ of
\eq~\eqref{eq:link_fluc_ft}. One gets
\begin{equation}
  \label{eq:selfconsistent_psi_BL}
  \teta_{\bk} (\bJ, \omega) = 
  - {(2 \pi)}^{d} \sum_{\bkp} \dint \bJp \, 
  \frac{\bkp \ccdot \partial F_{\rb} / \partial \bJp}{\omega - \bkp \ccdot \bO (\bJp)} \, 
  \psi_{\bk\bkp} (\bJ, \bJp) \, \teta_{\bkp} (\bJp, \omega) 
  - {(2 \pi)}^{d} \sum_{\bkp} \dint \bJp \, 
  \frac{\delta F_{\bkp} (\bJp, 0)}{\ri (\omega - \bkp \ccdot \bO (\bJp))} \, 
  \psi_{\bk\bkp} (\bJ, \bJp) .
\end{equation}
\Eq~\eqref{eq:selfconsistent_psi_BL} takes the form of a Fredholm equation
that has to be inverted in order to characterize the potential fluctuations
${ \teta_{\bk} (\bJ, \omega) }$. A first method to invert this relation is to
rely on the Kalnajs matrix method~\citep{Kalnajs1976II} and introduce a
biorthogonal set of density and potential basis elements. We briefly review
this method in Appendix~\ref{sec:BasisMethod}. As already noted
in~\cite{LucianiPellat1987}, \eq~\eqref{eq:selfconsistent_psi_BL} may also be
inverted implicitly without resorting to a set of basis elements. Inspired by
\eq~\eqref{eq:deltapsi_Landau}, let us assume that the potential fluctuations
follow the ansatz
\begin{equation}
  \label{eq:ansatz_psi}
  \teta_{\bk} (\bJ, \omega) = 
  - {(2 \pi)}^{d} \sum_{\bkp} \dint \bJp \, 
  \frac{\delta F_{\bkp} (\bJp, 0)}{\ri (\omega - \bkp \ccdot \bO (\bJp))} \, 
  \psi_{\bk\bkp}^{\rd} (\bJ, \bJp, \omega) ,
\end{equation}
which is simply the Laplace version of \eq~\eqref{eq:deltapsi_Landau}, where we
replaced the bare susceptibility coefficients,
${ \psi_{\bk\bkp} (\bJ, \bJp) }$, by their (yet unknown) dressed analogs
${ \psi_{\bk\bkp}^{\rd} (\bJ, \bJp, \omega) }$.\footnote{ Here we use the
  opposite sign convention for ${ \psi_{\bk\bkp}^{\rd} (\bJ, \bJp, \omega) }$
  from~\cite{Heyvaerts2010,Chavanis2012} so that in the limit where collective
  effects are neglected
  ${\psi_{\bk\bkp}^{\rd} (\bJ, \bJp, \omega) \to \psi_{\bk\bkp} (\bJ,
    \bJp)}$, as can be seen from
  \eq~\eqref{eq:selfconsistent_dressed_coefficients}.} Injecting this ansatz
into \eq~\eqref{eq:selfconsistent_psi_BL}, one gets a
self-consistent Fredholm equation of the second kind
\begin{equation}
  \label{eq:selfconsistent_dressed_coefficients}
  \psi_{\bk\bkp}^{\rd} (\bJ, \bJp, \omega) = 
  - {(2 \pi)}^{d} \sum_{\bkpp} \dint \bJpp \, 
  \frac{\bkpp \ccdot \partial F_{\rb} /\partial \bJpp}{\omega - \bkpp \ccdot \bO (\bJpp)} \, 
  \psi_{\bk\bkpp} (\bJ, \bJpp) \, 
  \psi_{\bkpp\bkp}^{\rd} (\bJpp, \bJp, \omega) 
  + \psi_{\bk\bkp} (\bJ, \bJp) .
\end{equation}
sourced by the bare susceptibility coefficients
${ \psi_{\bk\bkp} (\bJ, \bJp) }$~\citep{Chavanis2012}. As shown in
Appendix~\ref{sec:BasisMethod}, should one aim for an explicit expression of
the dressed susceptibility coefficients, one can rely on the basis method to
invert \eq~\eqref{eq:selfconsistent_dressed_coefficients}, which leads to the
explicit expression in \eq~\eqref{eq:explicit_psid}.

\begin{figure}
  \begin{center}
    \includegraphics[scale=1.2]{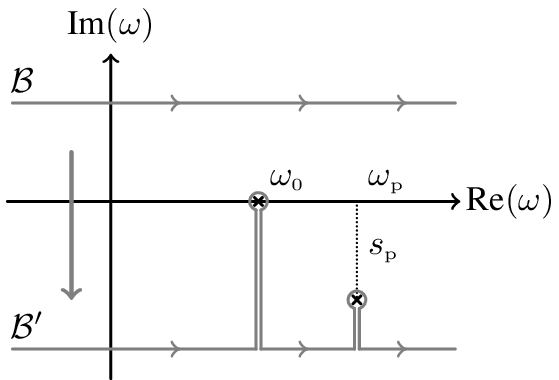}
    \caption{\label{fig:PolesDressing} Illustration of the computation of the
      inverse Laplace Transform from \eq~\eqref{eq:deltapsi_BL_I}. The
      integrand from \eq~\eqref{eq:deltapsi_BL_I} admits two different types of
      poles: either on the real axis in
      ${ \omega = \omega_{0} = \bkp \cdot \bO (\bJp) }$, or in the lower half
      complex plane in ${ \omega = \omega_{\rp} + \ri s_{\rp} }$, with
      ${ s_{\rp} \leq 0 }$. By distorting the contour $\mathcal{B}$ into the
      contour $\mathcal{B}'$ with large negative complex values, only the
      contributions from the residues of the poles remain. As
      ${ s_{\rp} \leq 0 }$, for $t$ large enough, the contributions from the
      poles in ${ \omega = \omega_{\rp} + \ri s_{\rp} }$ vanish, and only the
      pole on the real axis contributes.}
  \end{center}
\end{figure}

Once \eq~\eqref{eq:selfconsistent_dressed_coefficients} is inverted to obtain
the dressed susceptibility coefficients
${ \psi^{\rd}_{\bk\bkp} (\bJ, \bJp, \omega) }$, one may then take the inverse
Laplace transform of \eq~\eqref{eq:ansatz_psi} to obtain the time dependence of
the potential fluctuations ${ \eta_{\bk} (\bJ, t) }$ in the bath. Following
the convention from \eq~\eqref{eq:definition_Laplace}, it reads
\begin{equation}
  \label{eq:deltapsi_BL_I}
  \eta_{\bk} (\bJ, t) = 
  - {(2 \pi)}^{d} \sum_{\bkp} \dint \bJp \, 
  \delta F_{\bkp} (\bJp, 0) \, 
  \frac{1}{2 \pi} \dint[\mathcal{B}] \omega \, 
  \frac{\psi_{\bk\bkp}^{\rd} (\bJ, \bJp, \omega)}
  {\ri (\omega \!-\! \bkp \ccdot \bO (\bJp))} \, 
  \re^{- \ri \omega t} ,
\end{equation}
where the Bromwich contour $\mathcal{B}$ in the complex $\omega$-plane has to
pass above all the poles of the integrand. Figure~\ref{fig:PolesDressing}
illustrates how the integration over $\omega$ may be performed by deforming the
contour $\mathcal{B}$.
Assuming that the bath is linearly stable,
${\psi^{\rd}_{\bk\bkp} (\bJ, \bJp, \omega) }$ has poles only in the lower half
of the complex $\omega$ plane~\citep[see Section~5.3.2.\
in][]{BinneyTremaine2008}. These poles are of the generic form
${ \omega = \omega_{\rm p} + \ri s_{\rm p} }$ with ${ s_{\rp} \leq 0 }$ (see
the response matrix from \eq~\eqref{eq:Fourier_M}). Only one pole of the
integrand from \eq~\eqref{eq:deltapsi_BL_I} is on the real axis, namely in
${ \omega = \bkp \cdot \bO (\bJp) }$, while all the other poles are below the
real axis. Following Figure~\ref{fig:PolesDressing}, the contour $\mathcal{B}$
can be distorted into the contour $\mBp$, so that there remains only
contributions from the residues. Paying careful attention to the direction of
integration, each pole contributes a ${ - 2 \pi \ri \text{Res} [\ldots] }$, and
\eq~\eqref{eq:deltapsi_BL_I} becomes
\begin{equation}
  \label{eq:deltapsi_BL_II}
  \eta_{\bk} (\bJ, t) = {(2 \pi)}^{d} \sum_{\bkp} \dint \bJp \, \delta
  F_{\bkp} (\bJp, 0) \bigg[ \re^{- \ri \bkp \cdot \bO (\bJp) t} \,
  \psi^{\rd}_{\bk\bkp} (\bJ, \bJp, \bkp \ccdot \bO (\bJp)) + \sum_{\rp}
  \re^{s_{\rp} t} \, \re^{- \ri \omega_{\rp} t} \times ( \ldots )  \bigg] ,
\end{equation}
where the sum over ``$\rp$'' runs over all the poles of
${ \psi_{\bk\bkp}^{\rd} (\bJ, \bJp, \omega) }$. Since ${ s_{\rp} \leq 0 }$,
these modes are damped, and their contributions vanish for
${ t \gg 1/|s_{\rp}| }$. This implies, that a self-interacting bath,
initially uncorrelated, will develop correlations that will settle to a
steady state on timescale ${ t \gg 1/|s_{\rp}| }$, which can be assumed to be
of the order of the dynamical timescale ${ \sim 1/ \Omega }$. Once these damped
contributions have faded, \eq~\eqref{eq:deltapsi_BL_II} becomes
\begin{equation}
  \label{eq:deltapsi_BL}
  \eta_{\bk} (\bJ, t) = {(2 \pi)}^{d} \sum_{\bkp} \dint \bJp \,
  \psi_{\bk\bkp}^{\rd} (\bJ, \bJp, \bkp \ccdot \bO (\bJp)) \, 
  \re^{- \ri \bkp \cdot \bO (\bJp) t} \, \delta F_{\bkp} (\bJp, 0) .
\end{equation}
\Eq~\eqref{eq:deltapsi_BL} is the direct equivalent of
\eq~\eqref{eq:deltapsi_Landau}, when collective effects are accounted for. In
this context, considering the self-gravitating amplification amounts to
replacing the bare susceptibility coefficients
${ \psi_{\bk\bkp} (\bJ, \bJp) }$ from \eq~\eqref{eq:psi_basis} by their
dressed analogs ${ \psi_{\bk\bkp}^{\rd} (\bJ, \bJp, \bk \cdot \bO (\bJ)) }$
defined by the implicit relation from
\eq~\eqref{eq:selfconsistent_dressed_coefficients}. Because of the strong
analogies between \eqs~\eqref{eq:deltapsi_Landau} and~\eqref{eq:deltapsi_BL},
when self-gravity is accounted for, the statistics of the fluctuations in the
bath remain formally the same, except for a change in the pairwise interaction
potential which becomes dressed, i.e.\ one makes the change
${ \psi_{\bk\bkp} (\bJ, \bJp) \to \psi_{\bk\bkp}^{\rd} (\bJ, \bJp, \bk \cdot
  \bO (\bJ)) }$. In that sense, \eq~\eqref{eq:deltapsi_BL} could be interpreted
as describing the potential fluctuations present in a system where the bath
particles follow the mean field orbit, i.e.\
${ \bT (t) = \bT_{0} + \bO (\bJ) }$ and ${ \bJ (t) = \bJ (0)}$, but interact
via a different pairwise interaction potential dictated by the dressed
susceptibility coefficients
${ \psi_{\bk\bkp}^{\rd} (\bJ, \bJp, \bk \cdot \bO (\bJ)) }$. Starting from
the fluctuations from \eq~\eqref{eq:deltapsi_BL}, the correlation of the
potential fluctuations in the presence of collective effects becomes
\begin{align}
  \label{eq:Correl_psi_BL}
  C_{\bk \bk} (\bJ, \bJp, t - \tp) = {}
  &
    \left\langle \eta_{\bk} (\bJ, t) \, 
    \eta_{\bk}^{*} (\bJ, \tp) \right\rangle
    \nonumber  \\
  = {} 
  &
    \mb \, {(2 \pi)}^{d} \sum_{\bkp} \dint \bJp \, 
    {| \psi_{\bk\bkp}^{\rd} (\bJ, \bJp, \bkp \ccdot \bO (\bJp)) |}^{2} \, 
    F_{\rb} (\bJp) \, \re^{- \ri \bkp \cdot \bO (\bJp) (t - \tp)} .
\end{align}
Following \eq~\eqref{eq:diff_coeff_final}, one can immediately obtain the
\DCs\ of a test particle undergoing such stochastic
perturbations. They read
\begin{equation}
  \label{eq:Final_Diff_Coeff_BL}
  D_{ij} (\bJ) = m_{\rb} \, {(2 \pi)}^{d+1} \sum_{\bk, \bkp} k_{i} k_{j} 
  \dint \bJp \, 
  \deltaD (\bk \ccdot \bO (\bJ) - \bkp \ccdot \bO (\bJp)) \, 
  {| \psi_{\bk\bkp}^{\rd} (\bJ, \bJp) |}^{2} \, F_{\rb} (\bJp) .
\end{equation} 
In \eq~\eqref{eq:Final_Diff_Coeff_BL}, we recover the \DCs\ of
the inhomogeneous \BL\ equation, derived recently
in~\cite{Heyvaerts2010,Chavanis2012}. Let us emphasize in particular the
striking similarities between the dressed \DCs\ from
\eq~\eqref{eq:Final_Diff_Coeff_BL} and the bare ones obtained in
\eq~\eqref{eq:Final_Diff_Coeff_Landau}. The \DCs\ from
\eq~\eqref{eq:Final_Diff_Coeff_BL} describe the slow diffusion of a test
particle embedded in an external bath made of $N$ particles when collective
effects are accounted for, i.e.\ when the bath particles see the mean field
potential as well as the fluctuations they themselves generate (see
\eq~\eqref{eq:H_BL}). In the previous section, we discussed the physical
content of the bare \DCs\ of the inhomogeneous Landau
equation. Because of the fundamental similarities between
\eqs~\eqref{eq:Final_Diff_Coeff_Landau} and~\eqref{eq:Final_Diff_Coeff_BL},
this discussion directly translates to the dressed \DCs\ of
the inhomogeneous \BL\ equation. The only difference comes from the change in
the strength of the resonant couplings, which is now controlled by the squared
dressed susceptibility coefficients
${ {| \psi_{\bk\bkp}^{\rd} (\bJ, \bJp, \bk \cdot \bO (\bJ)) |}^{2} }$. Let us
note that in cold dynamical systems, i.e.\ systems able to strongly amplify
perturbations, dressing up the interactions might have a significant impact on
the long-term dynamics in the system, and could even lead to instability.
This was for example recently shown
in~\cite{FouvryPichonMagorrianChavanis2015} in the context of razor-thin
stellar disks. Collective effects, manifested in stellar disks by strong swing
amplification~\citep{Toomre1981}, can significantly hasten the
diffusion compared to the bare diffusion, leading in particular to the
formation of narrow resonant ridge features in action space. In previous
applications of the \BL\ formalism~\citep[see
e.g.,][]{FouvryPichonMagorrianChavanis2015,BenettiMarcos2017}, most of the
effort in the computation of the dressed \DCs\ was dedicated
to the computation of the dressed susceptibility coefficients
${ \psi_{\bk\bkp}^{\rd} (\bJ, \bJp, \omega) }$, which asked for the inversion
of the self-consistency definition from
\eq~\eqref{eq:selfconsistent_dressed_coefficients} via the basis method from
Appendix~\ref{sec:BasisMethod}. In Section~\ref{sec:HMF}, we illustrate how the
present $\eta$-formalism allows us to determine the dressed \DCs\ from
\eq~\eqref{eq:Final_Diff_Coeff_BL} in the case of the
one-dimensional inhomogeneous \HMF\ model. This is one of the main strengths of
the present $\eta$-formalism. It allows for the characterization of the dressed
\DCs\ without ever having to invert the self-consistency
relation from \eq~\eqref{eq:selfconsistent_dressed_coefficients}, i.e.\ without
ever having to compute explicitly the dressed susceptibility coefficients
${ \psi_{\bk\bkp}^{\rd} (\bJ, \bJp, \omega) }$.

Let us finally note that for a bath such that
${ \partial F_{\rb} / \partial \bJ = 0 }$,
\eq~\eqref{eq:selfconsistent_dressed_coefficients} gives
${ \psi_{\bk\bkp}^{\rd} (\bJ, \bJp, \omega) = \psi_{\bk\bkp} (\bJ, \bJp) }$.
In that limit, collective effects can be
neglected and the Landau and \BL\ \DCs\ are identical. As
will be show in Section~\ref{sec:DynFric}, in these systems, the friction force
by polarization also vanishes. Such a limit is of particular importance
to describe the scalar resonant relaxation of isotropic spherical
quasi-Keplerian systems~\citep{BarOr2014}.

\section{Dynamical Friction}
\label{sec:DynFric}

In the previous sections we assumed that the test particle has no influence on
the bath. In that limit, the potential fluctuations induced by the bath are
independent of the motion of the test particle. This led us to obtain a
diffusion equation (\eq~\eqref{eq:fp_non_markov}) for the test particle's \PDF\
with no advection term, i.e.\ with no drift term proportional to the mass of
the test particle~\citep{BinneyLacey1988,BinneyTremaine2008}. Let us note that
such a limit is correct for a bath such that 
${ \partial F_{\rb} / \partial \bJ = 0}$, or in the limit of a test particle of
zero mass, or for a purely external bath. Yet, if the test particle can
influence the bath particles, there will be an advection term associated with
the perturbations of the bath along the trajectory of the test particle. This
component is the so-called friction force by polarization, which captures in
particular the process of dynamical
friction~\citep{Chandrasekhar1943I,TremaineWeinberg1984,NelsonTremaine1999,BinneyTremaine2008}.
For a closed system, such a friction component is necessary because of the
constraint of energy conservation, i.e.\ any diffusion in the system must be
associated with a friction. We refer to~\cite{Heyvaerts2017} and references
therein for a thorough discussion of this contribution. The calculations
presented in this section follow~\cite{Weinberg1989,Chavanis2012,Heyvaerts2017}
and will therefore be presented in a concise manner.

We are interested in how the potential perturbations in the bath change because
of the influence of the test particle. In the presence of collective effects,
i.e.\ assuming that bath particles interact among themselves, and 
if the test particle can perturb the bath particles, the specific
Hamiltonian of the bath particles is
\begin{equation}
  \label{eq:H_BL_Fric}
  H_{\rb} (\bT, \bJ, t) = 
  H_{0} (\bJ) + \delta \phi_{\rt} (\bT, \bJ, t)
  + \delta \phi_{\rp} (\bT, \bJ, t) ,
\end{equation}
where here, ${ \delta \phi_{\rt} (\bT, \bJ, t) }$ is the
perturbation of the test particle on the bath particles and
${ \delta \phi_{\rp} (\bT, \bJ, t) }$ is the polarization response from the
bath, which captures the bath's self-gravity. The latter is
composed of two contributions: (i) the response of the bath to the finite-$N$
fluctuations associated with the discrete number of bath particles (i.e.\ the potential
perturbations ${ \eta (\bT, \bJ, t) }$ from \eq~\eqref{eq:H_BL}), (ii) the
response of the bath to the perturbation ${ \delta \phi_{\rt} }$ due to the
test particle.
Assuming that these potential perturbations are small compared to the mean
Hamiltonian $H_{0}$, similarly to \eq~\eqref{eq:Klim_BL}, the evolution of the bath
is given by the linearized Klimontovich equation reading
\begin{align}
  \label{eq:Klim_Fric}
  0  = {}
  & 
    \dpd{\delta F}{t} + \big[ \delta F, H_{0} \big] + 
    \big[ F_{\rb}, \delta \phi_{\rt} + \delta \phi_{\rp} \big]
  \nonumber  \\
  = {}
  & 
    \dpd{\delta F}{t} + \dpd{\delta F}{\bT} \cdot \bO (\bJ) 
    - \dpd{F_{\rb}}{\bJ} 
    \cdot \dpd{[ \delta \phi_{\rt} + \delta \phi_{\rp} ]}{\bT} ,
\end{align}
where ${ \delta F }$ stands for the fluctuations in the bath's \DF\@. 

Following \eq~\eqref{eq:Klim_BL_Laplace}, it is straightforward to write the
Laplace-Fourier transform of \eq~\eqref{eq:Klim_Fric} to obtain
\begin{equation}
  \label{eq:Klim_Fric_Laplace}
  \delta \tF_{\bk} (\bJ, \omega) = 
  - \frac{\delta F_{\bk} (\bJ, 0)}{\ri (\omega - \bk \ccdot \bO (\bJ))} 
  - \frac{\bk \ccdot \partial F_{\rb} / \partial \bJ}{\omega - \bk \ccdot \bO(\bJ)}
  \, \delta \tphi_{\bk}^{\rt} (\bJ, \omega)
  - \frac{\bk \ccdot \partial F_{\rb} / \partial \bJ}{\omega - \bk \ccdot \bO(\bJ)} \, 
  \delta \tphi_{\bk}^{\rp} (\bJ, \omega) .
\end{equation}
Similarly to \eq~\eqref{eq:link_fluc_ft}, the bath's polarization response
${ \delta \phi_{\rp} }$ results from the fluctuations of the bath's \DF,
${ \delta F }$, so that we can write
\begin{equation}
  \label{eq:link_fluctuations_II_Fric}
  \delta \tphi_{\bk}^{\rp} (\bJ, \omega) = 
  {(2 \pi)}^{d} \sum_{\bkp} \dint \bJp \, 
  \psi_{\bk\bkp} (\bJ, \bJp) \, \delta \tF_{\bkp} (\bJp, \omega) .
\end{equation}
Let us then act on both sides of \eq~\eqref{eq:Klim_Fric_Laplace} with the same
operator as in the r.h.s.\ of \eq~\eqref{eq:link_fluctuations_II_Fric}. Similarly to
\eq~\eqref{eq:selfconsistent_psi_BL}, we get the self-consistency relation
\begin{align}
  \label{eq:selfconsistent_psi_Fric}
  \delta \tphi_{\bk}^{\rp} (\bJ, \omega) = {}
  &  
    - {(2 \pi)}^{d} \sum_{\bkp} \dint \bJp \, 
    \frac{\bkp \ccdot \partial F_{\rb} / \partial \bJp}{\omega - \bkp\ccdot\bO(\bJp)} \,
    \psi_{\bk\bkp} (\bJ, \bJp) \, \delta \tphi_{\bkp}^{\rp} (\bJp, \omega)
    \nonumber \\
  & 
    - {(2 \pi)}^{d} \sum_{\bkp} \dint \bJp \, 
    \frac{\delta F_{\bkp} (\bJp, 0)}{\ri (\omega - \bkp \ccdot \bO (\bJp))} \, 
    \psi_{\bk\bkp} (\bJ, \bJp)
    \nonumber \\
  &
    - {(2 \pi)}^{d} \sum_{\bkp} \dint \bJp \, 
    \frac{\bkp \ccdot \partial F_{\rb} / \partial \bJ}{\omega - \bkp \ccdot \bO (\bJp)} \, 
    \psi_{\bk\bkp} (\bJ, \bJp) \, \delta \tphi_{\bkp}^{\rt} (\bJp, \omega) ,
\end{align}
which takes again the form of a Fredholm equation of the second kind for the
polarization response ${ \delta \tphi_{\bk}^{\rp} (\bJ, \omega) }$. In the
r.h.s.\ of that equation, the first line corresponds to the kernel of the
relation and captures the strength of the self-gravitating amplification in the
bath and is sourced by the gradients of the bath's \DF\@. This equation
possesses two source terms, namely the initial finite-$N$ fluctuations in the bath
(second term) and the potential perturbation from the test particle (third
term). In order to ease the inversion of this equation, and simplify the
discussion of its physical content, let us now rely on the basis method
introduced in Appendix~\ref{sec:BasisMethod}.

Relying on \eq~\eqref{eq:explicit_psi} to express the bare susceptibility
elements ${ \psi_{\bk\bkp} (\bJ, \bJp) }$ with the basis elements, we
introduce the vectors $\tbP$, $\tbT$ as
\begin{equation}
  \label{eq:definition_P_T}
  \delta \tphi_{\bk}^{\rp} (\bJ, \omega) = 
  \sum_{\alpha} \tbP_{\alpha} (\omega) \, \psi_{\bk}^{(\alpha)} (\bJ) 
  \;\;\; ;  \;\;\; 
  \delta \tphi_{\bk}^{\rt} (\bJ, \omega) = 
  \sum_{\alpha} \tbT_{\alpha} (\omega) \, \psi_{\bk}^{(\alpha)} (\bJ) .
\end{equation}
To decompose the fluctuating source term from
\eq~\eqref{eq:selfconsistent_psi_Fric}, we also introduce the vector $\tbS$
\begin{equation}
  \label{eq:definition_S}
  \tbS_{\alpha} (\omega) = 
  {(2 \pi)}^{d} \sum_{\bkp} \dint \bJp \, 
  \frac{\delta F_{\bkp} (\bJp, 0)}{\ri (\omega - \bkp \ccdot \bO (\bJp))} \, 
  \psi_{\bkp}^{(\alpha) *} (\bJp) .
\end{equation}
describing the initial finite-$N$ fluctuations in the system's \DF\@. The
self-consistency \eq~\eqref{eq:selfconsistent_psi_Fric} then takes the short
form
\begin{equation}
  \label{eq:selfconsistent_psi_Fric_short}
  \tbP (\omega) = \wbM (\omega) \ccdot \tbP (\omega) + 
  \tbS (\omega) + \wbM (\omega) \ccdot \tbT (\omega) ,
\end{equation}
where $\wbM$ is the bath's response matrix introduced in
\eq~\eqref{eq:Fourier_M}. Here, ${ \tbP (\omega) }$ characterizes the
polarization response of the bath, ${ \tbS (\omega) }$ the initial finite-$N$
fluctuations in the system's \DF, and ${ \tbT (\omega) }$ the perturbation from
the test particle. This third term is the one that captures the back-reaction of
the test particle on the bath and leads to the associated friction force by
polarization. Assuming that the bath is linearly stable (i.e.\
${ \omega \to {[\bI - \wbM (\omega)]}^{-1} }$ has no poles in the upper half
complex plane), \eq~\eqref{eq:selfconsistent_psi_Fric_short} is straightforward
to invert for ${ \tbP (\omega) }$. It becomes
\begin{equation}
  \label{eq:inversion_psi_Fric_short}
  \tbP (\omega) = 
  {\big[ \bI - \wbM (\omega) \big]}^{-1} \!\!\cdot \tbS  (\omega) 
  + \big\{ {\big[ \bI - \wbM (\omega) \big]}^{-1} - \bI \big\} 
  \ccdot \tbT (\omega) ,
\end{equation}
which in the absence of collective effects is simply
${ \tbP (\omega) = \tbS (\omega) + \wbM (\omega) \cdot \tbT (\omega) }$. 

Before writing down the time version of
\eq~\eqref{eq:inversion_psi_Fric_short}, we need first to express explicitly
the source term due to the test particle, ${ \tbT (\omega) }$. Noting the position of
the test particle at time $t$ with ${ (\bT_{\rt} (t), \bJ_{\rt} (t)) }$, we
can write ${ \delta \phi_{\bk}^{\rt} (\bJ, t) }$ as
\begin{equation}
  \label{eq:expression_deltaphi_t}
  \delta \phi_{\bk}^{\rt} (\bJ, t) = 
  m_{\rt} \sum_{\bkp} \psi_{\bk\bkp} (\bJ, \bJ_{\rt} (t)) \, 
  \re^{- \ri \bkp \cdot \bT_{\rt} (t)} ,
\end{equation}
where $m_{\rt}$ is the mass of the test particle. Relying on the timescale
separation between the fast timescale of the mean field orbital motion and the
slow timescale of diffusion, let us then replace in
\eq~\eqref{eq:expression_deltaphi_t} the motion of the test particle by its
mean field motion, so that
${ \bT_{\rt} (t) = \bT_{\rt}^{0} + \bO (\bJ_{\rt}) }$ and
${ \bJ_{\rt} (t) = \bJ_{\rt} }$, where $\bT_{\rt}^{0}$ is the initial phase of
the test particle. Following \eq~\eqref{eq:definition_P_T}, the coefficients
${ \tbT_{\alpha} (\omega) }$ are then straightforward to write, and one gets
\begin{equation}
  \label{eq:expression_T_Fric}
  \tbT_{\alpha} (\omega) = 
  m_{\rt} \sum_{\bkp} \frac{\re^{- \ri \bkp \cdot \bT_{\rt}^{0}}}
  {\ri (\omega - \bkp \ccdot \bO (\bJ_{\rt}))} \, 
  \psi_{\bkp}^{(\alpha) *} (\bJ_{\rt}) .
\end{equation}

Having specified all the terms appearing in
\eq~\eqref{eq:inversion_psi_Fric_short}, we may then take the inverse Laplace
transform of this equation. This calculation is essentially identical to the
one performed in \eq~\eqref{eq:deltapsi_BL_II}. In
\eq~\eqref{eq:inversion_psi_Fric_short}, there exist two kinds of poles: (i)
poles on the real axis coming from the source terms ${ \tbS (\omega) }$ and
${ \tbT (\omega) }$, (ii) poles below the real axis coming from the
susceptibility matrix ${ {[ \bI - \wbM (\omega) ]}^{-1} }$. We then consider
times long enough for the contributions from the damped modes to vanish. After
a straightforward calculation, we obtain from
\eq~\eqref{eq:inversion_psi_Fric_short} that
\begin{align}
  \label{eq:inversion_psi_Fric_time}
  \delta \phi_{\bk}^{\rp} (\bJ, t) = {}
  & 
    {(2 \pi)}^{d} \sum_{\bkp} \dint \bJp \, 
    \psi_{\bk\bkp}^{\rd} (\bJ, \bJp, \bkp \ccdot \bO (\bJp)) \, 
    \re^{- \ri \bkp \cdot \bO (\bJp) t} \, \delta F_{\bkp} (\bJp, 0)
    \nonumber \\
  &
    + m_{\rt} \sum_{\bkp} \big\{ \psi_{\bk\bkp}^{\rd} (\bJ, \bJ_{\rt}, \bkp \ccdot \bO (\bJ_{\rt})) 
    - \psi_{\bk\bkp} (\bJ, \bJ_{\rt}) \big\} \, \re^{- \ri \bkp \cdot \bT_{\rt} (t)}
    \nonumber \\
  = {} 
  & 
    \eta_{\bk} (\bJ, t) + h_{\bk}^{\rm fric} (\bJ, t) ,
\end{align}
where we used \eq~\eqref{eq:explicit_psid} to express the dressed
susceptibility coefficients without resorting to the basis elements. In
\eq~\eqref{eq:inversion_psi_Fric_time}, we introduced the two components of the
polarization response of the bath. Here, ${ \eta_{\bk} (\bJ, t) }$ was already
obtained in \eq~\eqref{eq:deltapsi_BL} and stands for the dressed potential
fluctuations present in the system as a result of the finite-$N$ fluctuations
from the bath (i.e.\ it is sourced by ${ \delta F_{\bkp} (\bJp, 0) }$). The
second contribution, ${ h_{\bk}^{\rm fric} (\bJ, t) }$, captures the
friction force by polarization, and describes the dressed potential
perturbations present in the bath as a result of the presence of the test
particle. Let us note that these two potential perturbations have some
fundamental differences. On the one hand, ${ \eta_{\bk} (\bJ, t) }$ is truly a
stochastic perturbation. It is of zero mean, depends on the bath's realization,
and its amplitude is proportional to $\sqrt{m_{\rb}}$. On the other hand,
because it only depends on the mean field parameters of the bath,
${ h_{\bk}^{\rm fric} (\bJ, t) }$ should be seen as a non-stochastic
perturbation. It is of non-zero mean, and its amplitude is proportional
to $m_{\rt}$. The more massive the test particle, the stronger the friction
force. In Section~\ref{sec:BLEq}, we have already shown how the
correlation of ${ \eta_{\bk} (\bJ, t) }$ leads to the \DC\ of
the inhomogeneous \BL\ equation. Let us now focus on the contribution of
${ h_{\bk}^{\rm fric} (\bJ, t) }$ to the diffusion equation for the test
particle. The associated contribution in \eq~\eqref{eq:F_dot} takes the form
\begin{align}
  \label{eq:comp_Fric_I}
  \ri \sum_{\bk} \bk \left\langle h_{\bk}^{\rm fric} (\bJ, t) \, \re^{\ri
  \bk \cdot \bT (t)} \, \varphi (\bJ, t) \right\rangle = {}
  & 
    \ri m_{\rt} \sum_{\bk, \bkp} \bk 
    \left\langle \big\{ \psi_{\bk\bkp}^{\rd} (\bJ, \bJ, \bkp \ccdot \bO (\bJ)) 
    - \psi_{\bk\bkp} (\bJ, \bJ) \big\} \, 
    \re^{\ri (\bk - \bkp) \cdot \bT (t)} \, 
    \varphi (\bJ, t) \right\rangle
    \nonumber \\
  = {} 
  &
    - \bF_{\rm pol} (\bJ, t) \, P (\bJ, t) .
\end{align}
To obtain the second line of \eq~\eqref{eq:expression_T_Fric}, we followed the
same assumption as in \eq~\eqref{eq:expression_T_Fric}, and assumed that the
motion of the test particle is given by ${ \bT (t) = \bT_{0} + \bO (\bJ) t }$
and ${ \bJ (t) = \bJ }$. The ensemble average from \eq~\eqref{eq:comp_Fric_I}
then only amounts to averaging over the initial phase $\bT_{0}$ of the test
particle: it is straightforward and imposes ${ \bk = \bkp }$. Finally, we also
used the definition of the test particle's \PDF,
${ P (\bJ, t) = \langle \varphi (\bJ, t) \rangle }$, and introduced the
friction force by polarization ${ \bF_{\rm pol} (\bJ, t) }$ defined as
\begin{align}
  \label{eq:def_Fpol}
  \bF_{\rm pol} (\bJ, t) = {} 
  &
    - \ri m_{\rt} \sum_{\bk} \bk \, 
    \big\{ \psi_{\bk\bk}^{\rd} (\bJ, \bJ, \bk \ccdot \bO (\bJ)) 
    - \psi_{\bk\bk} (\bJ, \bJ) \big\}
    \nonumber  \\
   = {}
  & 
    m_{\rt} \sum_{\bk} \bk \, 
    \text{Im} \big[ \psi_{\bk\bk}^{\rd} (\bJ, \bJ, \bk \ccdot \bO (\bJ)) \big] ,
\end{align}
where we used the fact that ${ \psi_{\bk\bk} (\bJ, \bJ) }$ and
${ \bF_{\rm pol} (\bJ, t) }$ are real.
In \Eq~\eqref{eq:def_Fpol}, we recover a result already obtained in~\eq~{(53)} of~\cite{Weinberg1989}.
We do not pursue further the
calculation of ${ \bF_{\rm pol} (\bJ, t) }$ and refer to \eq~(54)
of~\cite{Chavanis2012} to obtain an integral expression for
${ \text{Im} [ \psi_{\bk\bk}^{\rd} (\bJ, \bJ, \bk \cdot \bO (\bJ)) ] }$.
The friction force by polarization finally becomes
\begin{equation}
  \label{eq:final_Fpol}
  \bF_{\rm pol} (\bJ, t) = 
  m_{\rt} \pi {(2 \pi)}^{d} \sum_{\bk, \bkp} \bk \dint \bJp \, 
  \deltaD (\bk \ccdot \bO (\bJ) - \bkp \ccdot \bO (\bJp)) \, 
  {| \psi_{\bk\bkp}^{\rd} (\bJ, \bJp, \bk \ccdot \bO (\bJ)) |}^{2} \, 
  \bigg( \bkp \ccdot \dpd{F_{\rb} (\bJp, t)}{\bJp} \bigg) .
\end{equation}
Deriving in \eq~\eqref{eq:final_Fpol} the expression of the dressed friction
force by polarization is the main result of this section. Here, we recovered
the expression of the bare resonant dynamical friction in particular obtained in \eq~{(30)}
of~\cite{LyndenBellKalnajs1972}, and \eq~{(65)} of~\cite{TremaineWeinberg1984}, 
as well as its dressed generalisation obtained among others in \eq~{(53)} of~\cite{Weinberg1989},
\eq~{(24)} of~\cite{SeguinDupraz1994}, or in \eq~{(113)} of~\cite{Chavanis2012}.

\section{The Balescu-Lenard equation}
\label{sec:BLeq}

In the previous sections, we derived successively the two components involved
in the long-term evolution of a massive test particle embedded in a
self-gravitating discrete bath. First, in \eq~\eqref{eq:Final_Diff_Coeff_BL}, we derived
the diffusion coefficients, ${ D_{ij} (\bJ) }$, sourced by the temporal correlations of
the finite-$N$ dressed perturbations present in the bath. Second, in \eq~\eqref{eq:F_dot},
we derived the friction force by polarization, ${ F_{\rm pol} (\bJ) }$, which captures the
dressed back-reaction of the perturbations in the bath induced by the massive
test particle. Gathering these two components, the diffusion
equation~\eqref{eq:F_dot} takes the form
\begin{align}
  \label{eq:final_BL}
  \dpd{P(\bJ, t)}{t} = {} 
  & 
    \dpd{}{\bJ} 
    \ccdot \bigg[ - \bF_{\rm pol} (\bJ, t) \, P  (\bJ, t) + 
    \frac{1}{2} \bD (\bJ, t) \ccdot \dpd{P(\bJ, t)}{\bJ} \bigg]
    \nonumber \\
  = {}
  &  
    \pi {(2 \pi)}^{d} \dpd{}{\bJ} 
    \ccdot \bigg[ \sum_{\bk, \bkp} \! \int \!\! \rd \bJp \, 
    \deltaD (\bk \ccdot \bO (\bJ) - \bkp \ccdot \bO (\bJp)) \, 
    {| \psi_{\bk\bkp}^{\rd} (\bJ, \bJp, \bk \ccdot \bO (\bJ)) |}^{2}
    \nonumber \\
  & 
    \times \bigg( m_{\rb} \, \bk \ccdot \dpd{}{\bJ} 
    - m_{\rt} \, \bkp \ccdot \dpd{}{\bJp} \bigg) \, 
    P (\bJ, t) \, F_{\rb} (\bJp, t) \bigg] .
\end{align}
In \eq~\eqref{eq:final_BL}, we recover exactly the inhomogeneous \BL\
equation, already obtained in \eq~{(38)} of~\cite{Heyvaerts2010} and \eq~{(56)}
of~\cite{Chavanis2012}. This equation describes the evolution of a
given test particle embedded in a discrete system of $N$ particles. As
emphasized previously, in the limit where collective effects are not accounted
for, the \BL\ \eq~\eqref{eq:final_BL} becomes the inhomogeneous Landau
equation. Such a limit is obtained by replacing in \eq~\eqref{eq:final_BL} the
dressed susceptibility coefficients
${ \psi_{\bk\bkp}^{\rd} (\bJ, \bJp, \omega) }$ by their bare analogs
${ \psi_{\bk\bkp} (\bJ, \bJp) }$ introduced in \eq~\eqref{eq:psi_basis}. In
particular, for a bath such that
${ \partial F_{\rb} / \partial \bJ = 0 }$, these two equations are equivalent.

The \BL\ \eq~\eqref{eq:final_BL} is composed of two components. First, a
diffusion component associated with the term proportional to
${ m_{\rb} \bk \cdot \partial / \partial \bJ }$. This diffusion component is
sourced by the correlations in the bath potential fluctuations. It is
proportional to the mass of the bath particles ${ m_{\rb} = M / N }$, and
vanishes in the limit of a collisionless bath, i.e.\ in the limit
${ N \to + \infty }$. The second component of \eq~\eqref{eq:final_BL} is the
friction component and is proportional to
${ m_{\rt} \, \bkp \cdot \partial / \partial \bJp }$. This friction component is
sourced by the back-reaction of the test particle on the bath. It is therefore
proportional to the mass $m_{\rt}$ of the test particle, and is responsible for
mass segregation. It does not vanish in the limit of a collisionless bath, i.e.\
in the limit ${ N \to + \infty }$. This friction component also vanishes in the
limit of a bath satisfying ${ \partial F_{\rb} / \partial \bJ = 0 }$.

\Eq~\eqref{eq:final_BL} describes the evolution of the statistics of a
test particle (described via the \PDF\ ${ P (\bJ, t) }$), when embedded in a
bath (described by the \DF\ ${ F_{\rb} (\bJ, t) }$). It is then
straightforward to use \eq~\eqref{eq:final_BL} to obtain the self-consistent
evolution equation satisfied by the bath's \DF\ when diffusing on long-term
timescales. This only amounts to assuming that the statistics of the test
particle is given by the statistics of the bath particles, i.e.\ one performs
the replacement ${ P (\bJ, t) \to F_{\rb} (\bJ, t) }$. Such a replacement
transforms the differential \eq~\eqref{eq:final_BL} into a self-consistent
integro-differential equation for the bath's \DF\@. This is the self-consistent
inhomogeneous \BL\ equation~\citep{Heyvaerts2010, Chavanis2012}, reading
\begin{align}
  \label{eq:final_BL_selfconsistent}
  \dpd{F_{\rb} (\bJ, t)}{t} = {}
  &  
    \pi {(2 \pi)}^{d} \dpd{}{\bJ} 
    \ccdot \bigg[ \sum_{\bk, \bkp} \! \int \!\! \rd \bJp \, \deltaD (\bk \ccdot \bO (\bJ) 
    - \bkp \ccdot \bO (\bJp)) \, 
    {| \psi_{\bk\bkp}^{\rd} (\bJ, \bJp, \bk \ccdot \bO (\bJ)) |}^{2}
    \nonumber \\
  & 
    \times \bigg( m_{\rb} \, \bk \ccdot \dpd{}{\bJ} 
    - m_{\rb} \, \bkp \ccdot \dpd{}{\bJp} \bigg) \, 
    F_{\rb} (\bJ, t) \, F_{\rb} (\bJp, t) \bigg] .
\end{align}
As emphasized in~\cite{Heyvaerts2017}, let us finally recall that the
self-consistent \BL\ \eq~\eqref{eq:final_BL_selfconsistent} satisfies
a $H$-theorem for Boltzmann's entropy.
As a result, the \BL\ equation admits the Boltzmann's \DF\
${ F_{\rb} (\bJ) \propto \re^{- \beta H_{0} (\bJ)} }$ as an equilibrium
solution. Moreover, for such a thermal DF, it is straightforward to show that
the diffusion tensor, ${ D_{ij} (\bJ) }$, from \eq~\eqref{eq:Final_Diff_Coeff_BL} and the
friction force by polarization, ${ F_{\rm pol} (\bJ) }$, from \eq~\eqref{eq:final_Fpol}
satisfy a generalized fluctuation-dissipation relation of the form
\begin{equation}
{\big[ \bF_{\rm pol} (\bJ) \big]}_{i} = - \frac{1}{2} \beta \, \Omega_{j} (\bJ) \, D_{ij} (\bJ) ,
\label{Einstein_relation}
\end{equation}
where the sum over $j$ is implied. This equation was already put forward
in \eq~{(3.12)} of~\cite{BinneyLacey1988}, and \eq~{(119)} of~\cite{Chavanis2012}.

It is also straightforward to generalize
\eq~\eqref{eq:final_BL_selfconsistent} to a multi-mass bath. Indeed, let us
assume that the bath is composed of multiple components of individual mass
$m_{\alpha}$, $m_{\beta}$, etc. Each component is described by a
quasi-stationary \DF\ of the form ${ F_{\alpha} (\bJ, t) }$, following the
convention ${ \! \int \! \rd \bT \rd \bJ \, F_{\alpha} = M_{\alpha} }$, where
$M_{\alpha}$ is the total mass of the component
$\alpha$. \Eq~\eqref{eq:final_BL_selfconsistent} can then be generalized
to describe the self-consistent long-term evolution of the component $\alpha$,
under the effects of the stochastic perturbations from itself and all other
components. It reads
\begin{align}
  \label{eq:final_BL_multimass}
  \dpd{F_{\alpha} (\bJ, t)}{t} = {}
  &  
    \pi {(2 \pi)}^{d} \dpd{}{\bJ} \ccdot 
    \bigg[ \sum_{\bk, \bkp} \! \int \!\! \rd \bJp \, 
    \deltaD (\bk \ccdot \bO (\bJ) - \bkp \ccdot \bO (\bJp)) \, 
    {| \psi_{\bk\bkp}^{\rd} (\bJ, \bJp, \bk \ccdot \bO (\bJ)) |}^{2}
    \nonumber \\
  & 
    \times \sum_{\beta} \bigg( m_{\beta} \, \bk \ccdot \dpd{}{\bJ} 
    - m_{\alpha} \, \bkp \ccdot \dpd{}{\bJp} \bigg) \, 
    F_{\alpha} (\bJ, t) \, F_{\beta} (\bJp, t) \bigg] .
\end{align}
where the sum on ``$\beta$'' runs over all components. In the multi-component
case, the dressed susceptibility coefficients
${ \psi_{\bk\bkp}^{\rd} (\bJ, \bJp, \omega) }$ involve now all the active
components in the self-gravitating amplification. As a consequence, the
self-consistency definition from
\eq~\eqref{eq:selfconsistent_dressed_coefficients} becomes here
\begin{equation}
  \label{eq:selfconsistent_dressed_coefficients_multimass}
  \psi_{\bk\bkp}^{\rd} (\bJ, \bJp, \omega) = 
  - {(2 \pi)}^{d} \sum_{\bkpp} \dint \bJpp \, 
  \frac{\bkpp \ccdot \partial ( \sum_{\beta} F_{\beta} (\bJpp)) /\partial
    \bJpp}{\omega - \bkpp \ccdot \bO (\bJpp)} \, 
  \psi_{\bk\bkpp} (\bJ, \bJpp) \, 
  \psi_{\bkpp\bkp}^{\rd} (\bJpp, \bJp, \omega) 
  + \psi_{\bk\bkp} (\bJ, \bJp) .
\end{equation}
Let us note that the multi-mass
\eq~\eqref{eq:final_BL_multimass} allows for mass segregation between the
different components, because the friction force is proportional to the mass
of the considered component. We refer to~\cite{Heyvaerts2017} and references
therein for a detailed discussion of the physical content of
the inhomogeneous \BL\ equation.

As a final remark, as pointed out in~\cite{Weinberg1993,Heyvaerts2010,Chavanis2013},
one can assume local homogeneity in the inhomogeneous \BL\ \eq~\eqref{eq:final_BL_selfconsistent},
to recover a diffusion equation in velocity space, the so-called homogeneous Balescu-Lenard equation,
as given for example in \eq~{(47)} in~\cite{Heyvaerts2010}, and \eq~{(E.1)} in~\cite{Chavanis2013}.
In the limit where collective effects are not accounted for, this equation reduces to
the homogeneous Landau equation (see for example \eq~{(39)} in~\cite{Chavanis2013}).
This diffusion equation comes at the price of truncating the interactions on both large scales
(to account for the finite size of the system), and small scales (to account for strong collisions),
leading to the appearance of the Coulomb logarithm.
As shown in~\cite{Chavanis2013}, the homogeneous Landau equation is equivalent to
the classical Fokker-Planck diffusion coefficients sourced by weak encounters,
as given by the Rosenbluth potentials in \eq~{(7.83a)} in~\cite{BinneyTremaine2008}.
We do not pursue here the discussion of the homogeneous limit, and refer
to~\cite{Chavanis2013} for a detailed investigation of the equivalences
of these various approaches.

\section{Dressed diffusion by external perturbations}
\label{sec:dFPeq}

In the previous sections, we investigated the long-term evolution of a test
particle subject to (bare or dressed) stochastic potential perturbations from
an external bath constituted of a finite number of particles. This allowed us
to recover in \eqs~\eqref{eq:Final_Diff_Coeff_Landau}
and~\eqref{eq:Final_Diff_Coeff_BL} the \DCs\ of the inhomogeneous Landau and
\BL\ equations, and in \eq~\eqref{eq:final_BL_selfconsistent} the full
self-consistent kinetic equation. The associated \DCs\ are often said to
capture an internally induced long-term evolution, in the sense that finite-$N$
fluctuations can be seen as self-generated perturbations. Yet, as already
emphasized in the Introduction, collisionless systems (i.e.\ in the limit
${ N \to \infty }$) can also undergo a long-term diffusion as a result of
external potential fluctuations. Such diffusion was first
characterized in~\cite{BinneyLacey1988}, and generalized
in~\cite{Weinberg2001a} to account for collective effects. See also
e.g.,~\cite{PichonAubert2006,Chavanis2012EPJP,Nardini2012,FouvryPichonPrunet2015}
for a revisit of this equation. Let us now show how the present
$\eta$-formalism and the result from \eq~\eqref{eq:diff_coeff_final} allow for
a straightforward recovery of these \DCs\@.

We are interested in describing the long-term evolution of a collisionless
self-gravitating quasi-stationary system undergoing some external stochastic
perturbations ${ \delta \phi_{\rm ext} (\bx, t) }$. Assuming the mean field
system to be integrable, we may then expand the \DF\ and the specific
Hamiltonian of this collisionless system as
\begin{equation}
  \label{eq:DF_dFP}
  F_{\rm sys} (\bT, \bJ, t) = F (\bJ, t) + \delta F (\bT, \bJ, t),
\end{equation}
and
\begin{align}
  \label{eq:H_dFP}
  H_{\rm sys} (\bT, \bJ, t) = {} 
  & 
    H_{0} (\bJ) + \delta \phi_{\rm ext} (\bT, \bJ, t) 
    + \delta \phi_{\rp} (\bT, \bJ, t)
    \nonumber \\
  = {}
  & 
    H_{0} (\bJ) + \eta (\bT, \bJ, t) ,
\end{align}
where ${ H_{0} (\bJ) }$ is the mean field Hamiltonian of the system associated
with the mean field quasi-stationary \DF\ of the system ${ F (\bJ, t) }$. This
Hamiltonian defines the orbital frequencies
${ \bO (\bJ) = \partial H_{0} / \partial \bJ }$, driving the unperturbed
motions in the system. We also introduced two potential perturbations. Here,
${ \delta \phi_{\rm ext} (\bT, \bJ, t) }$ is the stochastic external
perturbation felt by the system. In order to rely on Novikov's theorem
(\eq~\eqref{eq:novikov}), we assume that these perturbations are small (i.e.
${ \delta \phi_{\rm ext} \ll H_{0} }$), of zero mean, Gaussian, and stationary
in time. \Eq~\eqref{eq:H_dFP} also involves
${ \delta \phi_{\rp} (\bT, \bJ, t) }$, the polarization response of the
self-gravitating system to the presence of the external perturbations. In
particular, the polarization perturbation satisfies the self-consistency
requirement
${ \delta \phi_{\rp} (\bx, t) = \! \int \! \rd \bxp \rd \bvelp \psi (\bx,
  \bx') \, \delta F (\bxp, \bvelp,t) }$. Following \eq~\eqref{eq:H_expn}, the
sum of the two potential perturbations
${ \eta = \delta \phi_{\rm ext} + \delta \phi_{\rp} }$ corresponds to the full
stochastic potential perturbations felt by the collisionless system of
interest. As given by the $\eta$-formalism, these fluctuations will drive a
long-term distortion of the system's orbital structure. In order to evaluate the
associated \DCs\ from \eq~\eqref{eq:diff_coeff_final}, one must therefore
characterize the properties of ${ \eta (\bT, \bJ, t) }$. The differences of
the present calculations with the previous calculations of the Landau and \BL\
\DCs\ are twofold. First, here the external perturbations
${ \delta \phi_{\rm ext} (\bx, t) }$ can be arbitrary (as long as they comply with
the requirements from Novikov's theorem), and do not need to be generated by a
discrete integrable bath composed of $N$ particles. Moreover, here the
dressing of the perturbations will be sourced by the response of the system to
the external perturbations (see \eq~\eqref{eq:Fourier_M_dFP}), and not by the
response of the bath to its own finite-$N$ fluctuations.

Similarly to \eq~\eqref{eq:Klim_BL}, at linear order, the evolution of the
fluctuations in the system's \DF\ is given by the linearized Vlasov equation
reading
\begin{align}
  \label{eq:Vlasov_dFP}
  0 & \, = \dpd{\delta F}{t} + \big[ \delta F, H_{0} \big] 
      + \big[ F, \delta \phi_{\rm ext} + \delta \phi_{\rp} \big]
  \nonumber
  \\
  & \, = \dpd{\delta F}{t} + \dpd{\delta F}{\bT} \cdot \bO (\bJ) 
    - \dpd{F}{\bJ} \cdot \dpd{[ \delta \phi_{\rm ext} + \delta \phi_{\rp} ]}{\bT} .
\end{align}
Let us note that \eq~\eqref{eq:Vlasov_dFP} is essentially the same as
\eq~\eqref{eq:Klim_Fric}, where one replaces the perturbation from the test
particle, ${ \delta \phi_{\rt} }$, with the external perturbation,
${ \delta \phi_{\rm ext} }$, bearing in mind that here
${ \delta \phi_{\rm ext} }$ is a stochastic perturbation, while
${ \delta \phi_{\rt} }$ in \eq~\eqref{eq:Vlasov_dFP} was more systematic than
noisy.

We may then follow the same approach as in Section~\ref{sec:DynFric} to solve
\eq~\eqref{eq:Vlasov_dFP}. Let us first decompose the Laplace-Fourier
transformed potential perturbations
${ \delta \tphi_{\bk}^{\rp} (\bJ, \omega) }$,
${ \delta \tphi_{\bk}^{\rm ext} (\bJ, \omega) }$ and
${ \teta_{\bk} (\bJ, \omega) }$, on the basis elements, so as to write
\begin{equation}
  \label{eq:decomposition_delta_dFP}
  \delta \tphi_{\bk}^{\rp} (\bJ, \omega) = 
  \sum_{\alpha} \tbP_{\alpha} (\omega) \, \psi_{\bk}^{(\alpha)} (\bJ) \;\;\; ; 
  \;\;\; \delta \tphi_{\bk} (\bJ, \omega) = \sum_{\alpha} \tbE_{\alpha}
  (\omega) \, 
  \psi_{\bk}^{(\alpha)} (\bJ) \;\;\; ; \;\;\; \teta_{\bk} (\bJ, \omega) = 
  \sum_{\alpha} \tbP_{\alpha}^{\rm tot} (\omega) \, \psi_{\bk}^{(\alpha)} (\bJ) . 
\end{equation}
Here, the vectors $\tbP$, $\tbE$, and $\tbP_{\rm tot}$ characterize the
polarization, external and total perturbations, when projected on the basis
elements. Following \eq~\eqref{eq:selfconsistent_psi_Fric_short}, one can
straightforwardly rewrite \eq~\eqref{eq:Vlasov_dFP} in the vector form
\begin{equation}
  \label{eq:vector_rel_dFP}
  \tbP (\omega) = \wbM (\omega) \ccdot \tbP (\omega) 
  + \wbM (\omega) \ccdot \tbE (\omega) + \tbS (\omega) ,
\end{equation}
In \eq~\eqref{eq:vector_rel_dFP}, as given by \eq~\eqref{eq:definition_S}, we
also introduced the source term ${ \tbS (\omega) \propto 1 / \sqrt{N} }$ (with
$N$ the number of particles in the considered system), describing the initial
fluctuations in the system's \DF\ at ${ t = 0 }$. Finally,
\eq~\eqref{eq:vector_rel_dFP} also involves the system's response matrix, which
describes the amplitude of the self-gravitating amplification carried by the
system. Similarly to \eq~\eqref{eq:Fourier_M}, it reads
\begin{equation}
  \label{eq:Fourier_M_dFP}
  \wbM_{\alpha \beta} (\omega) = 
  {(2 \pi)}^{d} \sum_{\bk} \dint \bJ \, 
  \frac{\bk \cdot \partial F / \partial \bJ}{\omega - \bk \cdot \bO (\bJ)} \, 
  \psi_{\bk}^{(\alpha) *} (\bJ) \, \psi_{\bk}^{(\beta)} (\bJ) ,
\end{equation}
where it is important to note that compared to \eq~\eqref{eq:Fourier_M}, we
replaced the bath's \DF, ${ F_{\rb} (\bJ) }$, by the system's mean \DF,
${ F (\bJ) }$. This emphasizes that in the present case, the support of the
self-gravitating amplification is the system's itself and not an external bath.

In order to focus only on the diffusion associated with the external
perturbations, let us place ourselves within the collisionless limit (i.e.\
${ N \to \infty }$), so that we neglect the initial finite-$N$ fluctuations in
the system. In that limit, \eq~\eqref{eq:vector_rel_dFP} can be easily inverted
to give the total perturbations, ${ \tbP_{\rm tot} = \tbP + \tbE }$, as a
function of the external perturbations. One has
\begin{equation}
  \label{eq:total_pert_dFP}
  \tbP_{\rm tot} (\omega) = {\big[ \bI - \wbM (\omega) \big]}^{-1} \!\cdot \tbE (\omega),
\end{equation}
where we recall that we assumed the mean field collisionless system to be
linearly stable, i.e.\ ${ {[ \bI - \wbM (\omega) ]}^{-1} }$ does not have any
pole in the upper complex plane. In \eq~\eqref{eq:total_pert_dFP}, we recover
that the total perturbations in the system, ${ \tbP_{\rm tot} (\omega) }$, are
given by the self-gravitating dressing, via
${ {\big[ \bI - \wbM (\omega) \big]}^{-1} }$, of the external perturbations,
${ \tbE (\omega) }$. Because of the absence of any instabilities, one can
neglect transient terms and bring the initial time to ${ - \infty }$ to focus
only on the forced regime of evolution. This amounts then to replacing the
Laplace transform in \eq~\eqref{eq:total_pert_dFP} by temporal Fourier
transform (as defined in \eq~\eqref{eq:convention_time_FT}), so that one
finally gets
\begin{equation}
  \label{eq:total_pert_dFP_Fourier}
  \wbP_{\rm tot} (\omega) = {\big[ \bI - \wbM (\omega) \big]}^{-1} \!\cdot \wbE (\omega) .
\end{equation}

Having characterized the potential fluctuations in the system, we may finally
determine the associated \DCs, as given by the $\eta$-formalism. Following
\eq~\eqref{eq:diff_coeff_final}, the externally induced evolution of the collisionless
system is given by the diffusion equation
\begin{equation}
  \label{eq:FP_dFP}
  \dpd{F(\bJ, t)}{t} = 
  \frac{1}{2} \dpd{}{J_{i}} D_{ij} (\bJ) \, \dpd{}{J_{j}} \, F (\bJ, t) ,
\end{equation}
Here, the basis method allows us to write the \DCs\ as
\begin{equation}
  \label{eq:Diff_dFP}
  D_{ij} (\bJ) = 
  \sum_{\bk} k_{i} k_{j} \sum_{\alpha, \beta} \psi_{\bk}^{(\alpha)} (\bJ) \, 
  \psi_{\bk}^{(\beta) *} (\bJ) \, 
  \wbC_{\alpha \beta}^{\rm tot} (\bk \ccdot \bO (\bJ)) ,
\end{equation}
where ${ \bC_{\rm tot} (t) }$ is the temporal correlation of the total
potential perturbations. It is defined as
\begin{equation}
  \label{eq:def_Ctot}
  \bC_{\alpha \beta}^{\rm tot} (t - \tp) = 
  \left\langle \bP_{\alpha}^{\rm tot} (t) \, 
    \bP_{\beta}^{\mathrm{tot} \, *} (\tp) \right\rangle .
\end{equation}
Assuming that the perturbations ${ \wbP_{\alpha}^{\rm tot} (t) }$ are
stationary in time, \eq~\eqref{eq:def_Ctot} can equivalently be rewritten in
Fourier space as
\begin{equation}
  \label{eq:rel_correl_Fourier}
  \left\langle \wbP_{\alpha}^{\rm tot} (\omega) \, 
    \wbP_{\beta}^{\mathrm{tot} \, *} (\omega') \right\rangle 
  = 2 \pi \, \deltaD (\omega - \omega') \, \wbC^{\rm tot}_{\alpha \beta} (\omega) .
\end{equation}
The \DCs\ from \eq~\eqref{eq:Diff_dFP} then become
\begin{equation}
  \label{eq:Diff_dFP_I}
  D_{ij} (\bJ) = 
  \sum_{\bk} k_{i} k_{j} \sum_{\alpha, \beta} \psi_{\bk}^{(\alpha)} (\bJ) \, 
  \psi_{\bk}^{(\beta) *} (\bJ) \, \!\!\int \!\! \frac{\rd \omega'}{2 \pi} \, 
  \left\langle \wbP_{\alpha}^{\rm tot} (\omega) \, 
    \wbP_{\beta}^{\mathrm{tot} \, *} (\omega') \right\rangle . 
\end{equation}
Following the amplification relation from \eq~\eqref{eq:total_pert_dFP}, these
\DCs\ can immediately be rewritten as a function of the correlation of the
external perturbations. Indeed, assuming once again that the external potential
perturbations are stationary in time, similarly to \eq~\eqref{eq:def_Ctot}, we
may define their correlation as ${ \bC_{\rm ext} (t) }$ as
\begin{equation}
  \label{eq:def_Cext}
  \bC_{\alpha \beta}^{\rm ext} (t - \tp) = 
  \left\langle \bE_{\alpha} (t) \, \bE_{\beta}^{*} (t') \right\rangle \;\;\; ; 
  \;\;\; \left\langle \wbE_{\alpha} (\omega) \, \wbE_{\beta}^{*} (\omega') \right\rangle 
  = 2 \pi \deltaD (\omega - \omega') \, \wbC^{\rm ext}_{\alpha \beta} (\omega) . 
\end{equation}
\Eq~\eqref{eq:total_pert_dFP} allows us finally to rewrite the \DCs\ from
\eq~\eqref{eq:Diff_dFP_I} as
\begin{equation}
  \label{eq:Diff_dFP_II}
  D_{ij} (\bJ) = \sum_{\bk} k_{i} k_{j} \sum_{\alpha, \beta}  \psi_{\bk}^{(\alpha)} (\bJ) \, 
  \psi_{\bk}^{(\beta) *} (\bJ) \, {\bigg[ {\big[ \bI - \wbM \big]}^{-1} 
    \!\cdot \wbC_{\rm ext} \ccdot 
    {\big[ \bI - \wbM^{\dagger} \big]}^{-1} \bigg]}_{\alpha \beta} (\omega = \bk \ccdot \bO (\bJ)) .
\end{equation}
In the absence of collective effects, i.e.\ in the absence of the amplification
of the external perturbations by the system, \eq~\eqref{eq:Diff_dFP_II}
immediately gives the bare \DCs\ reading
\begin{align}
  \label{eq:Diff_dFP_III}
  D_{ij} (\bJ) & \, = 
  \sum_{\bk} k_{i} k_{j} \sum_{\alpha, \beta}  \psi_{\bk}^{(\alpha)} (\bJ) \, 
  \psi_{\bk}^{(\beta) *} (\bJ) \, \wbC^{\rm ext}_{\alpha \beta} (\bk \ccdot \bO (\bJ))
  \nonumber
  \\
  & \, = \sum_{\bk} k_{i} k_{j} \, \wC^{\rm ext}_{\bk\bk} (\bJ , \bJ , \bk \cdot \bO (\bJ)) , 
\end{align}
where ${ \wC^{\rm ext}_{\bk\bk} (\bJ , \bJ , \omega) }$ stands for the temporal Fourier transform
of the correlation of the external potential fluctuations, ${ \delta \phi_{\bk}^{\rm ext} (\bJ , t) }$,
as defined in \eq~\eqref{eq:C_def}.
\Eqs~\eqref{eq:Diff_dFP_II} and~\eqref{eq:Diff_dFP_III} are the main
results of this section. These equations are identical to the bare \DCs\ first obtained
in \eq~{(3.9a)} of~\cite{BinneyLacey1988} and their dressed generalization obtained
in \eq~{(C.7)} of~\cite{Weinberg2001a}. As advocated by the present $\eta$-formalism, we note
once again that the diffusion occurring in the system is directly sourced by
the power spectrum of the correlation of the external perturbations, evaluated
at the local orbital frequency ${ \omega = \bk \cdot \bO (\bJ) }$. Finally,
when collective effects are accounted for, we find again that the external
perturbations have to be dressed by the system's self-gravity, as can be seen
from the factors ${ {[ \bI - \wbM (\omega) ]}^{-1} }$ in
\eq~\eqref{eq:Diff_dFP_II}.

\section{Application: The HMF model}
\label{sec:HMF}

The key input from the present $\eta$-formalism is that the \DCs\ of a test
particle are essentially given by the temporal correlation of the potential
fluctuations in the system, as can be seen in
\eq~\eqref{eq:diff_coeff_final}. To illustrate this point and to demonstrate
how the $\eta$-formalism can be used in practice, in this section we apply this
framework to the one-dimensional inhomogeneous \HMF\
model~\citep{Pichon1994,AntoniRuffo1995}, for which we will determine the bare
(i.e.\ Landau) and dressed (i.e.\ \BL) \DCs\@. In particular, we will compare
our predictions to the recent results of~\cite{BenettiMarcos2017}, hereafter
BM17. In BM17, the \DCs\ were computed in two ways: (i) via
the direct computation of the Landau and \BL\ \DCs\ from
\eqs~\eqref{eq:Final_Diff_Coeff_Landau} and~\eqref{eq:Final_Diff_Coeff_BL},
(ii) via direct \Nbody\ simulations to compute the second-order \DCs\
${ D_{2} (\bJ) = \lim_{\Delta t \to \infty} \big\langle {(\Delta \bJ)}^2
  \big\rangle / (\Delta t) }$. In approach (i), the computation of the \BL\
\DCs\ relied on solving the resonant condition
${ \bk\cdot\bO(\bJ) = \bkp\cdot\bO(\bJp) }$, as well as the self-consistency
relation in \eq~\eqref{eq:selfconsistent_dressed_coefficients}. As we will
show, both of these calculations are not needed in the $\eta$-formalism. In
the case of the \HMF\ model, because of the limited number of basis elements
(only two) and the existence of explicit expressions for their angular Fourier
transforms (see \eq~\eqref{eq:explicit_ck_sk_HMF}), the computation of the
dressed susceptibility coefficients is somewhat simplified. This calculation
can be much more cumbersome in more intricate self-gravitating systems such as, for
example, razor-thin stellar disks~\citep{FouvryPichonMagorrianChavanis2015}.

\subsection{The HMF model}
\label{sec:presHMF}

Let us first briefly present the \HMF\ model. The \HMF\ model is a
one-dimensional system where the phase-space coordinates are given by an angle
${ \phi \in \interval[open right]{0}{2\pi} }$ and a velocity $v$. Two particles
$i$ and $j$ interact via a pairwise potential of the form
\begin{equation}
  \label{eq:pot_HMF}
  \psi (\phi_{i}, \phi_{j}) = 
  - \cos (\phi_{i} - \phi_{j}) .
\end{equation}
We assume that the bath particles are of equal mass ${ m_{\rb} = 1 / N }$, so that the total
mass of the bath is ${ M \!=\! 1 }$. With such a convention, the specific
Hamiltonian of a test particle embedded in this system is given by
\begin{equation}
  \label{eq:H_test_HMF}
  H_\rt (\phi_{\rt}, v_{\rt}) = 
  \frac{v_{\rt}^{2}}{2} 
  + \sum_{i = 1}^{N} m_{\rb} \, \psi (\phi_{\rt}, \phi_{i} (t)) ,
\end{equation}
where ${ (\phi_{\rt}, v_{\rt}) }$ are the phase-space coordinates of the test
particle, and the sum over $i$ runs over all bath particles. Hamilton's
evolution equations for the test particle are then
\begin{equation}
  \label{eq:Hamilton_Eq_HMF}
  \dot{\phi}_{\rt} = \dpd{H_{\rt}}{v_{\rt}} 
  \;\;\; ; \;\;\; 
  \dot{v}_{\rt} = - \dpd{H_{\rt}}{\phi_{\rt}} .
\end{equation}
The pairwise potential from \eq~\eqref{eq:pot_HMF}, can be rewritten in the
separable form
${ \psi (\phi, \phip) \!=\! - \cos (\phi) \cos (\phi') - \sin (\phi) \sin
  (\phi') }$. As a result, the instantaneous potential, ${ \Phi (\phi , t) }$,
seen by the test particle takes the simple form
\begin{align}
  \label{eq:psi_hmf}
  \Phi(\phi, t) = {}
  &
    - \sum_{i=1}^N m_{\rb} \cos(\phi - \phi_i(t)) 
    \nonumber  \\
  = {}
  &
    - M_x(t) \, \cos(\phi) - M_y(t) \, \sin(\phi) ,
\end{align}
where
\begin{equation}
  \label{eq:M_xy}
  M_{x} (t) = 
  m_{\rb} \sum_{i=1}^{N} \, \cos(\phi_{i} (t)); \;\;\;   
  M_{y} (t) = m_{\rb} \sum_{i=1}^{N} \sin(\phi_{i} (t)), 
\end{equation}
are the elements of the magnetization vector. The mean potential,
${\Phi_{0} (\phi)}$, associated with a given quasi-stationary state is denoted
with
\begin{equation}
  \label{eq:Phi0_hmf}
  \Phi_{0} (\phi) = 
  \dint \phip \rd \vp \, F_{\rb} (\phip, \vp) \, \psi (\phi, \phip) ,
\end{equation}
where ${ F_{\rb} (\phi, v) }$ is the smooth \DF\ of the bath particles,
normalized so that ${ \!\int\!  \rd \phi \rd v \, F_{\rb} = M }$. Provided we
perform a shift in the definition of the angles, $\ophi = \phi - \alpha$, the
mean field potential can always be written as
\begin{equation}
  \label{eq:mean_pot_HMF}
  \Phi_{0} (\ophi) = - M_{0} \cos (\ophi),
\end{equation}
where ${ M_{0}^{2} = {\{ \cos \phi \}}^{2} + {\{ \sin \phi \}}^{2}}$ is the
system's mean magnetization and
${\alpha = \text{Arctan} [ \{ \cos (\phi) \}, \{ \sin (\phi) \} ] }$ is its
direction, with ${ \{ \, \cdot \, \} }$ standing for the average over all the
bath particles. The potential from \eq~\eqref{eq:mean_pot_HMF} is the one of a
pendulum and is therefore integrable. Following~BM17, one can construct
explicit angle-action coordinates $(\theta, J)$ for this Hamiltonian, as we
review in Appendix~\ref{sec:AA_HMF}. Rather than using $J$ to describe the
orbital space, it is easier from the computational perspective to describe the
orbits with another integral of motion, namely $\kappa$, which is related to
the mean field quantities by \eq~\eqref{eq:kappa_def}. As there is one-to-one
relation between $\kappa$ to $J$ (\eq~\eqref{eq:J_def_HMF}), we will use
$\kappa$ and $J$ interchangeably in the following.

\subsection{Measuring potential fluctuations}
\label{sec:flucHMF}

We consider a bath composed of $N$ particles sampled from a
smooth \DF\@, ${ F_{\rb} (v,\phi) }$, chosen to be a stable steady-state solution of
the Vlasov equation, that is satisfying
\begin{align}
  \label{eq:Vlasov-HMF}
  0 & \, = \big[ F_{\rb} (\phi, v), \tfrac{1}{2} v^{2} + \Phi (\phi) \big]
  \nonumber
  \\
  & \, = v \dpd{F_{\rb}}{\phi} - M_{0} \sin(\ophi)\dpd{F_{\rb}}{v} .
\end{align}
Even for such a steady state, the system's mean Hamiltonian can still undergo a
long-term evolution on timescales ${ \propto \sqrt{N} }$, due to the initial
deviations of the sampled \DF\ from the smooth sampled
one. This trend ${ \propto \sqrt{N} }$ will be zero under ensemble average, but
cannot be neglected for each individual realization. As a result, in order to
infer from simulations the \DCs\ of a test particle, it is better to use a
time-averaged Hamiltonian $H_{0}$ from which this long timescale trend
${ \propto \sqrt{N} }$ has been removed.

In the \HMF\ case, this can be done by using a time-dependent shifted angle
${ \ophi = \phi - \alpha (t) }$, where ${ \alpha (t) }$ changes on timescales
${ \propto \sqrt{N} }$, and is determined for each realization by time
averaging
${ \text{Arctan} [ \{ \cos(\phi_{i} (t)) \}, \{ \sin (\phi_{i} (t)) \} ] }$ on
long timescales. Within these shifted coordinates, the new Hamiltonian for the
test particle becomes
\begin{equation}
  \label{eq:Hbar_test_HMF}
  H = \oH_0(t) - (\oM_{x} (t) - M_{0}) \cos (\ophi) -
  \oM_y(t) \cos(\ophi) ,
\end{equation}
where the mean field Hamiltonian, ${ \oH_{0} (t) }$, is
\begin{equation}
  \label{eq:H0bar}
  \oH_{0} (t) = 
  \frac{v^{2}}{2} - M_{0} \cos(\ophi) - v \dpd{\alpha(t)}{t} ,
\end{equation}
In \eq~\eqref{eq:Hbar_test_HMF}, the (shifted) magnetization is characterized
by
\begin{equation}
  \label{eq:Mbar_xy}
  \oM_{x} (t) = m_{\rb} \sum_{i=1}^{N} \cos(\ophi_{i} (t)); \;\;\;   
  \oM_{y} (t) = m_{\rb} \sum_{i=1}^{N} \sin(\ophi_{i}(t)) ,
\end{equation}
which can be written in vector notation as
\begin{equation}
  \label{eq:Mbar_xy_vec}
  \left(\begin{array}{c}
          \oM_{x} (t) \\
          \oM_{y} (t) \\
        \end{array}
      \right)
      = 
      \left(\begin{array}{cc}
              \cos(\alpha(t)) & \sin(\alpha(t)) \\
              -\sin(\alpha(t)) & \cos(\alpha(t)) \\
            \end{array}
          \right)
          \cdot
          \left(\begin{array}{c}
                  M_{x} (t) \\
                  M_{y} (t) \\
                \end{array}
              \right).
\end{equation}
By construction, ${ \oH_{0} (t) }$ fluctuates only on short timescales while
the noise ${ \eta (\phi, t) }$ changes on long timescales. As a result,
${ \partial \alpha / \partial t}$ is ${1 / \sqrt{N}}$ smaller than ${\oH_0(t)}$
and can be neglected in \eq~\eqref{eq:H0bar}. The angle-action variables are
then constructed w.r.t.\ the new Hamiltonian $\oH_{0}$. By doing so, we have
removed the long timescale (${\propto \sqrt{N}}$) trends from the dynamics, which
will then allow for the computation of the \DCs\@.

Let us now specify how the noise term ${ \eta_{k} (J, t) }$ may be
computed. In the present case, the noise follows directly from the Fourier
expansion of ${ \cos (\phi) }$ and ${ \sin(\phi) }$ w.r.t.\ the canonical angle
$\theta$, using the Fourier coefficients ${ c_{k} (J) }$ and ${ s_{k} (J) }$
given in Appendix~\ref{sec:AA_HMF}. The noise induced by the bath can therefore
be written as
\begin{equation}
  \label{eq:eta_k_jmf}
  \eta_{k} (J, t) = c_{k} (J) \, \eta_{x} (t) + s_{k} (J) \, \eta_{y} (t),
\end{equation}
where ${ \eta_{x} (t) }$ and ${ \eta_{y} (t) }$ are the perturbations of the
magnetization
\begin{equation}
  \label{eq:eta_xy}
  \eta_{x} (t) = \oM_{x} (t) - M_{0} ; \;\;\;   \eta_{y} (t) = \oM_{y} (t) .
\end{equation}
Since ${ \cos (\ophi) }$ is symmetric and ${ \sin(\ophi) }$ is antisymmetric
w.r.t.\ $\theta$, the coefficients ${ c_{k} (J) }$ are real, while the
coefficients ${ s_{k} (J) }$ are imaginary. In addition, they satisfy
${ c_{-k} (J) = c_{k} (J) }$ and ${ s_{- k} (J) = - s_{k} (J) }$. We may then
write
\begin{equation}
  \label{eq:C_xy}
  \langle \eta_{k} (J, t) \, \eta_{k} (J, \tp) \rangle +  
  \langle \eta_{-k} (J, t) \, \eta_{-k} (J, \tp) \rangle   
  = 
  2 \, | c_{k} (J) |^{2} \, C_{x} (t - \tp) 
  + 2 \, | s_{k} (J) |^{2} \, C_{y} (t - \tp) ,
\end{equation}
where ${ C_{x} (t) }$ and ${ C_{y} (t) }$ are the correlation functions of the
fluctuations of the magnetization ${ \eta_{x} (t) }$ and ${ \eta_{y} (t) }$.
These two correlation functions fully characterize the statistical properties
of the noise induced by the bath.

Following \eq~\eqref{eq:diff_coeff_final}, the \DC\ of a test particle embedded
in this bath is
\begin{equation}
  \label{eq:D_HMF}
  D (J) = 
  \sum_{k > 1} k^{2} \big[
  | c_{k} (J) |^{2} \, \wC_{x} (k \, \Omega (J)) + 
  | s_{k} (J) |^{2} \, \wC_{y} (k \, \Omega (J)) 
  \big],
\end{equation}
where ${ \wC_{x} (\omega) }$ and ${ \wC_{y} (\omega) }$ are the temporal
Fourier transforms of the correlation functions of ${ C_{x} (t) }$ and
${ C_{y} (t) }$, and ${ \Omega (J) }$ is the mean field orbital frequency. Let
us note that because ${ C_{x} (t) }$ and ${ C_{y} (t) }$ are real even
functions, their Fourier transforms ${ \wC_{x} (\omega) }$ and
${ \wC_{y} (\omega) }$ are also real and even functions. In
\eq~\eqref{eq:D_HMF}, we also note that the Fourier number ${ k = 0 }$ never
contributes to the diffusion.

For generic self-gravitating systems, the noise terms
${ \eta_{\bk} (\bJ, t) }$ depend on the action $\bJ$, the considered
resonance vector $\bk$, and the time $t$. Fortunately, this dependence is made
simpler in the case of the \HMF\ model. Indeed, as can be seen in
\eq~\eqref{eq:eta_k_jmf}, the noise terms for the \HMF\ involve the
deterministic (i.e.\ time-independent) coefficients, ${ c_{k} (J) }$ and
${ s_{k} (J) }$, which depend on the action coordinate $J$ and Fourier number
$k$, as well as on the stochastic coefficients, ${ \eta_{x} (t) }$ and ${ \eta_{y} (t) }$, which
are time dependent. Having such a separated expression for the noise terms is
specific to the \HMF\ model and significantly simplifies the characterization
of the potential fluctuations in this system. Following \eqs~\eqref{eq:M_xy}
and~\eqref{eq:eta_xy}, the fluctuations of the magnetization are simple
functions of the positions ${ \phi_{i} (t) }$ of all the bath particles at time
$t$.

Given of set of simulations, one can directly calculate the correlation
functions by the following procedure. First, for each simulation, we compute
the magnetization ${ M_{x} (t) = \{ \cos(\phi_{i} (t)) \} }$ and
${ M_{y} (t) = \{ \sin(\phi_{i} (t))\} }$ at each time step. Then, we define a
slowly varying angle ${ \alpha (t) }$ which on timescales
${ \lesssim \sqrt{N / M_{0}} }$ can be obtained by fitting a second order
polynomial in time to ${ \text{Arctan} [ M_{x} (t), M_{y} (t) ] }$. Then,
${ \oM_{x} (t) }$ and ${ \oM_{y} (t) }$ are obtained from
\eq~\eqref{eq:Mbar_xy_vec} and the detrended noise terms computed from
\eq~\eqref{eq:eta_xy}. The correlation functions ${ C_{x;y} (t) }$ are
subsequently evaluated by averaging the discrete correlation functions
${ C_{x;y} (n \Delta T) }$ over different realizations. The subsequent step is
to compute the Fourier transforms ${ \wC_{x;y} (\omega) }$ of the correlation
functions. This is performed via the discrete Fourier transforms
${ \wC_{x;y} (2\pi n / (T - \Delta T)) = \wC_{x;y}^{n} }$. Finally, the \DC\
is obtained following \eq~\eqref{eq:D_HMF}, where ${ \wC_{x;y} (k \Omega(J)) }$
are evaluated by interpolating the discrete Fourier transforms
${ \wC_{x;y}^{n} (\omega) }$.

This procedure works for both non-interacting and self-interacting baths, i.e.\
it allows for the recovery of both the Landau and \BL\ inhomogeneous \DCs\@.
However, this requires running a large set of \Nbody\ simulations with ${ N \gg 1 }$
for few hundreds of dynamical times, in order to reach a statistical
convergence for the correlations. For a non-interacting bath, the motions of
the different bath particles are uncorrelated, which may be used to evaluate
the correlations of ${ \eta_{x} (t) }$ and ${ \eta_{y} (t) }$. Indeed, in that
simpler case, one only has to integrate (either numerically or analytically,
see Appendix~\ref{sec:AA_HMF}) a large set of particles driven by the mean
field. Owing to the absence of correlations between the bath particles, one can
compute for each bath particle the correlation function of ${ \sin (\phi) }$
and ${ \cos (\phi) - \langle \cos (\phi) \rangle_{t} }$ (with
${ \langle \; \cdot \; \rangle_{t} }$ a time average). The correlation
${ C_{x;y} (t) }$ of the fluctuations of the magnetization are then immediately
obtained by averaging over all the bath particles. Let us now work out in
detail one such computation of the \HMF\ \DCs\@.

\subsection{Diffusion coefficients}
\label{sec:DiffHMF}

\begin{figure}
  \begin{center}
    \includegraphics[width=0.4\textwidth]{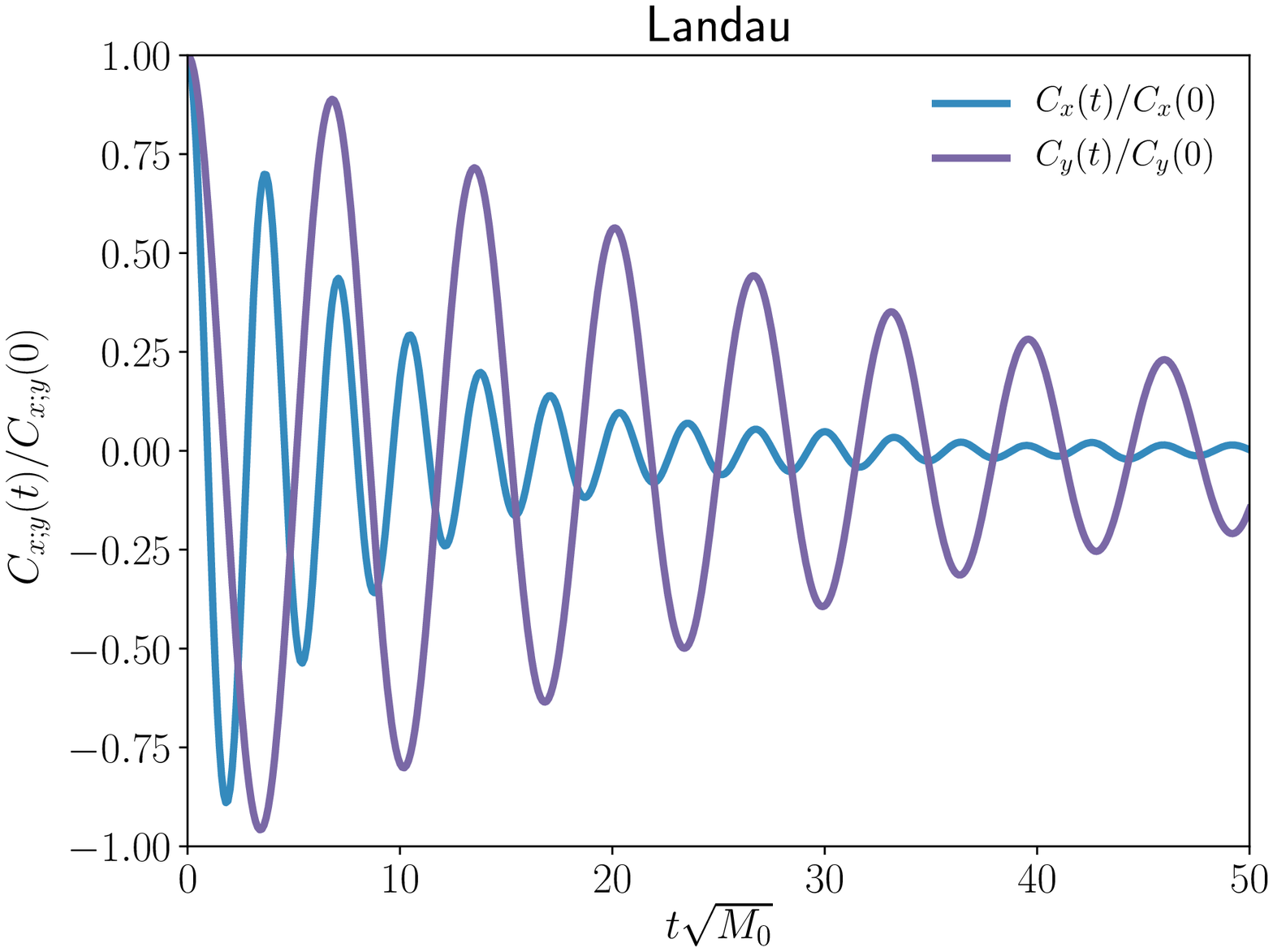}\includegraphics[width=0.4\textwidth]{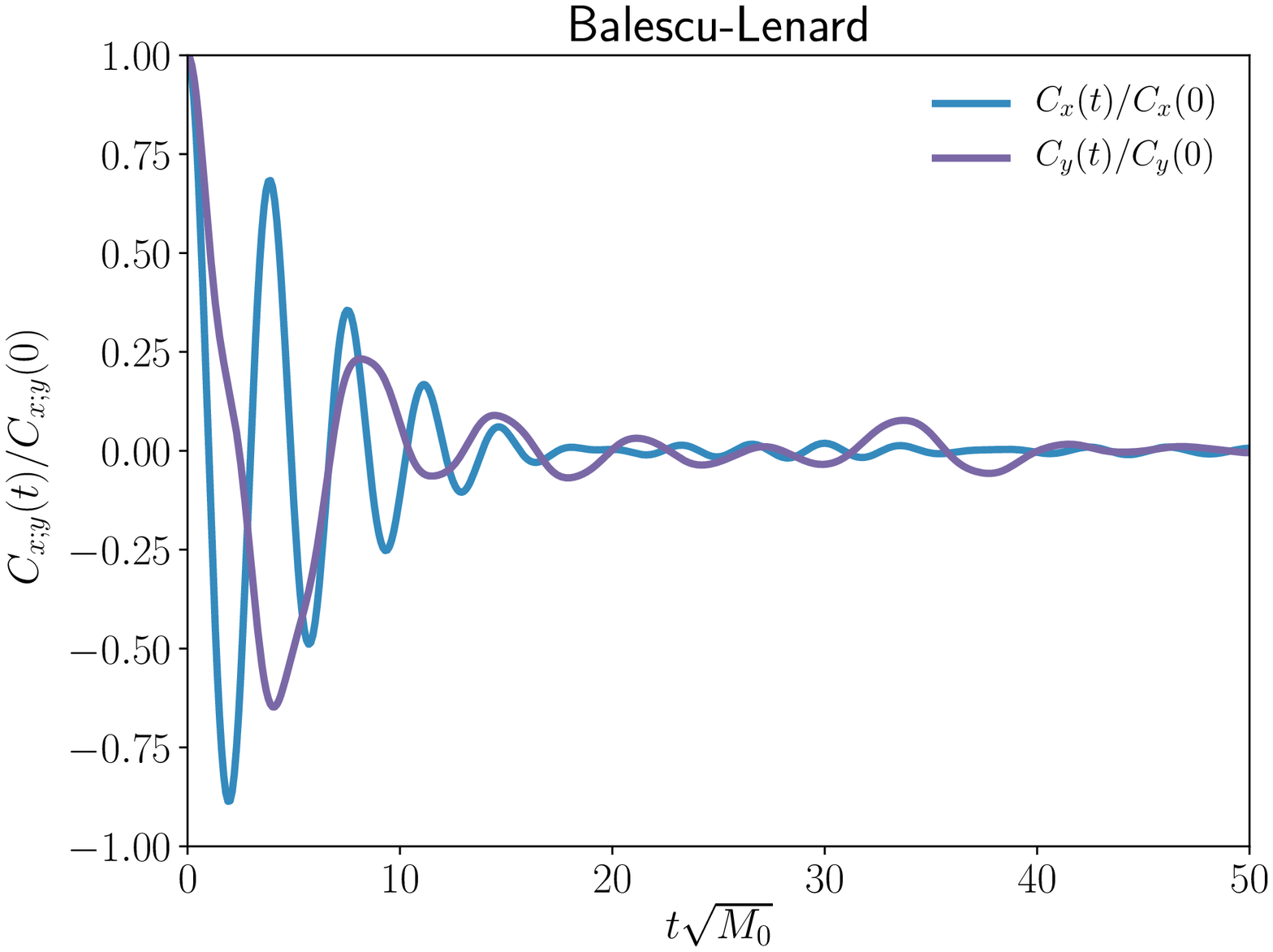}
    \includegraphics[width=0.4\textwidth]{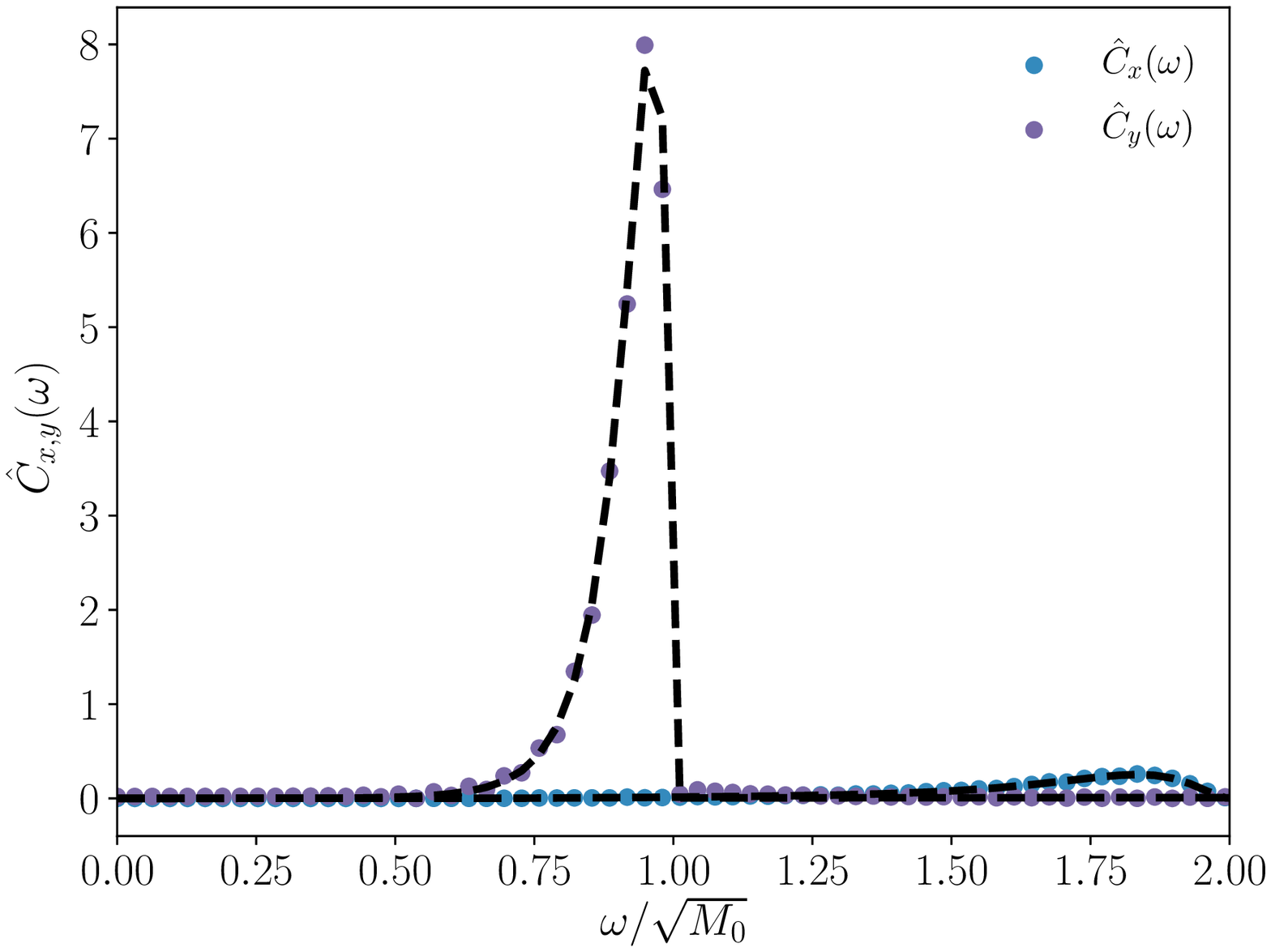}\includegraphics[width=0.4\textwidth]{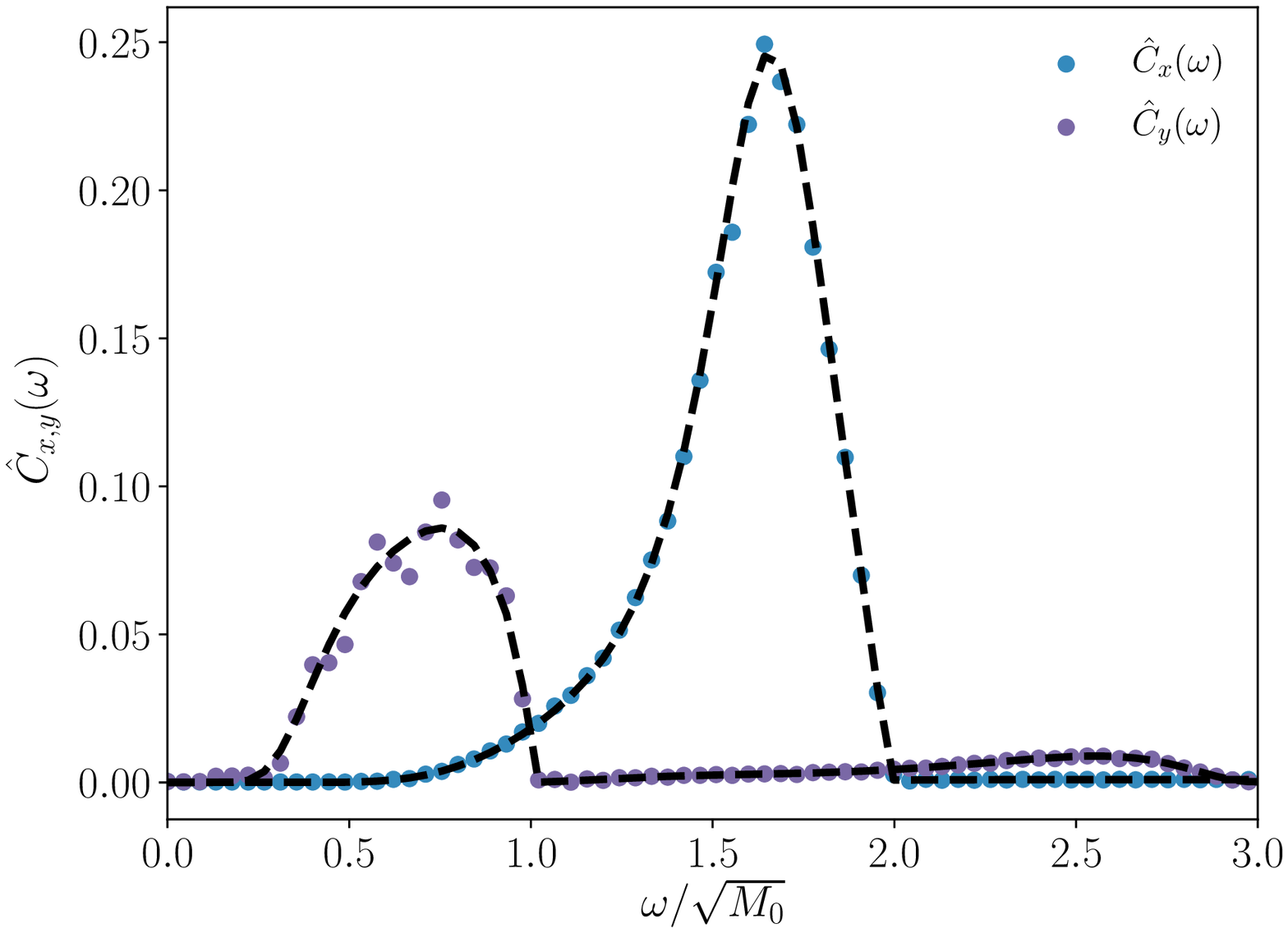}
    \caption{\label{fig:hmf_acf} Correlation functions ${ C_{x} (t) }$ and
      ${ C_{y} (t) }$ (upper panels) and their Fourier transforms
      ${ \wC_{x} (\omega) }$ and ${ \wC_{y} (\omega) }$ (lower panels) for a
      non-interacting (Landau) bath (left) and a self-interacting (\BL) bath
      (right) with magnetization ${ M_{0} = 0.816 }$. In the lower panels, the
      dashed lines show ${ \wC_{x} (\omega) }$ and ${ \wC_{y} (\omega) }$
      evaluated by \eq~\eqref{eq:hmf_Chat}.}
\end{center}
\end{figure}

Following~BM17, we consider a bath characterized by a thermal equilibrium \DF,
${ F_{\rb} (\phi, v) }$, reading
\begin{equation}
  \label{eq:feq}
  F_{\rb} (\phi, v) = F_{v} (v) \, F_{\phi} (\phi) ,
\end{equation}
where ${ F_{v} (v) }$ is a Normal distribution and ${ F_{\phi} (\phi) }$ is a
von Mises distribution, so that
\begin{equation}
  \label{eq:f_v_f_phi}
  F_{v} (v) = 
  \sqrt{\frac{\beta}{2 \pi}} \re^{-\beta p^{2} / 2}; \;\;
  F_{\phi} (\phi) = 
  \frac{\re^{\beta M_{0} \cos(\phi)}}{2 \pi I_{0} (\beta M_{0})} .
\end{equation}
Here, the mean magnetization $M_{0}$ and the inverse temperature $\beta$
satisfy a self-consistency relation of the form
${ M_{0} = I_{1}(\beta M_{0}) / I_{0} (\beta M_{0}) }$, with ${ I_{n} (x) }$
the modified Bessel function of the first kind of order $n$. Such a separated
\DF\ is straightforward to sample. Once initial conditions are generated, we
perform simulations of either non-interacting or self-interacting baths. For a
non-interacting bath, we may follow Appendix~\ref{sec:AA_HMF} to integrate
explicitly the motion of the bath particles in the mean field potential. For a
self-interacting bath, we follow the same procedure as in BM17, and use a
second-order symplectic integrator. As detailed in the previous sections, the bath
realizations allow us to compute the correlation functions ${ C_{x} (t) }$ and
${ C_{y} (t) }$, as well as their Fourier transforms ${ \wC_{x} (\omega) }$ and
${ \wC_{y} (\omega) }$. These measurements are illustrated in
Figure~\ref{fig:hmf_acf} for both a Landau and \BL\ baths, with
${ M_{0} = 0.816 }$, when computed via the procedure described above
($\eta$-formalism).
In order to compare the measurements of correlations made via bath
realizations, in Figure~\ref{fig:hmf_acf}, we also represented a direct
analytical calculation of the correlations ${ \wC_{x} (\omega) }$ and
${ \wC_{y} (\omega) }$. Indeed, following BM17, the Fourier transform of the
correlation functions giving rise to the \BL\ \DCs\ from
\eq~\eqref{eq:Final_Diff_Coeff_BL} read
\begin{equation}
  \label{eq:hmf_Chat}
  \wC_{x} (\omega) =
  \frac{2 \pi}{N}\sum_{k > 0} \sum_{\kappa^{\star}}
  \frac{|c_{k} (\kappa^{\star})|^2 F_{\kappa} (\kappa^{\star})}{|\veps_{cc} (k \Omega (\kappa^{\star}))|^{2}
    {| k \, \partial \Omega / \partial
    \kappa|}_{\kappa^\star}}  ;
  \;\;\;
  \wC_{y} (\omega) =
  \frac{2 \pi}{N} \sum_{n>0} \sum_{\kappa^{\star}}
  \frac{| s_{k} (\kappa^{\star})|^2 \, F_{\kappa} (\kappa^{\star})}{| \veps_{ss} (k \Omega (\kappa^{\star}))|^{2}
    {| k \, \partial \Omega / \partial \kappa|}_{\kappa^{\star}}} ,
\end{equation}
where ${ F_{\kappa} (\kappa) }$ is the \PDF\ of $\kappa$. In
\eq~\eqref{eq:hmf_Chat}, we also introduced the resonant locations
$\kappa^{\star}$, which are the solutions of
${ k \, \Omega (\kappa^{\star}) = \omega }$. Finally, we also introduced the
dressed susceptibility coefficients ${ \veps_{cc} (\omega) }$ and
${ \veps_{ss} (\omega) }$, capturing the amplitude of the self-gravitating
amplification in the system. Following BM17, and as generically given by the
response matrix from \eq~\eqref{eq:Fourier_M}, these coefficients read
\begin{equation}
  \label{exp_veps_cc_ss}
  \veps_{cc} (\omega) = 
  1 - 2 \pi \sum_{k} \dint \kappa \, 
  \frac{| c_{k} (\kappa) |^{2} \, k \, 
    \partial F_{\rb} / \partial \kappa}{\omega - k \Omega (\kappa) } ; \;\;\;
  \veps_{ss} (\omega) = 
  1 - 2 \pi \sum_{k} \dint \kappa \, 
  \frac{|s_{k} (\kappa)|^{2} \, k \, \partial F_{\rb} / \partial \kappa}
  {\omega - k \Omega (\kappa)} .
\end{equation}
In the absence of collective effects, the susceptibility coefficients become
${\varepsilon_{cc} (\omega) = \varepsilon_{ss} (\omega) = 1}$. Let us note
that there are two main difficulties associated with the computation of the
correlations from \eq~\eqref{eq:hmf_Chat}. First, in \eq~\eqref{eq:hmf_Chat},
one has to solve the non-local resonance condition, and determine the orbits
$\kappa^{\star}$ that may resonate with a given frequency ${ \omega / k
}$. Moreover, this expression also involves the dressed susceptibility
coefficients (see \eq~\eqref{exp_veps_cc_ss}), which ask for the delicate
computation of an integral over orbital space exhibiting a pole at the
resonance. While for the \HMF\ model, \eq~\eqref{exp_veps_cc_ss} takes the
form of a one-dimensional resonating integral, such expressions become more
cumbersome to evaluate for inhomogeneous systems of higher dimensions (see
e.g.,~\cite{FouvryPichonMagorrianChavanis2015} for ${2D}$ razor-thin disks).
Let us finally recall that this analytical approach is made simpler for the
\HMF\ model, owing to the constraint from \eq~\eqref{eq:eta_k_jmf}, which
separates the $J$, $k$, and $t$ dependence in the potential fluctuations.

In Figure~\ref{fig:hmf_acf}, we show that the computation of the noise
correlation either from bath realizations or from analytical calculations both
match. When the bath is self-interacting, Figure~\ref{fig:hmf_acf} illustrates
also the result from \eq~\eqref{eq:deltapsi_BL}. Indeed, for such a bath,
provided one waits sufficiently for the correlation to build up, the
fluctuations in the bath may be considered as being created by a
non-interacting bath for which the pairwise interaction potential has been
dressed, i.e.\ making the change
${ \psi_{\bk\bkp} (\bJ, \bJp) \to \psi_{\bk\bkp}^{\rd} (\bJ, \bJp, \omega)
}$.

\begin{figure}
  \begin{center}
    \includegraphics[width=0.4\textwidth]{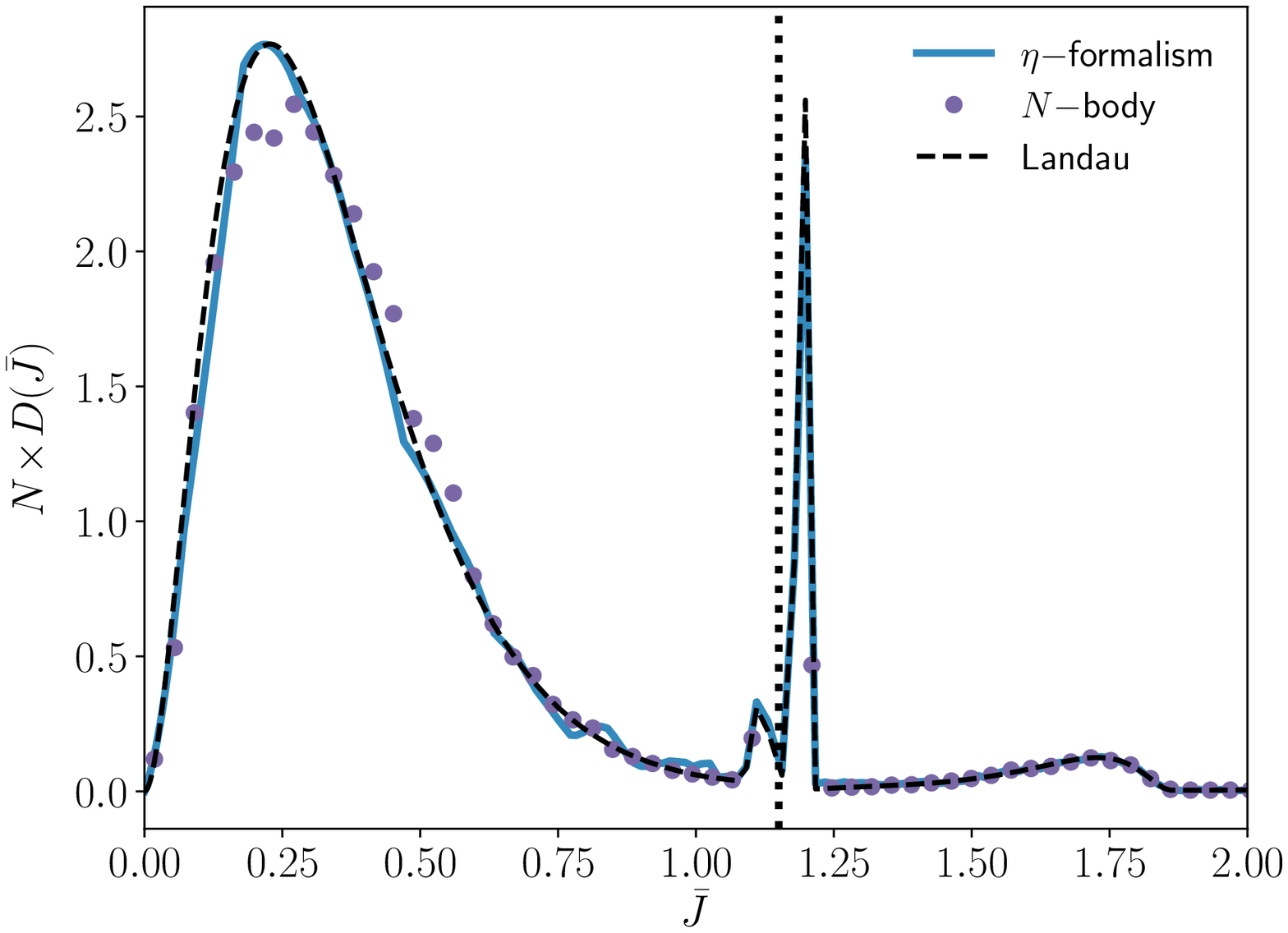}\includegraphics[width=0.4\textwidth]{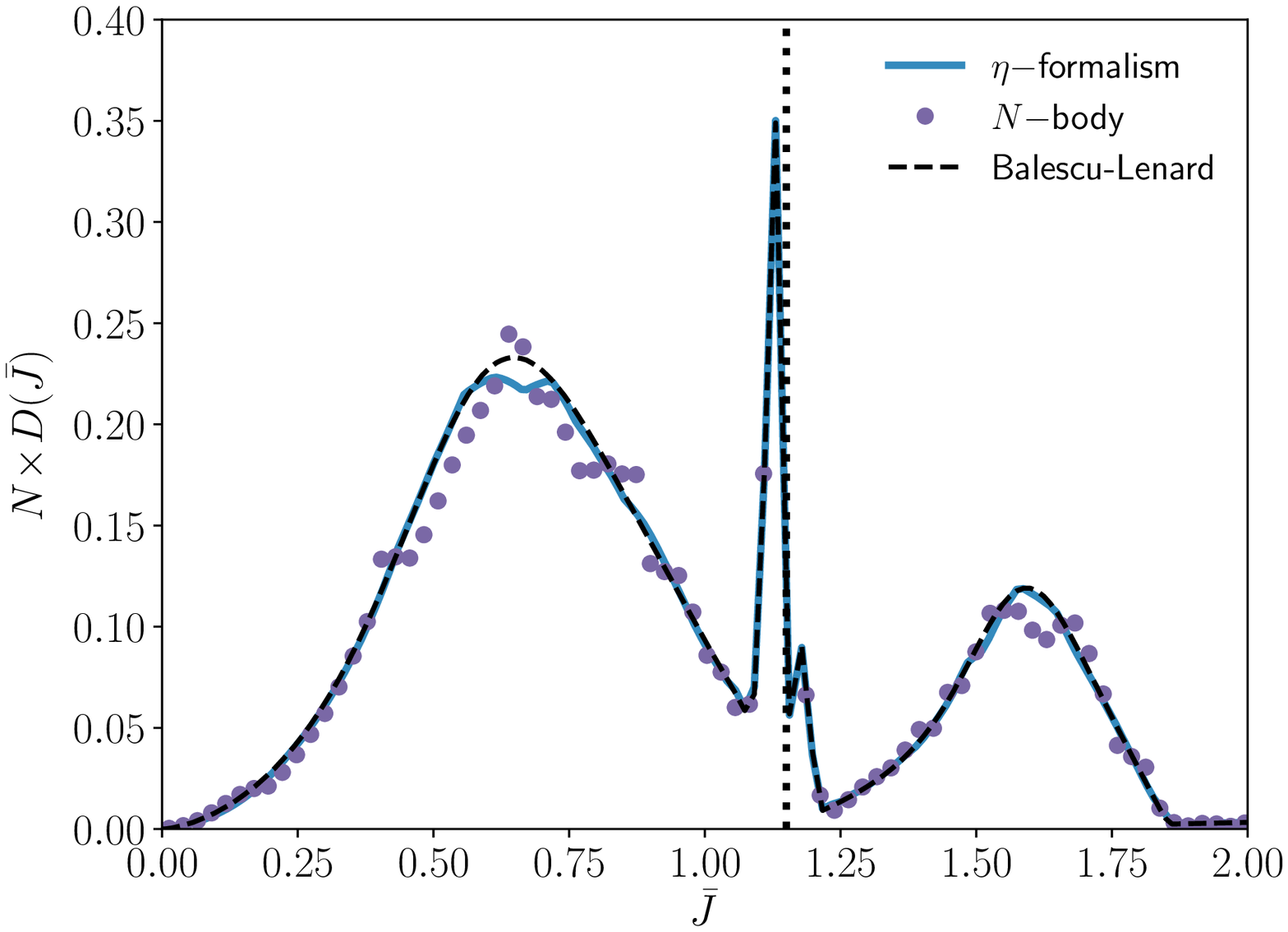}
    \caption{\label{fig:hmf_DC} Diffusion coefficients for a non-interacting
      bath (left) and an self-interacting bath (right) with magnetization
      ${ M_{0} = 0.816 }$, as a function of the rescaled action $\oJ$ (defined
      in \eq~\eqref{eq:Jbar}). The thick solid line shows the diffusion
      coefficients as obtained by the $\eta$-formalism, the dashed lines are
      obtained from \eqs~\eqref{eq:D_HMF} and~\eqref{eq:hmf_Chat}, while the
      dots reproduce the direct measurements of the diffusion coefficients
      of~BM17 performed from \Nbody\ simulations. The vertical dotted line
      indicates the location of the separatrix (i.e.\ ${ \kappa = 1 }$). }
  \end{center}
\end{figure}

Following the characterization of the correlation of the fluctuations in
Figure~\ref{fig:hmf_acf}, it is then straightforward to compute the associated
\DCs, as given by \eq~\eqref{eq:diff_coeff_final}. The \HMF\ \DCs\ are shown
in Figure~\ref{fig:hmf_DC} as a function of $\oJ$, as introduced in
\eq~\eqref{eq:Jbar}.
In this figure, we show different methods to compute the \DCs\@. The first
approach is based on the present $\eta$-formalism. As shown in
Figure~\ref{fig:hmf_acf}, it amounts to characterize the correlations of the
fluctuations in the system, over many bath realizations. Once these
perturbations are determined, the \DCs\ are immediately given by
\eq~\eqref{eq:D_HMF}. As already computed in~BM17, another approach is to
compute the system's dressed susceptibility coefficients,
${ \veps_{cc} (\omega) }$ and ${ \veps_{ss} (\omega) }$, from
\eq~\eqref{exp_veps_cc_ss}. Then, one may compute the correlation functions
from \eq~\eqref{eq:hmf_Chat}, which immediately give the \DCs\@. Finally, we
also reproduced in Figure~\ref{fig:hmf_DC} the data from~BM17, obtained by
measuring the \DCs\ directly from \Nbody\ simulations. As shown in
Figure~\ref{fig:hmf_DC}, all three approaches match, which illustrates the
versatility of the $\eta$-formalism. Let us emphasize that, contrary to the
analytical computation of the \DCs\ based on \eq~\eqref{eq:hmf_Chat}, the
$\eta$-formalism approach does not ask for the resolution of the resonant
condition ${ k \Omega(\kappa) - \kp \Omega(\kappap) = 0 }$, nor for the
computation of the dressed susceptibility coefficients
${ \veps_{cc} (\omega) }$ and ${ \veps_{ss} (\omega) }$ from
\eq~\eqref{exp_veps_cc_ss}, which can all prove to be cumbersome in generic
inhomogeneous self-gravitating systems. The only requirement of the
$\eta$-formalism is to be able to perform bath realizations, either non- or
self-interacting, from which the correlation of the potential perturbations can
be characterized.

To finish this section, let us briefly discuss some of the features of the
\HMF\ \DCs\ presented in Figure~\ref{fig:hmf_DC}. First, comparing the left and
right panels, one cannot that even if the \HMF\ pairwise interaction potential
from \eq~\eqref{eq:pot_HMF} is attractive, this does not necessarily translate
into the dressed (\BL) \DCs\ being larger than the bare
(Landau) ones. Here, for a \HMF\ model with a large magnetization, collective
effects tend to slow down the long-term diffusion. As shown in Figure~9 of~BM17,
should one consider a lower mean magnetization, this trend would invert and the
\BL\ \DCs\ would become larger than the Landau ones. In the case of razor-thin
cold stellar discs,~\cite{FouvryPichonMagorrianChavanis2015} similarly showed
how self-gravity could hasten the long-term diffusion by at least three other of
magnitudes.

In Figure~\ref{fig:hmf_DC}, surrounding the separatrix, one can also note the
presence of sharp peaks of enhanced diffusion. Such regions were not presented
in~BM17 and attributed to numerical instabilities of the computation near the
separatrix (B. Marcos, private communication). Here we show that these
peaks are not numerical artifacts, and are recovered in Figure~\ref{fig:hmf_DC}
by all three methods. Their origin may be understood as follows. First, we
note that these peaks are associated with orbits $\oJ$ with non vanishing
orbital frequencies ${ \Omega (\oJ) }$. In that region of orbital space, the
assumption of a dominating mean field motion in the Landau and \BL\ equations
therefore applies. Glancing back at the correlations represented in
Figure~\ref{fig:hmf_acf}, the origin of these peaks can be understood from the
$\eta$-formalism. Indeed, as given by \eq~\eqref{eq:Omega_HMF}, at the
separatrix, one has ${ \Omega (\kappa = 1) = 0 }$, and the frequency then grows
very rapidly away from the separatrix. As a result, when moving away from the
separatrix, the \DC\ ${ D (\oJ) \propto \wC (\Omega (\oJ)) }$ rapidly scans the
peak of the correlations represented in Figure~\ref{fig:hmf_acf}, leading to the
narrow peaks observed in Figure~\ref{fig:hmf_DC}. We finally note that the
present formalism cannot be naively applied at the separatrix, as particles
there have ${ \Omega (\oJ) \simeq 0 }$, so that the assumption of having a fast
orbital motion driven by the mean field potential does not apply
there. Characterizing the properties of the stochastic diffusion at the
separatrix requires therefore a more careful and elaborate study, which is
beyond the scope of this work.

\section{Conclusions}
\label{sec:conclusion}

A long-range interacting system in which the mean field potential is integrable
will evolve on the long-term due to the graininess of the potential associated with
the finite number of particles. For such a system in the statistical limit
${N \gg 1}$, the long-term evolution of the integrals of motion (conveniently
cast into actions $\bJ$) is stochastic and can be regarded as a diffusion
process.

It is then enlightening to consider the evolution of one subject particle whose
actions are undergoing a stochastic evolution sourced by the potential
fluctuations. This evolution can be described by a diffusion equation
containing a diffusion term and an advection (drift) term. The diffusion term
results from the stochastic forces induced by the potential fluctuations of the
\Nbody\ system of bath particles and is independent of the mass of the subject
particle. The advection term results from the response of the system to the
perturbation induced by the subject particle. As such,
this advection term results from the wake created by the subject
particle and gives rise to dynamical friction. Because the advection term is
proportional to the mass of the subject particle, it vanishes in the limit where
the mass of the subject particle is zero.

In this study, we presented a new approach, generalizing the so-called
$\eta$-formalism~\citep{BarOr2014}, to derive the diffusion equation of a test
particle forced by stochastic potential perturbations from bath
particles. Assuming that the mean potential is integrable and that the bath orbits
possess non-zero (and non-degenerate) orbital frequencies, the potential
fluctuations in the bath can be treated as a general correlated Gaussian noise.
In this work, we showed that the \DCs\ are proportional to the Fourier
transform of the temporal correlation (i.e.\ spectral density) of the noise,
evaluated at the test particle's local orbital frequency
${ \omega = \bk \cdot \bO (\bJ) }$ (see \eq~\eqref{eq:diff_coeff_final}). The
calculation of the \DCs\ is therefore reduced to the task of
obtaining the spectral density of the potential fluctuations generated by the
bath. In the absence of additional simplifying assumptions, this spectral
density can be obtained from numerical simulations.

One limit for which the correlation functions can be computed more easily is the
limit in which the bath particles are assumed to be non-interacting, so that
they are only driven by the mean field potential. This limit, which amounts to
neglecting collective effects, is valid in particular when the mean field
\DF\ satisfies
${ \partial F_{\rb} (\bJ) / \partial \bJ = 0 }$. This is for example relevant
to describe the scalar resonant relaxation of three-dimensional
isotropic quasi-Keplerian systems~\citep{BarOr2014}. We showed explicitly in
\eq~\eqref{eq:Final_Diff_Coeff_Landau}, that in the limit of a non-interacting
bath, the \DCs\ from the $\eta$-formalism are equivalent to the ones of the
inhomogeneous Landau equation.

In the more general case where the bath particles can interact one with
another, i.e.\ when accounting for collective effects, we showed in
\eq~\eqref{eq:Final_Diff_Coeff_BL} that the $\eta$-formalism allows for a
straightforward recovery of the inhomogeneous \BL\ \DCs\@. These coefficients
rely on the inversion of the self-consistency amplification relation in
\eq~\eqref{eq:selfconsistent_dressed_coefficients} defining the dressed
susceptibility coefficients.

In Section~\ref{sec:DynFric}, we lifted the assumption of treating the bath
as external, and recovered the friction force by polarization associated with
the back-reaction of the test particle on the bath particles' motion. This
allowed us to fully recover the Landau and \BL\ equations in Section~\ref{sec:BLeq}.
Finally, in Section~\ref{sec:dFPeq}, we also briefly illustrated how the $\eta$-formalism
allows for the recovery of the externally induced dressed \DCs\ previously
obtained in~\cite{BinneyLacey1988,Weinberg2001a}. These various diffusion
equations are now being increasingly used to investigate complex regimes of
long-term evolution of self-gravitating systems.

Finally, in order to emphasize the alternative point of view provided by the
$\eta$-formalism, we illustrated in Section~\ref{sec:HMF} how the
$\eta$-formalism allows for the computation of the \DCs\ in the one-dimensional
inhomogeneous \HMF\ model. In particular, we showed how the characterization of
the potential fluctuations makes it possible to determine the dressed
susceptibility coefficients without having to invert their self-consistency
definition from \eq~\eqref{eq:selfconsistent_dressed_coefficients} required in
the \BL\ calculation.

Together with the \BL\ equation, the $\eta$-formalism offers an equivalent but
alternative framework, in which it is now possible to tackle the question of
the long-term evolution of self-gravitating systems on a whole range of nested
astrophysical systems. As an example, such methods may be used to investigate
the long-term (scalar) resonant relaxation of galactic nuclei, generalizing the
previous descriptions of~\cite{BarOr2014, Bar-Or+2016}.

\section*{Acknowledgements}
We warmly thank Fernanda Benetti and Bruno Marcos for providing us with the
detailed data from their article. We are grateful to Pierre-Henri Chavanis,
Bruno Marcos, and Christophe Pichon for fruitful discussions. We also thank
Scott Tremaine for his useful comments on an earlier version of this work. JBF
acknowledges support from Program number HST-HF2--51374 which was provided by
NASA through a grant from the Space Telescope Science Institute, which is
operated by the Association of Universities for Research in Astronomy,
Incorporated, under NASA contract NAS5--26555. BB acknowledges support from the
Schmidt Fellowship and from the Institute for Advanced Study.

\bibliographystyle{mnras}
\bibliography{main}

\appendix

\section{Initial correlations of the fluctuations}
\label{sec:FlucDF}

Following~\cite{Chavanis2012}, let us briefly characterize the statistical
properties of the initial fluctuations in the \DF\ of a bath composed of $N$
identical particles. As introduced in \eq~\eqref{eq:decomposition_DF}, the
\DF's fluctuations are given by ${ \delta F = F_{\rd} - F_{\rb} }$, where
$F_{\rd}$ is the discrete \DF\ from \eq~\eqref{eq:properties_Fd} and $F_{\rb}$
is the smooth mean field \DF\ of the bath. To shorten the notations, we
temporarily drop the time dependence, ${ t = 0 }$. We can write
\begin{equation}
  \label{eq:Fluc_DF_I}
  \left\langle \delta F (\bT, \bJ) \, \delta F (\bTp, \bJp) \right\rangle = 
  \mb^{2} \sum_{i, j}^{N} \left\langle \deltaD  (\bT - \bT_{i}) \, 
    \deltaD  (\bJ - \bJ_{i}) \, \deltaD  (\bTp - \bT_{j}) \, 
    \deltaD  (\bJp - \bJ_{j}) \right\rangle - F_{\rb} (\bJ)  \, F_{\rb} (\bJp) ,
\end{equation}
where we relied on the fact that the fluctuations are of zero mean so that
${ \left\langle \delta F \right\rangle = 0 }$. One can straightforwardly
evaluate the first term from \eq~\eqref{eq:Fluc_DF_I}. It reads
\begin{align}
  \label{eq:Fluc_DF_II}
  \mb^{2} \sum_{i, j}^{N} \left\langle \deltaD  (\bT - \bT_{i}) \, \deltaD
  (\bJ - \bJ_{i}) \, \deltaD  (\bTp - \bT_{j}) \, \deltaD  (\bJp - \bJ_{j})
  \right\rangle = {}
  &
    \mb^{2} \sum_{i}^{N} \left\langle  \deltaD  (\bT - \bT_{i}) \, 
    \deltaD  (\bJ - \bJ_{i}) \, \deltaD  (\bT - \bTp) \, 
    \deltaD  (\bJ - \bJp) \right\rangle
    \nonumber \\
  & 
    +  \mb^{2} \sum_{i \neq j}^{N} \left\langle \deltaD  (\bT - \bT_{i}) \, 
    \deltaD  (\bJ - \bJ_{i}) \, \deltaD  (\bTp - \bT_{j}) \, 
    \deltaD  (\bJp - \bJ_{j}) \right\rangle
    \nonumber \\
  = {}
  & 
    \mb \, F_{\rb} (\bJ) \, \deltaD (\bT - \bTp) \, 
    \deltaD  (\bJ - \bJp) + F_{\rb} (\bJ) \, F_{\rb} (\bJp) ,
\end{align}
where to obtain the last line, we assumed that the particles were initially
uncorrelated and following \eq~\eqref{eq:decomposition_DF} used the relation
${ \left\langle F_{\rd} \right\rangle \!=\! F_{\rb} }$. Injecting
\eq~\eqref{eq:Fluc_DF_II} into \eq~\eqref{eq:Fluc_DF_I}, we immediately get
\begin{equation}
  \label{eq:Fluc_DF_time}
  \left\langle \delta F (\bT, \bJ) \, \delta F (\bTp, \bJp) \right\rangle = 
  \mb \, F_{\rb} (\bJ) \, \deltaD  (\bT - \bTp) \, \deltaD  (\bJ - \bJp) .
\end{equation}
When Fourier transformed w.r.t.\ the angles, one finally gets the needed
correlation of the initial fluctuations
\begin{equation}
  \label{eq:Fluc_DF_Fourier}
  \left\langle \delta F_{\bk} (\bJ, 0) \, \delta F_{\bkp} (\bJp, 0)  \right\rangle = 
  \frac{\mb}{{(2 \pi)}^{d}} \, \delta_{-\bk\bkp} \, \deltaD  (\bJ - \bJp) \, F_{\rb} (\bJ) .
\end{equation}

\section{The basis method}
\label{sec:BasisMethod}

In this Appendix, we briefly present the matrix method first introduced
in~\cite{Kalnajs1976II}, which allows for an easier inversion of
\eq~\eqref{eq:selfconsistent_psi_BL}, which defines implicitly the dressed
susceptibility coefficients. Let us introduce a representative biorthogonal
basis of potentials and densities ${ \psi^{(\alpha)} (\bx) }$ and ${ \rho^{(\alpha)} (\bx) }$
satisfying
\begin{equation}
  \label{eq:definition_basis}
  \psi^{(\alpha)} (\bx) = 
  \dint \bxp \, \psi (\bx, \bxp) \, \rho^{(\alpha)} (\bxp) \;\;\; ; 
  \;\;\; \dint \bx \, \psi^{(\alpha)} (\bx) \, \rho^{(\beta)}  (\bx) 
  = - \delta_{\alpha\beta} ,
\end{equation}
where once again, ${ \psi (\bx, \bxp) }$ is the pairwise interaction
potential, i.e.\ ${ \psi (\bx, \bxp) \!=\! - G / |\bx - \bxp| }$ in the
gravitational context. These basis elements can then be used to represent the
potential and density perturbations in the system. Following
\eq~\eqref{eq:psi_basis}, we define the Fourier transform of the potential
elements as
\begin{equation}
  \label{eq:FT_basis}
  \psi_{\bk}^{(\alpha)} (\bJ) = 
  \!\! \int \!\! \frac{\rd \bT}{{(2 \pi)}^{d}} \, 
  \psi^{(\alpha)} (\bx [\bT, \bJ]) \, \re^{- \ri \bk \cdot \bT} .
\end{equation}
These basis elements allow then for the explicit inversion of
\eq~\eqref{eq:selfconsistent_dressed_coefficients}, which defines
self-consistently the dressed susceptibility coefficients
${ \psi_{\bk\bkp}^{\rd} (\bJ, \bJp, \omega) }$. Indeed, one has (see
e.g.,~\cite{Chavanis2012})
\begin{equation}
  \label{eq:explicit_psid}
  \psi_{\bk\bkp}^{\rd} (\bJ, \bJp, \omega) = 
  - \sum_{\alpha, \beta} \psi_{\bk}^{(\alpha)} (\bJ) \, 
  {\big[ \bI - \wbM (\omega) \big]}_{\alpha\beta}^{-1} \, \psi_{\bkp}^{(\beta)} (\bJp) .
\end{equation}
In \eq~\eqref{eq:explicit_psid}, the sums over $\alpha$ and $\beta$ run over
all the considered basis elements.\footnote{ We also recall that the minus sign
  in \eq~\eqref{eq:explicit_psid} comes from our convention in
  \eq~\eqref{eq:selfconsistent_dressed_coefficients} for the definition of the
  dressed susceptibility coefficients
  ${ \psi_{\bk\bkp}^{\rd} (\bJ, \bJp, \omega) }$. This is the opposite sign
  from the convention used in~\cite{Heyvaerts2010,Chavanis2012}, but allows us
  to simply have
  ${ \lim_{\rm bare} \psi_{\bk\bkp}^{\rd} (\bJ, \bJp, \omega) =
    \psi_{\bk\bkp} (\bJ, \bJp) }$. } In this equation, $\bI$ is the identity
matrix, while ${ \wbM (\omega) }$ is the response matrix of the bath. It reads
\begin{equation}
  \label{eq:Fourier_M}
  \wbM_{\alpha\beta} (\omega) = 
{(2 \pi)}^{d} \sum_{\bk} \dint \bJ \, 
\frac{\bk \ccdot \partial F_{\rb} / \partial \bJ}{\omega - \bk \ccdot \bO (\bJ)} \, \psi_{\bk}^{(\alpha) *} (\bJ) \, \psi_{\bk}^{(\beta)} (\bJ) .
\end{equation}
In the limit where collective effects are not accounted for,
\eq~\eqref{eq:explicit_psid} recovers the expression of the bare susceptibility
coefficients reading
\begin{equation}
  \label{eq:explicit_psi}
  \psi_{\bk\bkp} (\bJ, \bJp) = - \sum_{\alpha} \psi_{\bk}^{(\alpha)} (\bJ) \, \psi_{\bkp}^{(\alpha) *} (\bJp) .
\end{equation}
The response matrix $\wbM$ is an essential dynamical quantity that
characterizes the properties of the self-gravitating amplification of
perturbations in the bath. As an example, the bath is linearly unstable if
there exists $\omega$ with ${ \text{Im} [\omega] > 0 }$, such that
${ \det [ \bI - \wbM (\omega) ] = 0 }$. One should note from
\eq~\eqref{eq:Fourier_M}, that the response matrix is a global mean field
quantity, in the sense that it only depends on the mean field properties of the
bath (via~$F_{\rb}$ and ${ \bO (\bJ) }$) and involves an integration over all
action space. As such, numerical calculations of unstable modes of
self-gravitating systems are cumbersome tasks. In astrophysics, it has only
been made for a small number of razor-thin disks~\citep[see
e.g.,][]{Zang1976,Kalnajs1977,VauterinDejonghe1996,PichonCannon1997,EvansRead1998II,JalaliHunter2005,Polyachenko2005,Jalali2007,Jalali2010,FouvryPichonMagorrianChavanis2015,DeRijcke2016}
or three-dimensional spherical clusters~\citep[see
e.g.,][]{PolyachenkoShukhman1981,Saha1991,Weinberg1991}.

\section{Angle-action coordinates for the HMF model}
\label{sec:AA_HMF}

In this Appendix, we follow~\cite{Barre2010,BenettiMarcos2017} and present
angle-action coordinates for the inhomogeneous \HMF\ model considered in
Section~\ref{sec:HMF}. Let us first introduce the quantity $\kappa$ conserved
for the mean field dynamics
\begin{equation}
  \label{eq:kappa_def}
  \kappa = \sqrt{\frac{H_\rt \!+\! M_{0}}{2 M_{0}}} ,
\end{equation}
where the one-particle Hamiltonian, $H_\rt$, stands for the specific energy of
the particle, and was introduced in \eq~\eqref{eq:H_test_HMF}. The quantity
$\kappa$ is a useful parameter to separate the librating particles from the
circulating ones. The separatrix lies in ${ \kappa = 1 }$, so that particles
with ${ \kappa < 1 }$ are trapped and librate, while particles with ${ \kappa > 1 }$
are circulating. The action $J$ may then be defined as
\begin{equation}
  \label{eq:J_def_HMF}
  J (\kappa) = \frac{4 \sqrt{M_{0}}}{\pi}
  \begin{cases}
    \displaystyle 2 [  E (\kappa^{2}) - (1 \!-\! \kappa^{2}) K (\kappa^{2})]
    \;\;\; 
    \text{if} \;\;\; \kappa < 1,
    \\
    \displaystyle \kappa E (1 / \kappa^{2}) \;\;\; 
    \text{if} \;\;\; \kappa > 1 .
  \end{cases}
\end{equation}
In \eq~\eqref{eq:J_def_HMF}, we introduced the complete elliptic integrals of
the first and second kind, ${ K (m) }$ and ${ E (m) }$, with the convention
\begin{equation}
  \label{eq:definition_K_E}
  K (m) = \!\! \int_{0}^{\pi / 2} \!\!\!\! \rd x \, 
  {\big[ 1 \!-\! m \sin^{2} (x) \big]}^{-1/2} 
  \;\;\; ; \;\;\; 
  E (m) = \!\! \int_{0}^{\pi / 2} \!\!\!\! \rd x \, 
  {\big[ 1 \!-\! m \sin^{2} (x) \big]}^{1/2} .
\end{equation}
In \eq~\eqref{eq:J_def_HMF}, one should pay attention to the fact that the
action $J$ is discontinuous at the separatrix ${ \kappa \!=\! 1 }$. To avoid
this issue, we introduce a rescaled action $\oJ$ as
\begin{equation}
  \label{eq:Jbar}
  \oJ = 
  \begin{cases}
    J / 2 & \kappa < 1,
    \\
    J & \kappa > 1 .
\end{cases}
\end{equation}
Finally, the orbital frequency
${ \Omega (\kappa) }$ is given by
\begin{equation}
  \label{eq:Omega_HMF}
  \Omega (\kappa) = \pi \sqrt{M_{0}}
  \begin{cases}
    \displaystyle 1 / (2 K (\kappa^{2})) \;\;\; \text{if} \;\;\; \kappa < 1,
    \\
    \displaystyle \kappa / (K (1 / \kappa^{2})) \;\;\; \text{if} \;\;\; \kappa > 1 .
  \end{cases}
\end{equation}

In \eq~\eqref{eq:feq}, we assumed that the quasi-stationary \DF\ of the bath
takes the form of a thermal \DF\ that can be written as
\begin{equation}
  \label{eq:thermal_DF_HMF}
  F_{0} (\kappa) = 
  C \, \exp \big[ - \beta M_{0} (2 \kappa^{2} \!-\! 1) \big] \;\;\; 
  \text{with} \;\;\; 
  C = \sqrt{\frac{\beta}{{(2 \pi)}^3}} \frac{1}{I_{0} (\beta M_{0})} ,
\end{equation}
In \eq~\eqref{eq:thermal_DF_HMF}, the \DF\ satisfies the normalization
convention ${ \dint \theta \rd J F_{0} (J) \!=\! M \!=\! 1 }$. We
also introduced the inverse temperature $\beta$ which is determined
self-consistently from the mean magnetization $M_{0}$ by imposing
${ \{ \cos (\phi) \} \!=\! M_{0} }$, so that one has
\begin{equation}
  \label{eq:link_M0_beta}
  M_{0} = \frac{I_{1} (\beta M_{0})}{I_{0} (\beta M_{0})} .
\end{equation}

The explicit angle-action coordinates from \eq~\eqref{eq:J_def_HMF} allow us to
perform the needed Fourier transforms w.r.t.\ the angle $\theta$.
Following~\cite{BenettiMarcos2017}, these are characterized by two quantities
${ c_{k} (\kappa) }$ and ${ s_{k} (\kappa) }$ defined as
\begin{equation}
  \label{eq:definition_ck_sk_HMF}
  c_{k} (\kappa) = \!\! \int_{- \pi}^{\pi} \! \frac{\rd \theta}{2 \pi} \, 
  \cos [\phi (\theta, \kappa)] \, \re^{- \ri k \theta} \;\;\; ; \;\;\; s_{k} (\kappa) = 
  \!\! \int_{- \pi}^{\pi} \! \frac{\rd \theta}{2 \pi} \, 
  \sin [\phi (\theta, \kappa)] \, \re^{- \ri k \theta} .
\end{equation}
Fortunately, these integrations can be performed explicitly and one gets
\begin{equation}
  \label{eq:explicit_ck_sk_HMF}
  c_{k} (\kappa) =
  \begin{cases}
    \begin{aligned}
      & \frac{\pi^2}{K^2 (\kappa^{2})} \frac{|k| \, {q (\kappa^2)}^{|k|/2}}{1
        \!-\! {q (\kappa^2)}^{|k|}} && \kappa < 1, \; k \; \text{even},
      \\
      & 0 && \kappa < 1, \; k \; \text{odd},
      \\
      & \frac{2 \pi^2 \kappa^2}{K^2(1 / \kappa^2)} \frac{|k| \, 
      {q(1 /\kappa^2)}^{|k|}}{1 - {q (1/ \kappa^2)}^{2 |k|}} && \kappa > 1 ,
  \end{aligned}
\end{cases}
\!\!\! ; \;
s_{k} (\kappa) =
\begin{cases}
  \begin{aligned}
    & 0 && \kappa < 1, \; k \; \text{even},
    \\
    & - \ri \frac{\pi^2}{K^2 (\kappa^2)} \frac{k \, {q (\kappa^2)}^{|k|/2}}{1
      + {q (\kappa^2)}^{|k|}} && \kappa < 1, \; k \; \text{odd},
    \\
    & - \ri \frac{2 \pi^{2} \kappa^{2}}{K^2 (1 / \kappa^{2})} \frac{k \, {q (1
      / \kappa^2)}^{|k|}}{1 + {q (1 / \kappa^2)}^{2 |k|}} && \kappa > 1, \; v > 0
   ,
    \\
    & \ri \frac{2 \pi^2 \kappa^2}{K^2 (1 / \kappa^{2})} \frac{k \, {q (1 /
      \kappa^2)}^{|k|}}{1 + {q (1 / \kappa^2)}^{2 |k|}} && \kappa > 1, \; v < 0,
  \end{aligned}
\end{cases}
\end{equation}
where ${ q (m) \!=\! \exp [ - \pi K (1 \!-\! m) / K (m) ] }$ is the elliptic nome.
The coefficients ${ c_{k} (J) }$ and ${ s_{k} (J) }$ are the ones appearing
in \eq~\eqref{eq:eta_k_jmf} to describe the noise induced by the bath.

Finally, following~\cite{Brizard2013}, one can integrate explicitly the motion
of a given particle when driven by the mean field HMF potential. Such explicit
time integrations are used in Section~\ref{sec:HMF} to characterize the
fluctuations of the magnetization in a non-interacting HMF bath. Let us first
consider a trapped particle, i.e.\ for which ${ \kappa < 1 }$, characterized by
the initial coordinates ${ (\phi_{0}, v_{0}) }$. We introduce
$\varepsilon_{0}$ as
\begin{equation}
  \label{eq:eps_HMF}
  \varepsilon_{0} = \text{sign} [ v_{0} ],
\end{equation}
and the phase of the pendulum ${ \phi_{\rP} }$ as
\begin{equation}
  \label{eq:phiP_HMF}
  \phi_{\rP} = \varepsilon_{0} \, \text{sn}^{-1} \big[ \sin (\phi_{0}/2) / \kappa, \kappa^{2} \big] .
\end{equation}
In \eq~\eqref{eq:phiP_HMF}, we defined the inverse Jacobi elliptic function
${ \text{sn}^{-1} (u, m) }$ with the convention
\begin{equation}
  \label{eq:convention_elliptic}
  \text{sn} [ u \,, \, m ] = \sin ( \varphi ) \;\;\; \text{with} \;\;\; \varphi = 
  \text{am} [ u \,, \, m ] \;\; ; \;\; u = F [ \varphi \,, \, m ] = 
  \dint[0][\varphi] \vartheta \, {\big[ 1 \!-\! m \sin^{2} (\vartheta) \big]}^{-1/2} .
\end{equation}
The motion of a librating particle can then be explicitly integrated in time
and reads
\begin{equation}
  \label{eq:explicit_HMF_trapped}
  \phi (t) = 
  \varepsilon_{0} \, 2 \, \text{Arcsin} \big[ \kappa \, 
  \text{sn} \big[ \phi_{\rP} + t \sqrt{M_{0}} \,, \, \kappa^{2} \big] \big] .
\end{equation}
For a circulating orbit, i.e.\ with ${ \kappa > 1 }$, similarly to
\eqs~\eqref{eq:eps_HMF} and~\eqref{eq:phiP_HMF}, we introduce
${ \varepsilon_{0} }$ and ${ \phi_{\rP} }$ as
\begin{equation}
  \label{eq:eps_phiP_circ_HMF}
  \varepsilon_{0} = \text{sign} [ v_{0} ] \;\;\; ; 
  \;\;\; \phi_{\rP} = \varepsilon_{0} \, 
  \text{sn}^{-1} \big[ \sin (\phi_{0}/2), 1/\kappa^{2} \big] / \kappa .
\end{equation}
We also introduce the times $t_{0}$ and the period ${ \Delta t }$ as
\begin{equation}
  \label{eq:t0_Deltat_HMF}
  t_{0} = \frac{1}{\sqrt{M_{0}}} \bigg( \frac{\text{sn}^{-1} 
    \big[ 1, 1/ \kappa^{2} \big]}{\kappa} - \phi_{\rP} \bigg) \;\;\; ; 
  \;\;\; 
  \Delta t = \frac{2 \, K [1 / \kappa^{2}]}{\sqrt{M_{0}} \kappa} ,
\end{equation}
where, following \eq~\eqref{eq:convention_elliptic}, we introduced the complete
elliptic function ${ K [m] = F [\pi / 2, m ] }$. The motion of the circulating
particle can then be explicitly integrated in time and reads
\begin{equation}
  \label{eq:explicit_HMF_circulating}
  \phi (t) =
  \begin{cases}
    \displaystyle \varepsilon_{0} \, 2 \, \text{Arcsin} \big[ \text{sn} \big[
    \kappa (\phi_{\rP} + t \sqrt{M_{0}}), 1 / \kappa^{2} \big] \big] \;\;\;
    \text{if} \;\;\; \text{Mod} \big[ t - t_{0}, 2 \Delta t \big] - \Delta t
    \leq 0,
    \\
    - \varepsilon_{0} \, 2 \, \text{Arcsin} \big[ \text{sn} \big[ \kappa
    (\phi_{\rP} + t \sqrt{M_{0}}), 1 / \kappa^{2} \big] \big] \;\;\;
    \text{otherwise}, \displaystyle
  \end{cases}
\end{equation}
with ${ \text{Mod} [x, y] \!=\! y \, \text{frac} (x / y) }$. The explicit
expressions from \eqs~\eqref{eq:explicit_HMF_trapped}
and~\eqref{eq:explicit_HMF_circulating} allow us to easily perform realizations
of non-interacting HMF bath, as presented in Section~\ref{sec:HMF}.

\end{document}